\documentclass[prd,showpacs,superscriptaddress,nofootinbib,floatfix,showkeys,10pt]{revtex4-2}
\usepackage{bm}
\usepackage{times}
\usepackage{braket}
\usepackage{amsfonts,amssymb,stmaryrd,latexsym,amsmath}
\usepackage[usenames,dvipsnames]{color}
\usepackage{epsfig}
\usepackage{slashed}
\usepackage{hyperref}
\usepackage{subfigure}
\usepackage{graphics}
\usepackage{slashed}
\usepackage{orcidlink}
\usepackage{multirow}

\allowdisplaybreaks[1]

\definecolor{nicered}{rgb}{0.7,0.1,0.1}
\definecolor{nicegreen}{rgb}{0.1,0.5,0.1}
\definecolor{emph}{rgb}{1,0,0}
\definecolor{doub}{rgb}{0.7,0.2,1.0}
\definecolor{navyblue}{RGB}{0, 110, 184}

\hypersetup{colorlinks,citecolor=nicegreen,linkcolor=nicered,urlcolor=navyblue}

\begin{document}
	
\title{Electromagnetic polarizabilities of the spin-$\frac{3}{2}$  baryons in heavy baryon chiral perturbation theory} 

\author{Liang-Zhen Wen\,\orcidlink{0009-0006-8266-5840}}
\affiliation{School of Physics, Peking University, Beijing 100871, China}

\author{Yan-Ke Chen\,\orcidlink{0000-0002-9984-163X}}
\email{chenyanke@stu.pku.edu.cn}
\affiliation{School of Physics, Peking University, Beijing 100871, China}

\author{Lu Meng\,\orcidlink{0000-0001-9791-7138}}\email{lmeng@seu.edu.cn}
\affiliation{School of Physics, Southeast University, Nanjing 210094, China}
\affiliation{Institut f\"ur Theoretische Physik II, Ruhr-Universit\"at Bochum,  D-44780 Bochum, Germany }

\author{Shi-Lin Zhu\,\orcidlink{0000-0002-4055-6906}}\email{zhusl@pku.edu.cn}
\affiliation{School of Physics and Center of High Energy Physics,
Peking University, Beijing 100871, China}

\begin{abstract}
We employ Heavy Baryon Chiral Perturbation Theory (HB$\chi$PT), a non-relativistic effective field theory that treats baryons as heavy static sources, to calculate the electromagnetic polarizabilities of spin-3/2 baryons in two sectors: the light-flavor decuplet baryons and singly heavy sextet baryons. We derive the analytical expressions up to $\mathcal{O}\left(p^3\right)$. Our results indicate that the long-range chiral corrections provide substantial contributions to the polarizabilities. In addition, magnetic dipole (M1) transitions of the baryons can significantly affect the magnetic polarizabilities and may even reverse their signs. For the decuplet baryons, the $\Delta^+$ and $\Delta^0$ exhibit the largest electric polarizabilities. Their values, $\alpha_E(\Delta^+) = (17.5 \pm 9.5)\times 10^{-4} \, \mathrm{fm}^3$ and $\alpha_E(\Delta^0) = (17.0 \pm 9.3)\times 10^{-4} \, \mathrm{fm}^3$, significantly exceed those typically observed for nucleons. Meanwhile, the electric polarizabilities of spin-3/2 singly heavy baryons are comparable to those of their spin-1/2 partners.

\end{abstract}
 
\maketitle


\section{Introduction}


Revealing the internal structure of hadrons is a central goal in hadron physics, closely related to the non-perturbative dynamics of Quantum Chromodynamics (QCD), such as color confinement and chiral symmetry breaking. Electromagnetic processes offer a powerful probe due to the well-understood nature of electromagnetic interactions. For example, the electromagnetic form factors (EMFFs) of hadrons can characterize their internal charge and magnetization distributions. The concept of nucleon EMFFs was initially introduced through the Rosenbluth formula in electron-nucleon scattering~\cite{Rosenbluth:1950yq}. Since then, the proton charge radius has been determined using various techniques, including spectroscopy of the 2S-2P transition in hydrogen~\cite{Bezginov:2019mdi}, electron-proton scattering~\cite{Xiong:2019umf}, and measurements of the Lamb shift in muonic hydrogen~\cite{Antognini:2013txn}. Furthermore, the proton magnetic moment has been precisely measured using the double–Penning trap technique~\cite{Schneider:2017lff}, and its magnetic radius has been extracted from high-precision electron–proton scattering data~\cite{Lin:2021umk}.

The $\Delta(1232)$, with spin-parity $J^P = \frac{3}{2}^+$, is the lowest-lying excited state of the nucleon and plays a prominent role in the physics of strong interactions~\cite{Pascalutsa:2006up}. On the theoretical side, a wide range of approaches has been employed to study the EMFFs of $\Delta(1232)$ and other decuplet baryons. These include quark model (QM)~\cite{Krivoruchenko:1991pm,Schlumpf:1993rm,Buchmann:1996bd,Berger:2004yi,Ramalho:2008dc,Kim:1997ip,Ledwig:2008es,Fu:2022rkn,Wang:2023bjp}, the bag model~\cite{Kaelbermann:1983zb,Lu:1996rj}, the large $N_c$ expansion~\cite{Luty:1994ub,Jenkins:1994md}, the Skyrme model~\cite{Oh:1995hn}, QCD sum rules~\cite{Lee:1997jk,Aliev:2000rc,Aliev:2009pd}, the general parametrization method~\cite{Buchmann:2002xq}, the chiral perturbation theory ($\chi$PT)~\cite{Butler:1993ej,Banerjee:1995wz,Arndt:2003we,Pascalutsa:2004je,Hacker:2006gu,Tiburzi:2009yd,Geng:2009ys,Li:2016ezv}, and lattice QCD simulations~\cite{Nozawa:1990gt,Leinweber:1992hy,Cloet:2003jm,Lee:2005ds,Alexandrou:2010jv,Boinepalli:2009sq,Aubin:2008qp}. Experimentally, the magnetic moments of the $\Delta^{++}$ and $\Delta^{+}$ have been measured via pion-induced ($\pi^+p \to \pi^+p\gamma$) and photon-induced ($\gamma p \to \pi^0p\gamma'$) processes, respectively. The latest experimental results yield $\mu_{\Delta^+} = (2.7_{-1.3}^{+1.0} \pm 1.5 \pm 3)\,\mu_N$~\cite{Kotulla:2002cg} and $\mu_{\Delta^{++}} = (4.52 \pm 0.50 \pm 0.45)\,\mu_N$~\cite{Bosshard:1991zp}. Furthermore, the recent review~\cite{Endrodi:2024cqn} provides a comprehensive overview of the electromagnetic properties of light hadrons, while Refs.~\cite{Can:2021ehb,Meng:2022ozq} focus on recent progress in understanding the EMFFs of heavy baryons.

Another class of important electromagnetic properties is the electric and magnetic polarizabilities, denoted as $\alpha_E$ and $\beta_M$, respectively. These quantities characterize the response of a physical system to external electromagnetic fields. Classically, $\alpha_E$ and $\beta_M$ describe the induced electric and magnetic dipole moments in the presence of constant external electric and magnetic fields:
\begin{equation}
\vec{d}= 4\pi \alpha_E \vec{E}, \qquad \vec{m}=4\pi \beta_M \vec{B}.
\end{equation}
The interaction between matters and external electromagnetic field give rise to a potential energy,
\begin{equation}
U_E=-\frac{1}{2} 4 \pi \alpha_E \boldsymbol{E}^2, \quad U_H=-\frac{1}{2} 4 \pi \beta_M \boldsymbol{H}^2.
\end{equation}
For the hadronic systems, electromagnetic polarizabilities can be defined through the second-order effective Hamiltonian, 
\begin{equation}
H^{(2)}= -\frac{1}{2}4\pi\alpha_E \vec{E}^2 -\frac{1}{2} 4\pi\beta_M \vec{B}^2,
\end{equation}
and can be extracted from the Compton scattering amplitude. A general expression for the leading-order electromagnetic polarizabilities of a particle in the state $|0\rangle$ is
\begin{equation}
\begin{aligned}
\bar{\alpha}_E=  2 \alpha_{em}\sum_{n \neq 0} \frac{\left.\left|\langle n| D_z\right| 0\right\rangle\left.\right|^ 2}{E_n-E_0}, \qquad  
\bar{\beta}_M=   2 \alpha_{em}\sum_{n \neq 0} \frac{\left.\left|\langle n| M_z\right| 0\right\rangle\left.\right|^ 2}{E_n-E_0}.
\end{aligned}
\end{equation}
While $\left\langle n|D_z|0 \right\rangle$ and $\left\langle n|M_z|0 \right\rangle$ correspond to the electric dipole and magnetic dipole transitions, respectively. It is evident that near-degenerate energy levels can lead to large polarizabilities, which formally diverge in the case of exact degeneracy.


The lightest pseudoscalar mesons are regarded as Goldstone bosons associated with spontaneous breaking of chiral symmetry. In the chiral limit, Goldstone bosons are strictly massless. A single-particle state can become nearly degenerate with states containing additional pions, leading to strong pion cloud fluctuations in the vicinity of matter fields. Therefore, the  chiral dynamics is expected to contribute significantly to the electromagnetic polarizabilities of matter fields. Chiral perturbation theory ($\chi$PT) provides a powerful, model-independent framework for studying these effects. It is built on a systematic power counting scheme based on small momenta or Goldstone boson masses. 

When baryons are included in $\chi$PT, the baryon mass scale $M_B$, which remains non-zero in the chiral limit, introduces power counting breaking (PCB). Several approaches have been developed to resolve PCB, including heavy baryon chiral perturbation theory (HB$\chi$ PT)~\cite{Jenkins:1990jv,Hemmert:1997ye}, infrared regularization~\cite{Becher:1999he}, and Extended On-Mass-Shell schemes~\cite{Gegelia:1999gf,Fuchs:2003qc}. Among these, HB$\chi$PT provides a valid description of the physics at momenta much smaller than the baryon mass $M_B$ and $\Lambda_\mathrm{QCD}$. In this framework, the baryon fields are decomposed as follows:
\begin{equation}\label{eq:heavy_baryon_field}
	\mathcal{B}_n(x)=e^{i M_{\mathcal{B}_n} v \cdot x} \frac{1+\slashed v}{2} \psi_n, \quad \mathcal{H}_n(x)=e^{i M_{\mathcal{B}_n} v \cdot x} \frac{1-\slashed v}{2} \psi_n,
\end{equation}
where $\mathcal{B}_n$ ($\mathcal{H}_n$) represents the ``light'' (``heavy'') component of the corresponding heavy baryon field. Here, $M_{\mathcal{B}_n}$ denotes the baryon mass, and $v^\mu=(1, \bm{0})$ represents the static velocity. The heavy field $\mathcal{H}_n(x)$ corresponds to excitations with a mass of $2M_B$, representing the creation of a heavy baryon-antibaryon pair. $\mathcal{H}_n(x)$ is subsequently integrated out of the baryon chiral perturbation theory (B$\chi$PT) Lagrangian to construct the HB$\chi$PT Lagrangian.

For the nucleon polarizabilities, Bernard, Kaiser, Schmidt, and Mei\ss ner (BKSM) performed calculations based on HB$\chi$PT~\cite{Bernard:1991rq,Bernard:1991ru,Bernard:1993bg,Bernard:1993ry}. They provided results at both $\mathcal{O}(p^3)$ and $\mathcal{O}(p^4)$ orders, which quantitatively agree with experimental data and recent Lattice QCD results~\cite{Wang:2023omf}. In their studies, the pion cloud effect dominates. Other investigations of nucleon polarizabilities have also been conducted using HB$\chi$PT~\cite{Butler:1992ci,Babusci:1996jr,Beane:2004ra,Choudhury:2007qiz} and covariant B$\chi$PT~\cite{Lensky:2014efa,Lensky:2015awa,Thurmann:2020mog}. However, research on electromagnetic polarizabilities of spin-3/2 baryons and charmed(bottom) baryons remains limited. Although the short lifetimes of excited baryons make direct measurements of their electromagnetic polarizabilities challenging, we expect that Lattice QCD will provide reliable estimates in the future. 
In our previous work~\cite{Chen:2024xks}, we first calculate electromagnetic polarizabilities for spin-1/2 singly heavy baryons with HB$\chi$PT. We now systematically extend this approach to spin-3/2 baryons.

In this work, we investigate the spin-averaged Compton scattering of $J^{P}=\frac{3}{2}^+$ baryons using HB$\chi$PT up to $\mathcal{O}(p^3)$. 
We analyze the electromagnetic polarizabilities in two sectors: light-flavor decuplet baryons and singly heavy sextet baryons.

The paper is arranged as follows: we present the general theoretical framework for electromagnetic polarizabilities in Compton scattering in Sec.~\ref{sec:theoretical_framework}, including the derivation of electromagnetic polarizabilities from the Compton amplitude in Sec.~\ref{subsec:spin-averaged_forward_Compton_ tensor} and the effective Lagrangian used in Sec.~\ref{subsec:effective_Lagrangian}. We then provide the analytical and numerical results for decuplet baryons in Sec.~\ref{sec:T}, followed by the analytical and numerical results for singly charmed baryons in Sec.~\ref{sec:B}. In Sec.~\ref{sec:DISCUSSION_AND_CONCULSION}, we offer a brief summary. The results for bottom baryons are listed in Appendix~\ref{appendix:b_baryon_polarizabilities}.

\section{theoretical framework}\label{sec:theoretical_framework}

\subsection{Spin-averaged forward Compton scattering tensor}\label{subsec:spin-averaged_forward_Compton_ tensor}

The electromagnetic polarizabilities are determined by the spin-averaged forward Compton scattering tensor $\Theta_{\mu \nu}$~\cite{Bernard:1991ru,Bernard:1991rq}. For a spin-3/2 baryon, the spin-averaged forward Compton scattering tensor is  
\begin{equation}  
	\Theta^{3/2}_{\mu \nu}=-\frac{e^2}{4} \operatorname{Tr}\left[(\slashed p + m) P_{\beta \alpha}^{3/2} T^{\alpha} {}_{ \mu \nu}{}^{\beta}(p, k)\right],  
\end{equation}  
where $k$ and $p$ denote the momentum of the photon and baryon, respectively. $P_{\beta \alpha}^{3/2}$ represents the on-mass-shell spin-3/2 projection, which arises from the spin averaging of the wave function of the Rarita–Schwinger field $u_\beta^s$ and $\bar{u}_\alpha^s$~\cite{Rarita:1941mf},
\begin{equation}  
    \sum_s u_\beta^s \bar{u}_\alpha^s = -(\slashed p + m) P_{\beta \alpha}^{3/2}.
\end{equation}  
The term $T^{\alpha} {}_{ \mu \nu}{}^{\beta}(p, k)$ corresponds to the Fourier-transformed matrix element of two time-ordered electromagnetic currents for the spin-3/2 baryons $\psi^*$,
\begin{equation}
\bar{u}_{\alpha}(p) T^{\alpha} {}_{ \mu \nu}{}^{\beta}(p, k) u_{\beta}(p)=\int d^4 x e^{i k \cdot x}\left\langle \psi^*(p)\left|T\left[J_\mu^{\mathrm{e m}}(x) J_\nu^{\mathrm{e m}}(x)\right]\right| \psi^*(p)\right\rangle.
\end{equation}
The forward spin-averaged Compton tensor $\Theta_{\mu \nu}$ can be expanded in terms of a few Lorentz structures. In the heavy baryon formalism, 
\begin{equation}
\begin{aligned}
\Theta_{\mu \nu}^{3/2} &= -\frac{e^2}{8} \operatorname{Tr}\left[(1+ \slashed v) P_{\beta \alpha}^{3/2} T^{\alpha} {}_{ \mu \nu}{}^{\beta}(v, k)\right] \\
&=U(\omega) g_{\mu \nu}+V(\omega) k_\mu k_\nu+W(\omega)\left(k_\mu v_\nu+v_\mu k_\nu\right)+X(\omega) v_\mu v_\nu,
\end{aligned}
\end{equation}
where $\omega=v \cdot k$ is the energy of photon. We adopt the ``Coulomb gauge" in this work, where $\epsilon \cdot v = 0$ for the photon polarization vector $\epsilon$. Then the auxiliary function $\epsilon^{\prime \mu} \Theta_{\mu \nu} \epsilon^\nu$ reads:
\begin{equation}\label{eq:4-dimensional compton}
	\epsilon^{\prime \mu} \Theta_{\mu \nu} \epsilon^\nu=\left(\epsilon^{\prime} \cdot \epsilon\right) U(\omega)+\left(\epsilon^{\prime} \cdot k \epsilon \cdot k \right)V(\omega).
\end{equation}
The spin-averaged forward Compton amplitude is associated with two form factors, $U(\omega)$ and $V(\omega)$. In Appendix~\ref{appendix:General_Spin-averaged_Compton_Tensor}, we present the general form of the spin-averaged Compton tensor. The factors $U(\omega)$ and $V(\omega)$ correspond to $B_1\left(0, M_B \omega\right)$ and $B_2\left(0, M_B \omega\right)$ in Eq.~\eqref{eq:general compton}, respectively. The general expression for the Compton scattering amplitude in three dimensions is given by~\cite{Hemmert:1996rw}:
\begin{equation}\label{eq:3-dimensional compton}
\operatorname{Amp}=  \hat{\epsilon} \cdot \hat{\epsilon}^{\prime}\left(-\frac{Q^2}{M}+\omega \omega^{\prime} 4 \pi \alpha_E\right)+\hat{\epsilon} \times \vec{k} \cdot \hat{\epsilon}^{\prime} \times \vec{k}^{\prime} 4 \pi \beta_M +O\left(\omega^4\right),
\end{equation}
where $\hat{\epsilon}$ and $\hat{\epsilon}^{\prime}$ are the three-dimensional polarization vectors of the initial and final photons, and $\vec{k}$ and $\vec{k}^{\prime}$ denote their respective momenta. By setting $\vec{k}=\vec{k'}$ in Eq.~\eqref{eq:3-dimensional compton} and comparing it with Eq.~\eqref{eq:4-dimensional compton}, the electric and magnetic polarizabilities are derived as~\cite{Bernabeu:1976jq,Llanta:1979kj,Bernard:1993bg,Bernard:1993ry}:
\begin{equation}\label{eq:def_alpha_beta}
\alpha_E+\beta_M  =-\left.\frac{1}{8 \pi} \frac{\partial^2}{\partial \omega^2} U(\omega)\right|_{\omega=0},\qquad
\beta_M  =-\frac{1}{4 \pi} V(\omega=0).
\end{equation}
The subsequent task is to calculate all contributions to $U(\omega)$ and $V(\omega)$ up to $\mathcal{O}(p^3)$ within HB$\chi$PT.

\subsection{Effective Lagrangians}\label{subsec:effective_Lagrangian}

The chiral symmetry is realized through a nonlinear representation,
\begin{equation}\label{eq:nonlinear_realization}
	U=\exp \left(i \frac{\phi}{F_0}\right),
\end{equation}
where $\phi$ is the matrix for octet Goldstone bosons,
\begin{equation}\label{eq:octet_meson}
\phi=\sum_{a=1}^{8}\lambda_a\phi_a=\left(\begin{array}{ccc}
\pi^0+\frac{1}{\sqrt{3}} \eta & \sqrt{2} \pi^{+} & \sqrt{2} K^{+} \\
\sqrt{2} \pi^{-} & -\pi^0+\frac{1}{\sqrt{3}} \eta & \sqrt{2} K^0 \\
\sqrt{2} K^{-} & \sqrt{2} \bar{K}^0 & -\frac{2}{\sqrt{3}} \eta
\end{array}\right).
\end{equation}
$F_0$ is the decay constant of the pseudoscalar meson in chiral limit. We adopt $F_\pi=92.4~ \mathrm{MeV}, F_K=113~\mathrm{MeV}$ and $F_\eta=116~\mathrm{MeV}$ in this work.
Under the $\mathrm{SU}(3)_L \times \mathrm{SU}(3)_R$ chiral transformation, $U$ is transformed as
\begin{equation}\label{eq:transformation1}
U  \rightarrow R U L^{\dagger}, 
\end{equation}
where $R$ and $L$ are $\mathrm{SU}(3)_R$ and $\mathrm{SU}(3)_L$ transformation matrices, respectively.

When external fields are introduced, the effective Lagrangian is constructed by promoting its global symmetries to local ones and introducing couplings to external fields in the same way as in QCD. In this framework, the electromagnetic fields are introduced as left- and right-handed external fields:
\begin{equation}
	r_\mu  =l_\mu =-e Q_{m(\mathcal{B})} A_\mu.
\end{equation}
where $A_\mu$ is the electromagnetic field and $Q_{m(\mathcal{B})}$ represents the charge matrix. $Q_m = \operatorname{diag}\left(\frac{2}{3}, -\frac{1}{3}, -\frac{1}{3}\right)$ and $Q_{\mathcal{B}} = \operatorname{diag}(1, 0, 0)$ represent the charge matrices for light Goldstone mesons and singly charmed baryons, respectively. The covariant derivatives of Goldstone boson fields are then defined as,
\begin{equation}
    \nabla_\mu U=\partial_\mu U-i r_\mu U+i U l_\mu,
\end{equation}

In order to introduce the chiral symmetry breaking effect, we define $\chi$,
\begin{equation}
\chi=2 B_0 \operatorname{diag}\left(m_u, m_d, m_s\right).
\end{equation}
where $B_0$ is a parameter related to the quark condensate and $m_{u, d, s}$ is the current quark mass.

The leading order (LO) chirally and gauge invariant pure-meson Lagrangian is  
\begin{equation}
	\mathcal{L}_{\phi \phi}^{(2)}=\frac{F_0^2}{4}\left\langle\nabla_\mu U\left(\nabla^\mu U\right)^{\dagger}\right\rangle+\frac{F_0^2}{4}\left\langle\chi U+ U\chi^\dagger\right\rangle,
\end{equation}
where the superscript denotes the chiral order. The $\langle \cdots \rangle$ denotes the trace in the flavor space.

For convenience in constructing the effective baryonic Lagrangian, it is useful to define several building blocks in advance. We denote the square root of $U$ by $u$, $u^2=U$, and define the SU(3)-valued function $K(L,R,U)= \sqrt{RUL^\dagger}^{-1} R \sqrt{U}$. Under the $\mathrm{SU}(3)_L \times \mathrm{SU}(3)_R$ chiral transformation, $u$ is transformed as
\begin{equation}\label{eq:transformation2}
u  \rightarrow R u K^{\dagger} = K u L^{\dagger}.
\end{equation}
The chiral connection and vielbein are defined as~\cite{Bernard:1995dp,Scherer:2002tk}:
\begin{align}
	\Gamma_\mu=\frac{1}{2}\left[u^{\dagger}\left(\partial_\mu-i r_\mu\right) u+u\left(\partial_\mu-i l_\mu\right) u^{\dagger}\right],\\
	u_\mu=\frac{i}{2}\left[u^{\dagger}\left(\partial_\mu-i r_\mu\right) u-u\left(\partial_\mu-i l_\mu\right) u^{\dagger}\right].
\end{align}
The light-flavor octect baryons $\mathcal{N}$ and decuplet baryons $\mathcal{T}$ reads,
\begin{equation}\label{eq:octet_baryon}
\mathcal{N}=\left(\begin{matrix}
\frac{1}{\sqrt{2}}\Sigma^0+ \frac{1}{\sqrt{6}}\Lambda & \Sigma^+ & p \\
\Sigma^- & -\frac{1}{\sqrt{2}}\Sigma^0+ \frac{1}{\sqrt{6}}\Lambda & n \\
\Xi^- & \Xi^0 & -\frac{2}{\sqrt{6}} \Lambda
\end{matrix}\right), 
\end{equation}

\begin{equation}\label{eq:decuplet_baryon}
\mathcal{T}=\left(\left(\begin{array}{ccc}
\Delta^{++} & \frac{\Delta^{+}}{\sqrt{3}} & \frac{\Sigma^{*+}}{\sqrt{3}} \\
\frac{\Delta^{+}}{\sqrt{3}} & \frac{\Delta^0}{\sqrt{3}} & \frac{\Sigma^{* 0}}{\sqrt{6}} \\
\frac{\Sigma^{*+}}{\sqrt{3}} & \frac{\Sigma^{* 0}}{\sqrt{6}} & \frac{\Xi^{* 0}}{\sqrt{3}}
\end{array}\right),\left(\begin{array}{ccc}
\frac{\Delta^{+}}{\sqrt{3}} & \frac{\Delta^0}{\sqrt{3}} & \frac{\Sigma^{* 0}}{\sqrt{6}} \\
\frac{\Delta^0}{\sqrt{3}} & \Delta^{-} & \frac{\Sigma^{*-}}{\sqrt{3}} \\
\frac{\Sigma^{* 0}}{\sqrt{6}} & \frac{\Sigma^{*-}}{\sqrt{3}} & \frac{\Xi^{*-}}{\sqrt{3}}
\end{array}\right),\left(\begin{array}{ccc}
\frac{\Sigma^{*+}}{\sqrt{3}} & \frac{\Sigma^{* 0}}{\sqrt{6}} & \frac{\Xi^{* 0}}{\sqrt{3}} \\
\frac{\Sigma^{* 0}}{\sqrt{6}} & \frac{\Sigma^{*-}}{\sqrt{3}} & \frac{\Xi^{*-}}{\sqrt{3}} \\
\frac{\Xi^{* 0}}{\sqrt{3}} & \frac{\Xi^{*-}}{\sqrt{3}} & \Omega^{-}
\end{array}\right)\right).
\end{equation}
A singly heavy baryon contains one heavy quark and two light quarks. Under SU(3) flavor symmetry, the light diquark can form either an antisymmetric flavor antitriplet $\bar{3}_f$ or a symmetric flavor sextet $6_f$. According to the spin-statistics theorem, the light diquark in the antitriplet has spin 0, while that in the sextet has spin 1. As a result, the total spin of the antitriplet baryon is $S_{\bar{3}} = \frac{1}{2}$, and that of the sextet baryon can be either $S_6 = \frac{1}{2}$ or $S_6 = \frac{3}{2}$. We denote the spin-$\frac{1}{2}$ antitriplet, spin-$\frac{1}{2}$ sextet, and spin-$\frac{3}{2}$ sextet baryons as $\mathcal{B}_{\bar{3}}$, $\mathcal{B}_6$, and $\mathcal{B}_6^*$, respectively~\cite{Yan:1992gz}:
\begin{equation}\label{eq:baryon_multi}
\mathcal{B}_{\bar{3},c}=\left(\begin{matrix}
0 & \Lambda_c^{+} & \Xi_c^{+} \\
-\Lambda_c^{+} & 0 & \Xi_c^0 \\
-\Xi_c^{+} & -\Xi_c^0 & 0
\end{matrix}\right), \quad \mathcal{B}_{6,c}=\left(\begin{matrix}
\Sigma_c^{++} & \frac{\Sigma_c^{+}}{\sqrt{2}} & \frac{\Xi_c^{\prime+}}{\sqrt{2}} \\
\frac{\Sigma_c^{+}}{\sqrt{2}} & \Sigma_c^0 & \frac{\Xi_c^{\prime 0}}{\sqrt{2}} \\
\frac{\Xi_c^{\prime+}}{\sqrt{2}} & \frac{\Xi_c^{\prime 0}}{\sqrt{2}} & \Omega_c^0
\end{matrix}\right), \quad \mathcal{B}_{6,c}^{* \mu}=\left(\begin{matrix}
\Sigma_c^{*++} & \frac{\Sigma_c^{*+}}{\sqrt{2}} & \frac{\Xi_c^{*+}}{\sqrt{2}} \\
\frac{\Sigma_c^{*+}}{\sqrt{2}} & \Sigma_c^{* 0} & \frac{\Xi_c^{* 0}}{\sqrt{2}} \\
\frac{\Xi_c^{*+}}{\sqrt{2}} & \frac{\Xi_c^{* 0}}{\sqrt{2}} & \Omega_c^{* 0}
\end{matrix}\right)^\mu.
\end{equation}
\begin{equation}\label{eq:baryon_multi_2}
\mathcal{B}_{\bar{3},b}=\left(\begin{matrix}
0 & \Lambda_b^{0} & \Xi_b^{0} \\
-\Lambda_b^{0} & 0 & \Xi_b^- \\
-\Xi_b^{0} & -\Xi_b^- & 0
\end{matrix}\right), \quad \mathcal{B}_{6,b}=\left(\begin{matrix}
\Sigma_b^{+} & \frac{\Sigma_b^{0}}{\sqrt{2}} & \frac{\Xi_b^{\prime 0}}{\sqrt{2}} \\
\frac{\Sigma_b^{0}}{\sqrt{2}} & \Sigma_b^- & \frac{\Xi_b^{\prime-}}{\sqrt{2}} \\
\frac{\Xi_b^{\prime 0}}{\sqrt{2}} & \frac{\Xi_b^{\prime-}}{\sqrt{2}} & \Omega_b^-
\end{matrix}\right), \quad \mathcal{B}_{6,b}^{* \mu}=\left(\begin{matrix}
\Sigma_b^{*+} & \frac{\Sigma_b^{* 0}}{\sqrt{2}} & \frac{\Xi_b^{* 0}}{\sqrt{2}} \\
\frac{\Sigma_b^{* 0}}{\sqrt{2}} & \Sigma_b^{* -} & \frac{\Xi_b^{* -}}{\sqrt{2}} \\
\frac{\Xi_b^{* 0}}{\sqrt{2}} & \frac{\Xi_b^{* -}}{\sqrt{2}} & \Omega_b^{* -}
\end{matrix}\right)^\mu.
\end{equation}
Under the $\mathrm{SU}(3)_L \times \mathrm{SU}(3)_R$ chiral transformation, the baryon fields $\mathcal{N}$,$\mathcal{T}$, and $\mathcal{B}_n$ are transformed as follows:
\begin{equation}
\begin{aligned}
\mathcal{N} & \rightarrow K \mathcal{N} K^\dagger,\\
\mathcal{T}^{abc} & \rightarrow K^a_i K^b_j K^c_k \mathcal{T}^{ijk}, \\
\mathcal{B}_n & \rightarrow K \mathcal{B}_n K^T.
\end{aligned}
\end{equation}
Thus the covariant derivatives of these baryon fields are defined as follows:  
\begin{equation}
\begin{aligned}
    D_\mu \mathcal{N}=&\partial_\mu \mathcal{N}+\Gamma_\mu \mathcal{N}-\mathcal{N} \Gamma_\mu,\\
        D_\mu \mathcal{T}^{abc}_\nu=&\partial_\mu \mathcal{T}^{abc}_\nu+(\Gamma_\mu)^a_i \mathcal{T}^{ibc}_\nu+(\Gamma_\mu)^b_i \mathcal{T}^{aic}_\nu+(\Gamma_\mu)^c_i \mathcal{T}^{abi}_\nu, \\
        D_\mu \mathcal{B}_n=&\partial_\mu \mathcal{B}_n+\Gamma_\mu \mathcal{B}_n+\mathcal{B}_n \Gamma_\mu^{\mathrm{T}}.
\end{aligned}
\end{equation}
The chiral covariant electromagnetic field strength tensors $F_{\mu \nu}^{ \pm}$ are defined as
\begin{equation}
F_{\mu \nu}^{ \pm}  =u^{\dagger} F_{\mu \nu}^R u \pm u F_{\mu \nu}^L u^{\dagger}.
\end{equation}
The covariant spin-operator is defined as
\begin{equation}
	S^\mu=\frac{i}{2} \gamma_5 \sigma^{\mu \nu} v_\nu=-\frac{1}{2} \gamma_5\left(\gamma^\mu \slashed v-v^\mu\right), \quad S^{\mu \dagger}=\gamma_0 S^\mu \gamma_0.
\end{equation}
The spin-3/2 projector in d-dimension is defined as
\begin{equation}
P_{\mu \nu}^{3 / 2}=g_{\mu \nu}-v_{\mu} v_{ \nu}+\frac{4}{d-1} S_\mu S_\nu.
\label{eq:3/2 project operator}
\end{equation}


In the HB$\chi$PT formalism, the LO baryon Lagrangian reads~\cite{Ou-Yang:2024neq,Wang:2018gpl},
\begin{equation}
\begin{aligned}
\mathcal{L}_{\mathcal{T/N}\phi}^{(1)}=&\left\langle\overline{\mathcal{N}}(i v \cdot D) \mathcal{N}\right\rangle-i\left\langle \overline{\mathcal{T}}^\mu(v \cdot D-\delta) \mathcal{T}_\mu\right\rangle   \\
&+\mathcal{C}\left\langle\overline{\mathcal{T}}^\mu u_\mu \mathcal{N}\right\rangle+\text { H.c. }+2 \mathcal{H} \left\langle\overline{\mathcal{T}}^\mu S^\nu u_\nu \mathcal{T}_\mu\right\rangle,    
\end{aligned}
\end{equation}
\begin{equation}\label{eq:L_BPhi_1}
\begin{aligned}
\mathcal{L}_{\mathcal{B} \phi}^{(1)} = &\frac{1}{2}\left\langle\bar{\mathcal{B}}_{\bar{3}} i v \cdot D \mathcal{B}_{\bar{3}}\right\rangle+\left\langle\bar{\mathcal{B}}_6\left(i v \cdot D-\delta_2\right) \mathcal{B}_6\right\rangle-\left\langle\bar{\mathcal{B}}_6^*\left(i v \cdot D-\delta_3\right) \mathcal{B}_6^*\right\rangle \\
& +2 g_1\left\langle\bar{\mathcal{B}}_6 S \cdot u \mathcal{B}_6\right\rangle+2 g_2\left\langle\bar{\mathcal{B}}_6 S \cdot u \mathcal{B}_{\bar{3}}\right\rangle+\text { H.c. }+g_3\left\langle\bar{\mathcal{B}}_{6 \mu}^* u^\mu \mathcal{B}_6\right\rangle+\text { H.c. } \\
& +g_4\left\langle\bar{\mathcal{B}}_{6 \mu}^* u^\mu \mathcal{B}_{\bar{3}}\right\rangle+\text { H.c. }+2 g_5\left\langle\bar{\mathcal{B}}_6^* S \cdot u \mathcal{B}_6^*\right\rangle+2 g_6\left\langle\bar{\mathcal{B}}_{\bar{3}} S \cdot u \mathcal{B}_{\bar{3}}\right\rangle,
\end{aligned}
\end{equation}
where $\delta, \delta_1, \delta_2$, and $\delta_3$ represent the average mass differences between different baryon multiplets:
\begin{equation}
\begin{aligned}
&\delta=M_{\mathcal{T}}-M_{\mathcal{N}}= 294 ~\mathrm{MeV}, \\
	&\delta_1=M_{\mathcal{B}_{6}^*}-M_{\mathcal{B}_{6}}= 67~\mathrm{MeV}, \\
	&\delta_2=M_{\mathcal{B}_{6}}-M_{\mathcal{B}_{\bar{3}}}= 127~\mathrm{MeV}, \\
	&\delta_3=M_{\mathcal{B}_{6}^*}-M_{\mathcal{B}_{\bar{3}}}= 194 ~\mathrm{MeV},
\end{aligned}
\end{equation}
where the masses $M$ refer to the average values within each baryon multiplet. In this work, we neglect the mass splittings among states within the same multiplet. The parameters $\mathcal{C}, \mathcal{H}$, and $g_i$ denote the coupling constants.

The next-to-leading order (NLO) baryon Lagrangian reads:
\begin{equation}\label{eq:L_TPhi_2}
    \mathcal{L}_{\mathcal{T}/\mathcal{N} \phi}^{(2)}=-\left\langle \overline{\mathcal{T}}^\mu\frac{\left(v\cdot D\right)^2-D^2}{2M_{\mathcal{T}}}\mathcal{T}_\mu\right\rangle+b_2 \frac{-i}{2 M_\mathcal{N}} \left\langle \overline{\mathcal{T}}^\mu F_{\mu \nu}^{+} S^\nu \mathcal{N}\right\rangle+ \text{H.c.},
\end{equation}
\begin{equation}\label{eq:L_BPhi_2}
\begin{aligned}
\mathcal{L}_{\mathcal{B} \phi}^{(2)} = & \frac{1}{2}\left\langle \bar{\mathcal{B}}_{\bar{3}} \frac{\left(v\cdot D\right)^2-D^2}{2M_{\bar{3}}}\mathcal{B}_{\bar{3}}\right\rangle+\left\langle \bar{\mathcal{B}}_6 \frac{\left(v\cdot D\right)^2-D^2}{2M_{6}}\mathcal{B}_6\right\rangle-\left\langle \bar {\mathcal{B}}_6^{*\mu} \frac{\left(v\cdot D\right)^2-D^2}{2M_{6^*}}\mathcal{B}_{6\mu}^*\right\rangle\\
& -\frac{i d_2}{4 M_N}\left\langle\bar{\mathcal{B}}_3\left[S^\mu, S^\nu\right] \hat{F}_{\mu \nu}^{+} \mathcal{B}_3\right\rangle-\frac{i d_3}{4 M_N}\left\langle\bar{\mathcal{B}}_3\left[S^\mu, S^\nu\right] \mathcal{B}_3\right\rangle\left\langle F_{\mu \nu}^{+}\right\rangle-\frac{i d_5}{4 M_N}\left\langle\bar{\mathcal{B}}_6\left[S^\mu, S^\nu\right] \hat{F}_{\mu \nu}^{+} \mathcal{B}_6\right\rangle \\
& -\frac{i d_6}{4 M_N}\left\langle\bar{\mathcal{B}}_6\left[S^\mu, S^\nu\right] \mathcal{B}_6\right\rangle\left\langle F_{\mu \nu}^{+}\right\rangle-\frac{i f_2}{4 M_N}\left\langle\bar{\mathcal{B}}_3\left[S^\mu, S^\nu\right] \hat{F}_{\mu \nu}^{+} \mathcal{B}_6\right\rangle+\text { H.c. }+\frac{i f_4}{4 M_N}\left\langle\bar{\mathcal{B}}_3 \hat{F}_{\mu \nu}^{+} S^\nu \mathcal{B}_6^{* \mu}\right\rangle+\text { H.c. } \\
& +\frac{i f_6}{4 M_N}\left\langle\bar{\mathcal{B}}_6 \hat{F}_{\mu \nu}^{+} S^\nu \mathcal{B}_6^{* \mu}\right\rangle+\text { H.c. }+\frac{i f_7}{4 M_N}\left\langle\bar{\mathcal{B}}_6 S^\nu \mathcal{B}_6^{* \mu}\right\rangle\left\langle F_{\mu \nu}^{+}\right\rangle+\text {H.c. }+\frac{i f_9}{4 M_N}\left\langle\bar{\mathcal{B}}_6^{* \mu} \hat{F}_{\mu \nu}^{+} \mathcal{B}_6^{* \nu}\right\rangle \\
& +\frac{i f_{10}}{4 M_N}\left\langle\bar{\mathcal{B}}_6^{* \mu} \mathcal{B}_6^{* \nu}\right\rangle\left\langle F_{\mu \nu}^{+}\right\rangle,
\end{aligned}
\end{equation}
where $b_2,\,d_i$, and $f_i$ are the coupling constants. The traceless part of the field strength tensor, $\hat{f}_{\mu \nu}^{+}=f_{\mu \nu}^{+}-\frac{1}{3} \left\langle f_{\mu \nu}^{+}\right\rangle$ is related to the traceless charge matrix of the light quarks $Q_l=\operatorname{diag}\left(\frac{2}{3},-\frac{1}{3},-\frac{1}{3}\right)$. The trace part $\left\langle f_{\mu \nu}^{+}\right\rangle$ is related to the charge matrix of the charm quark $Q_c=\operatorname{diag}\left(\frac{1}{3}, \frac{1}{3}, \frac{1}{3}\right)$. We use the nucleon mass to render the LECs dimensionless. This choice is particularly convenient, as many physical quantities are often expressed in units of the nuclear magneton. In the following Feynman diagrams, we use black dots ({\large $\bullet$}) to represent the vertices from $\mathcal{L}_{\mathcal{T}/\mathcal{N} \phi}^{(2)}$ and $\mathcal{L}_{\mathcal{B} \phi}^{(2)}$.

The $\mathcal{L}_{\mathcal{T}/\mathcal{N} \phi }^{(3)}$ and $\mathcal{L}_{\mathcal{B} \phi}^{(3)}$ generate Compton scattering amplitudes that are odd in the photon momentum. However, the forward Compton scattering amplitudes are even in photon momentum due to crossing symmetry. Thus $\mathcal{L}_{\mathcal{T}/\mathcal{N} \phi }^{(3)}$ and $\mathcal{L}_{\mathcal{B} \phi}^{(3)}$ do not contribute to the polarizabilities and do not need to be considered explicitly~\cite{Bernard:1991rq,Bernard:1991ru,Bernard:1992qa,Bernard:1993bg,Bernard:1993ry}.

According to the standard power counting~\cite{Bernard:1995dp,Scherer:2002tk}, the chiral order $D_\chi$ of a Feynmen digram is
\begin{equation}
	D_{\chi}=2L+1+\sum_d (d-2)N_d^{\phi}+\sum_d(d-1)N_d^{\phi B},
\end{equation}
where $L$, $N_d^{\phi}$ and $N_d^{\phi B}$ are the numbers of loops, pure meson vertices and meson-baryon vertices, respectively. $d$ is the chiral dimension. 

\begin{figure}[htbp]
\centering
\includegraphics[width=17cm,height=10cm]{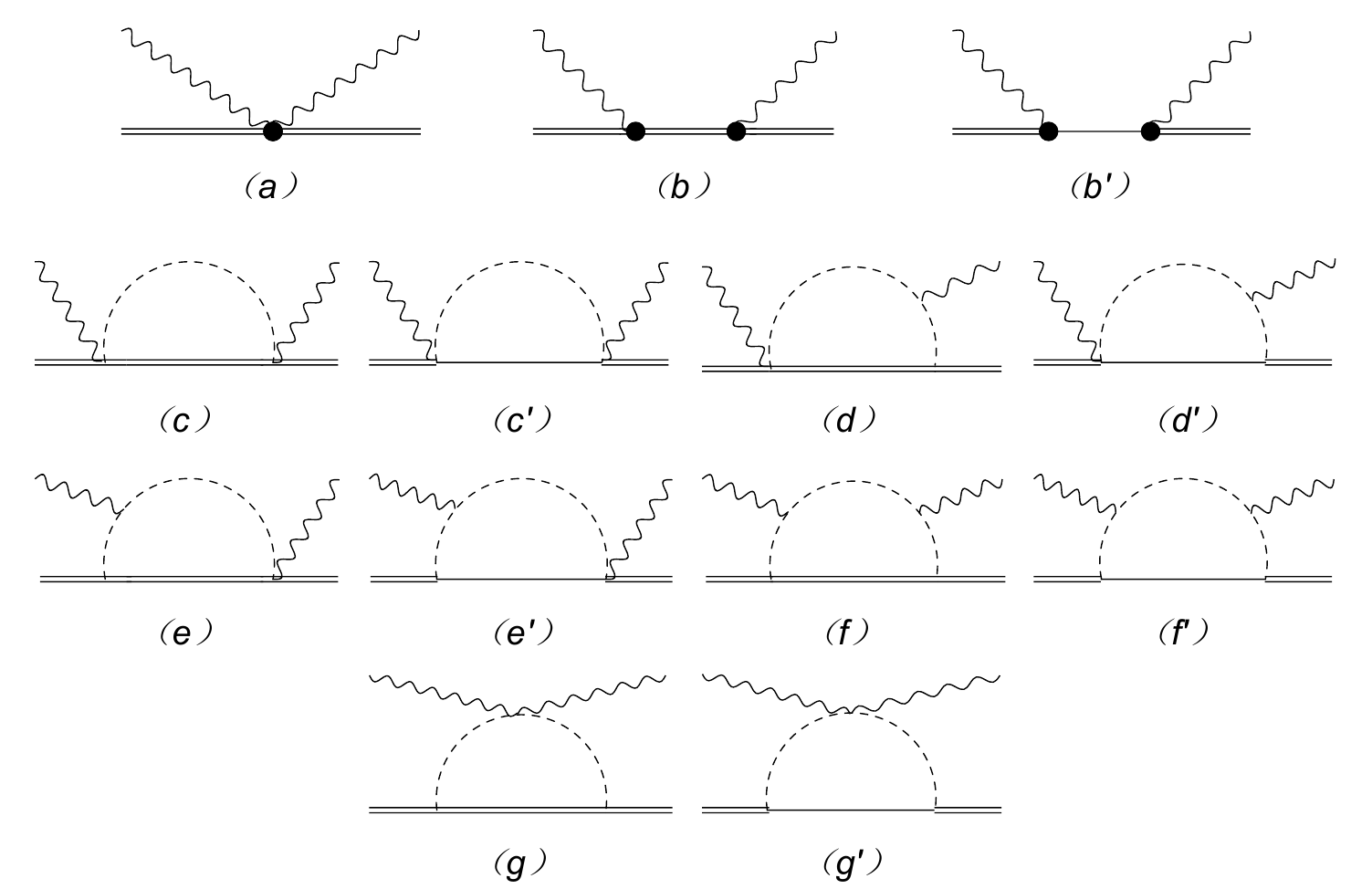}
\caption{The Born and loop diagrams contribute to the electromagnetic polarizabilities up to $\mathcal{O}\left(p^3\right)$. The solid dots denote the $\mathcal{L}_{\mathcal{T}/\mathcal{N} \phi }^{(2)}$ and $\mathcal{L}_{\mathcal{B} \phi }^{(2)}$ vertices. The single and double lines represent the spin-$\frac{3}{2}$ and spin-$\frac{1}{2}$ baryons, respectively. Crossed diagrams are not shown.}
\label{fig:fmdiagrams_full}
\end{figure}

In the Coulomb gauge,  the photon-baryon-baryon vertices derived from $\mathcal{L}_{\mathcal{T}/\mathcal{N} \phi}^{(1)}$ and $\mathcal{L}_{\mathcal{B} \phi}^{(1)}$ are proportional to $\epsilon \cdot v=0$. This condition significantly reduces the number of Feynman diagrams that need to be calculated. The tree and loop Feynman diagrams contributing to the electromagnetic polarizabilities up to \(\mathcal{O}(p^3)\) are shown in Fig.~\ref{fig:fmdiagrams_full}.


\section{Electromagnetic polarizabilities of decuplet baryons   }\label{sec:T}

\subsection{ANALYTICAL EXPRESSIONS}\label{subsec:ANALYTICAL EXPRESSIONS_T}
\subsubsection{Tree diagrams}
The $\mathcal{O}(p^2)$ seagull diagram in Fig.~\ref{fig:fmdiagrams_full}($a$) yields the Thomson amplitude:
\begin{equation}\label{eq:UV_T_ab}  
    U_{\xi}^{(a)}(\omega)=\frac{Q_\xi^2 e^2}{M_{\mathcal{T}}}, \quad V_{\xi}^{(a)}(\omega)=0.  
\end{equation}  
where $\xi$ denotes a specific baryon. The $\mathcal{O}(p^3)$ tree diagram in Fig.~\ref{fig:fmdiagrams_full}($b$) cancels exactly with its crossed counterpart. As a result, these diagrams do not contribute to the electromagnetic polarizabilities:
\begin{equation}
	\alpha_E^{(a+b)}\left(\xi\right)=\beta_M^{(a+b)}\left(\xi\right)=0.
\end{equation}
The $\mathcal{O}(p^3)$ tree diagram with an intermediate octet baryon, shown in Fig.~\ref{fig:fmdiagrams_full}($b^\prime$), corresponds to a cascade $M_1$ transition. The resulting form factors are,  
\begin{equation}\label{eq:UV_fig_b_2_TN}  
    U_{\xi}^{(b^\prime)}(\omega) = \frac{e^2 b_2^2 Q_{\mathcal{TN}}^2 \omega^2}{6 M_\mathcal{N}^2} \frac{\delta}{\delta^2 - \omega^2}, \quad  
    V_{\xi}^{(b^\prime)}(\omega) = \frac{e^2 b_2^2 Q_{\mathcal{TN}}^2}{6 M_\mathcal{N}^2} \frac{\delta}{\delta^2 - \omega^2},  
\end{equation}  
where $Q_{\mathcal{TN}}$ is the flavor Clebsch-Gordan coefficient for each decuplet baryon, as listed in Table~\ref{tab:C_xiTN}. The coefficient $b_2$ is related to the transition magnetic moment $\mu_{\xi\rightarrow\xi'+\gamma}$, which can be obtained by comparing the matrix elements at the hadron and quark levels. At the quark level, the magnetic interaction reads 
\begin{equation}\label{44}
\mathcal{L}_{\mathrm{quark}}=-\frac{e}{4}\left(\frac{2}{3 m_u} \bar{u} \sigma^{\mu \nu} u-\frac{1}{3 m_d} \bar{d} \sigma^{\mu \nu} d-\frac{1}{3 m_s} \bar{s} \sigma^{\mu \nu} s\right) F_{\mu \nu}.
\end{equation}
At the hadron level, the corresponding interaction is given by Eq.~\eqref{eq:L_TPhi_2}, which can be rewritten to separate the Lorentz and flavor structures:
\begin{equation}\label{45}
\mathcal{L}_{\mathcal{T/N}\phi}^{(2)}=-iQ_{\mathcal{TN}} \frac{2 b_2}{2 M_\mathcal{N}}  \overline{\mathcal{T}}^\mu S^\nu \mathcal{N} F_{\mu \nu}+ \text{H.c.}.
\end{equation}
This yields  
\begin{equation}\label{eq:transition_LO}  
    b_2~Q_{\mathcal{TN}}\frac{e}{2m_\mathcal{N}} = -\frac{\sqrt{6}}{2} \mu^{\mathrm{LO}}_{\xi \to \xi'+ \gamma}.  
\end{equation}

\begin{table}[htbp]
    \centering
    \caption{The non-zero coefficients $Q_{\mathcal{TN}}$ of the diagram in Fig.~\ref{fig:fmdiagrams_full}($b^\prime$)}
    \label{tab:C_xiTN}
    \setlength{\tabcolsep}{2.5mm}
    \begin{tabular}{c|ccccccccccc}
    \hline\hline
         &  $\Delta^{+}\rightarrow p \gamma$ & $\Delta^0\rightarrow n \gamma$ &  $\Sigma^{*+}\rightarrow \Sigma^+ \gamma$ & $\Sigma^{* 0}\rightarrow \Sigma\gamma$ & $\Sigma^{*0}\rightarrow\Lambda\gamma$ &  $\Xi^{*0}\rightarrow\Xi\gamma$ &  \\ 
         \hline
        $Q_{\mathcal{TN}}$  & $\frac{1}{\sqrt{3}}$ & $\frac{1}{\sqrt{3}}$ & $\frac{1}{\sqrt{3}}$ & $\frac{1}{2 \sqrt{3}}$ & $\frac{1}{2}$  & $\frac{1}{\sqrt{3}}$ &\\
        \hline\hline
    \end{tabular}
\end{table}

The corresponding electromagnetic polarizabilities extracted from Eq.~\eqref{eq:UV_fig_b_2_TN} are  
\begin{equation}\label{eq:ab_b6_b2_TN}  
    \alpha^{(b^\prime)}_{E}(\xi) = 0, \quad \beta^{(b^\prime)}_{M}(\xi) = -\frac{\alpha_{\mathrm{em}} b_2^2 Q_{\mathcal{TN}}^2}{6 M_\mathcal{N}^2} \frac{1}{\delta}.  
\end{equation}  
The diagram in Fig.~\ref{fig:fmdiagrams_full}($b^\prime$) represents purely magnetic polarizability through the inclusion of a $\delta$ pole term. A comparable effect is observed when evaluating the contribution of $\Delta$-baryons to nucleon polarizabilities, as analyzed in~\cite{Schumacher:2005an,Holstein:2013kia,Hemmert:1996rw,Pascalutsa:2002pi}.

\subsubsection{Loop diagrams}
Using the $\mathcal{J}$-function defined in Appendix~\ref{appendix:loop_integrals}, we obtain the form factors for the $\mathcal{T}\phi$-loop diagrams in Figs.~\ref{fig:fmdiagrams_full}($c$)--($g$):
\begin{align}
U^{(c)}_{\xi}(\omega)= & C\sum_{\chi}\frac{D^{(c)}_{\xi,\chi}}{F_\chi^2}\mathcal{J}_0(\omega,0,M_{\chi}^2),\label{eq:U_6_c_T}\\
	U^{(d+e)}_{\xi}(\omega)= & C\left[\sum_{\chi} \frac{D_{\xi,\chi}^{(d+e)}}{F_\chi^2}\int_0^1 dx~\mathcal{J}^{\prime}_2(\omega x,0,M_{\chi}^2)\right],\\
	U^{(f)}_{\xi}(\omega)= & C\left[\sum_{\chi} \frac{D_{\xi,\chi}^{(f)}}{F_\chi^2}\int_0^1dx~ (1-x)(d+1)\mathcal{J}_6^{\prime\prime}(\omega x,0,M_{\chi}^2)\right.\notag \\
	&\quad ~ -\left. \sum_{\chi} \frac{D_{\xi,\chi}^{(f)}}{F_\chi^2} \int_0^1dx~\omega^2(1-x)x^2\mathcal{J}_2^{\prime\prime}(\omega x,0,M_{\chi}^2)\right],\\
	U^{(g)}_{\xi}(\omega)= &C\sum_{\chi}\frac{D^{(g)}_{\xi,\chi}}{F_\chi^2}(d-1)\mathcal{J}_2^{\prime}(0,0,M_{\chi}^2),\\
	V^{(c)}_{\xi}(\omega)= & 0,\\
	V^{(d+e)}_{\xi}(\omega)= & C\left[\sum_{\chi} -\frac{1}{2 F_\chi^2}D_{\xi,\chi}^{(d+e)}\int_0^1 dx~x(1-2x)\mathcal{J}^{\prime}_0(\omega x,0,M_{\chi}^2)\right],\\
	V^{(f)}_{\xi}(\omega)= & C\left[\sum_{\chi} \frac{1}{4 F_\chi^2}D^{(f)}_{\xi,\chi}\int_0^1dx~(1-x)\left[8x(2x-1)+(2x-1)^2(d-1)\right]\mathcal{J}_2^{\prime\prime}(\omega x,0,M_{\chi}^2)\right.\notag \\
	 &\quad ~ -\left. \sum_{\chi} \frac{1}{4 F_\chi^2}D^{(f)}_{\xi,\chi} \int_0^1 dx~ \omega^2(1-x)x^2(2x-1)^2 \mathcal{J}_0^{\prime\prime}(\omega x, 0,M_{\chi}^2)\right],\\
	 V^{(g)}_{\xi}(\omega)= & 0.\label{eq:V_6_g_T}
\end{align}
In Eqs.~\eqref{eq:U_6_c_T}--\eqref{eq:V_6_g_T}, we have considered the contributions from the crossed diagrams. $\chi$ denotes a specific meson in Eq.~\eqref{eq:octet_meson}, and $M_\chi$ represents its mass. $D_{\xi,\chi}$ are the coefficients of the loops, which are given in Table~\ref{tab:loop_cofficients_TT}. $C$ is a common factor
\begin{equation}
	C=i\frac{(d^3-4d^2+d+6)e^2\mathcal{H}^2}{4(d-1)^2}.
\end{equation}

\begin{table}[htbp]
    \centering
    \caption{The coefficients for loop diagrams involving decuplet baryons $\mathcal{T}$ as intermediate states.}
    \label{tab:loop_cofficients_TT}
    \setlength{\tabcolsep}{3mm}
    \begin{tabular}{c|cccccccccc}
    \hline\hline
& $\Delta^{++}$ & $\Delta^{+}$ & $\Delta^0$ & $\Delta^{-}$ & $\Sigma^{*+}$ & $\Sigma^{* 0}$ & $\Sigma^{*-}$ & $\Xi^{*0}$ & $\Xi^{*-}$ & $\Omega^{-}$ \\
\hline
$D_\pi^{(c)}$ & -$\frac{1}{3}$ & -$\frac{7}{9}$ & -$\frac{7}{9}$ & -$\frac{1}{3}$ & -$\frac{2}{9}$ & -$\frac{4}{9}$ & -$\frac{2}{9}$ & -$\frac{1}{9}$ & -$\frac{1}{9}$ & 0\\
$D_K^{(c)}$ & -$\frac{1}{3}$ & -$\frac{2}{9}$ & -$\frac{1}{9}$ & 0 & -$\frac{7}{9}$ & -$\frac{4}{9}$ & -$\frac{1}{9}$ & -$\frac{7}{9}$ & -$\frac{2}{9}$ & -$\frac{1}{3}$ \\
$D_\pi^{(d+e)}$ & $\frac{4}{3}$ & $\frac{28}{9}$ & $\frac{28}{9}$ & $\frac{4}{3}$ & $\frac{8}{9}$ & $\frac{16}{9}$ & $\frac{8}{9}$ & $\frac{4}{9}$ & $\frac{4}{9}$ & 0\\
$D_K^{(d+e)}$ & $\frac{4}{3}$ & $\frac{8}{9}$ & $\frac{4}{9}$ & 0 & $\frac{28}{9}$ & $\frac{16}{9}$ & $\frac{4}{9}$ & $\frac{28}{9}$ & $\frac{8}{9}$ & $\frac{4}{3}$ \\
$D_\pi^{(f)}$ & -$\frac{4}{3}$ & -$\frac{28}{9}$ & -$\frac{28}{9}$ & -$\frac{4}{3}$ & -$\frac{8}{9}$ & -$\frac{16}{9}$ & -$\frac{8}{9}$ & -$\frac{4}{9}$ & -$\frac{4}{9}$ & 0\\
$D_K^{(f)}$ & -$\frac{4}{3}$ & -$\frac{8}{9}$ & -$\frac{4}{9}$ & 0 & -$\frac{28}{9}$ & -$\frac{16}{9}$ & -$\frac{4}{9}$ & -$\frac{28}{9}$ & -$\frac{8}{9}$ & -$\frac{4}{3}$ \\
$D_\pi^{(g)}$ & $\frac{1}{3}$ & $\frac{7}{9}$ & $\frac{7}{9}$ & $\frac{1}{3}$ & $\frac{2}{9}$ & $\frac{4}{9}$ & $\frac{2}{9}$ & -$\frac{1}{9}$ & -$\frac{1}{9}$ & 0\\
$D_K^{(g)}$ & $\frac{1}{3}$ & $\frac{2}{9}$ & $\frac{1}{9}$ & 0 & $\frac{7}{9}$ & $\frac{4}{9}$ & $\frac{1}{9}$ & -$\frac{7}{9}$ & $\frac{2}{9}$ & $\frac{1}{3}$ \\
        \hline\hline
    \end{tabular}
\end{table}

Expanding Eqs.~\eqref{eq:U_6_c_T}--\eqref{eq:V_6_g_T} into a power series in $\omega$ and combining with Eq.~\eqref{eq:def_alpha_beta}, we can obtain the electromagnetic polarizabilities from the $\mathcal{T}\phi$-loop diagrams in Figs.~\ref{fig:fmdiagrams_full}($c$)--($g$):
\begin{align}
	\alpha_E^{(c-g)}(\Delta^{++})=& \frac{25 \alpha_{\mathrm{em}}\mathcal{H}^2}{2592 \pi M_\pi F_\pi^2}+ \frac{25 \alpha_{\mathrm{em}}\mathcal{H}^2}{2592 \pi M_K F_K^2},\\
	\alpha_E^{(c-g)}(\Delta^+)=& \frac{175 \alpha_{\mathrm{em}}\mathcal{H}^2}{7776 \pi M_\pi F_\pi^2}+ \frac{25 \alpha_{\mathrm{em}}\mathcal{H}^2}{3888 \pi M_K F_K^2},\\
	\alpha_E^{(c-g)}(\Delta^0)=& \frac{175 \alpha_{\mathrm{em}}\mathcal{H}^2}{7776 \pi M_\pi F_\pi^2}+ \frac{25 \alpha_{\mathrm{em}}\mathcal{H}^2}{7776 \pi M_K F_K^2},\\
	\alpha_E^{(c-g)}(\Delta^-)=& \frac{25 \alpha_{\mathrm{em}}\mathcal{H}^2}{2592 \pi M_\pi F_\pi^2},\\
	\alpha_E^{(c-g)}(\Sigma^{*+})=& \frac{25 \alpha_{\mathrm{em}}\mathcal{H}^2}{3888 \pi M_\pi F_\pi^2}+ \frac{175 \alpha_{\mathrm{em}}\mathcal{H}^2}{7776 \pi M_K F_K^2},\\
	\alpha_E^{(c-g)}(\Sigma^{*0})=& \frac{25 \alpha_{\mathrm{em}}\mathcal{H}^2}{1944 \pi M_K F_K^2}+ \frac{25 \alpha_{\mathrm{em}}\mathcal{H}^2}{1944 \pi M_K F_K^2},\\
    	\alpha_E^{(c-g)}(\Sigma^{*-})=& \frac{25 \alpha_{\mathrm{em}}\mathcal{H}^2}{3888 \pi M_K F_K^2}+ \frac{25 \alpha_{\mathrm{em}}\mathcal{H}^2}{7776 \pi M_K F_K^2},\\
        \alpha_E^{(c-g)}(\Xi^{*0})=& \frac{25 \alpha_{\mathrm{em}}\mathcal{H}^2}{7776 \pi M_K F_K^2}+ \frac{175 \alpha_{\mathrm{em}}\mathcal{H}^2}{7776 \pi M_K F_K^2},\\
        \alpha_E^{(c-g)}(\Xi^{*-})=& \frac{25 \alpha_{\mathrm{em}}\mathcal{H}^2}{7776 \pi M_K F_K^2}+ \frac{25 \alpha_{\mathrm{em}}\mathcal{H}^2}{3888 \pi M_K F_K^2},\\
        \alpha_E^{(c-g)}(\Omega^{-})=&\frac{25 \alpha_{\mathrm{em}}\mathcal{H}^2}{2592 \pi M_K F_K^2},\\
	\beta_M^{(c-g)}(\xi) =& \frac{1}{10} \alpha_E^{(c-g)}(\xi).
\end{align}
The $\mathcal{O}(p^3)$ $\mathcal{T}\phi$ loop diagrams exhibit a characteristic $1/m_\chi$ dependence. This underscores the importance of including meson cloud contributions in the calculation of electromagnetic polarizabilities.

Similarly, the form factors of the $\mathcal{N}\phi$-loop diagrams in Figs.~\ref{fig:fmdiagrams_full}($c^\prime$)--($g^\prime$) are:
\begin{align}
	U^{(c^\prime)}_{\xi}(\omega)= & D\sum_{\chi}\frac{D^{(c')}_{\xi,\chi}}{F_\chi^2}\mathcal{J}_0(\omega,\delta,M_{\chi}^2)\label{eq:U_6s_c_N},\\
	U^{(d^\prime+e^\prime)}_{\xi}(\omega)= & D\left[\sum_{\chi} \frac{D_{\xi,\chi}^{(d'+e')}}{F_\chi^2}\int_0^1 dx~\mathcal{J}^{\prime}_2(\omega x,\delta,M_{\chi}^2)\right],\\
	U^{(f^\prime)}_{\xi}(\omega)= & D\left[\sum_{\chi} \frac{D_{\xi,\chi}^{(f')}}{F_\chi^2}\int_0^1dx~ (1-x)(d+1)\mathcal{J}_6^{\prime\prime}(\omega x,\delta,M_{\chi}^2)\right.\notag \\
	&\quad ~ -\left. \sum_{\chi} \frac{D_{\xi,\chi}^{(f')}}{F_\chi^2} \int_0^1dx~\omega^2(1-x)x^2\mathcal{J}_2^{\prime\prime}(\omega x,\delta,M_{\chi}^2)\right],\\
	U^{(g^\prime)}_{\xi}(\omega)= &D\sum_{\chi}\frac{D^{(g')}_{\xi,\chi}}{F_\chi^2}(d-1)\mathcal{J}_2^{\prime}(0,\delta,M_{\chi}^2),\\
	V^{(c^\prime)}_{\xi}(\omega)= & 0,\\
	V^{(d^\prime+e^\prime)}_{\xi}(\omega)= & D\left[\sum_{\chi} -\frac{1}{2 F_\chi^2}D_{\xi,\chi}^{(d'+e')}\int_0^1 dx~x(1-2x)\mathcal{J}^{\prime}_0(\omega x,\delta,M_{\chi}^2)\right],\\
	V^{(f^\prime)}_{\xi}(\omega)= & D\left[\sum_{\chi} \frac{1}{4 F_\chi^2}D^{(f')}_{\xi,\chi}\int_0^1dx~(1-x)\left[8x(2x-1)+(2x-1)^2(d-1)\right]\mathcal{J}_2^{\prime\prime}(\omega x,\delta,M_{\chi}^2)\right.\notag \\
	 &\quad ~ -\left. \sum_{\chi} \frac{1}{4 F_\chi^2}D^{(f')}_{\xi,\chi} \int_0^1 dx~ \omega^2(1-x)x^2(2x-1)^2 \mathcal{J}_0^{\prime\prime}(\omega x,\delta,M_{\chi}^2)\right],\\
	 V^{(g^\prime)}_{\xi}(\omega)= & 0,\label{eq:V_6s_g_N}
\end{align}
where the common factor $D$ is
\begin{equation}
	D=i\frac{e^2\mathcal{C}^2}{4}\frac{d-2}{d-1},
\end{equation}
and the coefficients for the $\mathcal{N} \phi$-loop diagrams are listed in Table~\ref{tab:loop_cofficients_TN}.

\begin{table}[htbp]
    \centering
    \caption{The coefficients for loop diagrams involving octet baryons $\mathcal{N}$ as intermediate states.}
    \label{tab:loop_cofficients_TN}
    \setlength{\tabcolsep}{3mm}
    \begin{tabular}{c|cccccccccc}
    \hline\hline
& $\Delta^{++}$ & $\Delta^{+}$ & $\Delta^0$ & $\Delta^{-}$ & $\Sigma^{*+}$ & $\Sigma^{* 0}$ & $\Sigma^{*-}$ & $\Xi^{*0}$ & $\Xi^{*-}$ & $\Omega^{-}$ \\
\hline
$D_\pi^{(c')}$ & -1 & -$\frac{1}{3}$ & -$\frac{1}{3}$ & -1 & -$\frac{2}{3}$ & -$\frac{1}{3}$ & -$\frac{2}{3}$ & -$\frac{1}{3}$ & -$\frac{1}{3}$ & 0\\
$D_K^{(c')}$ & -1 & -$\frac{2}{3}$ & -$\frac{1}{3}$ & 0 & -$\frac{1}{3}$ & -$\frac{1}{3}$ & -$\frac{1}{3}$ & -$\frac{1}{3}$ & -$\frac{2}{3}$ & -1\\
$D_\pi^{(d'+e')}$ & 4 & $\frac{4}{3}$ & $\frac{4}{3}$ & 4 & $\frac{8}{3}$ & $\frac{4}{3}$ & $\frac{8}{3}$ & $\frac{4}{3}$ & $\frac{4}{3}$ & 0\\
$D_K^{(d'+e')}$ & 4 & $\frac{8}{3}$ & $\frac{4}{3}$ & 0 & $\frac{4}{3}$ & $\frac{4}{3}$ & $\frac{4}{3}$ & $\frac{4}{3}$ & $\frac{8}{3}$ & 4\\
$D_\pi^{(f')}$ & -4 & -$\frac{4}{3}$ & -$\frac{4}{3}$ & -4 & -$\frac{8}{3}$ & -$\frac{4}{3}$ & -$\frac{8}{3}$ & -$\frac{4}{3}$ & -$\frac{4}{3}$ & 0\\
$D_K^{(f')}$ & -4 & -$\frac{8}{3}$ & -$\frac{4}{3}$ & 0 & -$\frac{4}{3}$ & -$\frac{4}{3}$ & -$\frac{4}{3}$ & -$\frac{4}{3}$ & -$\frac{8}{3}$ & -4\\
$D_\pi^{(g')}$ & 1 & $\frac{1}{3}$ & $\frac{1}{3}$ & 1 & $\frac{2}{3}$ & $\frac{1}{3}$ & $\frac{2}{3}$ & $\frac{1}{3}$ & $\frac{1}{3}$ & 0\\
$D_K^{(g')}$ & 1 & $\frac{2}{3}$ & $\frac{1}{3}$ & 0 & $\frac{1}{3}$ & $\frac{1}{3}$ & $\frac{1}{3}$ & $\frac{1}{3}$ & $\frac{2}{3}$ & 1\\
        \hline\hline
    \end{tabular}
\end{table}

Expanding Eq.~\eqref{eq:U_6s_c_N}--\eqref{eq:V_6s_g_N} into a power series in $\omega$ and combining with Eq.~\eqref{eq:def_alpha_beta}, we can obtain the electromagnetic polarizabilities from the $\mathcal{N}\phi$-loop diagrams in Figs.~\ref{fig:fmdiagrams_full}($c^\prime$)--($g^\prime$):

\begin{align}
	\alpha_E^{(c^\prime-g^\prime)}\left(\Delta^{++}\right)=&\frac{\alpha_{\mathrm{em}} \mathcal{C}^2 S_\pi}{288\pi^2 F_\pi^2 \left(M_\pi^2-\delta^2\right)^2}+\frac{\alpha_{\mathrm{em}} \mathcal{C}^2 S_K}{288\pi^2 F_K^2 \left(M_K^2-\delta^2\right)^2},\\
	\alpha_E^{(c^\prime-g^\prime)}\left(\Delta^{+}\right)=&\frac{\alpha_{\mathrm{em}} \mathcal{C}^2 S_\pi}{864\pi^2 F_\pi^2 \left(M_\pi^2-\delta^2\right)^2}+\frac{\alpha_{\mathrm{em}} \mathcal{C}^2 S_K}{432\pi^2 F_K^2 \left(M_K^2-\delta^2\right)^2},\\
	\alpha_E^{(c^\prime-g^\prime)}\left(\Delta^{0}\right)=&\frac{\alpha_{\mathrm{em}} \mathcal{C}^2 S_\pi}{864\pi^2 F_\pi^2 \left(M_\pi^2-\delta^2\right)^2}+\frac{\alpha_{\mathrm{em}} \mathcal{C}^2 S_K}{864\pi^2 F_K^2 \left(M_K^2-\delta^2\right)^2},\\
	\alpha_E^{(c^\prime-g^\prime)}\left(\Delta^-\right)=&\frac{\alpha_{\mathrm{em}} \mathcal{C}^2 S_\pi}{288\pi^2 F_\pi^2 \left(M_\pi^2-\delta^2\right)^2},\\
	\alpha_E^{(c^\prime-g^\prime)}\left(\Sigma^{*+}\right)=&\frac{\alpha_{\mathrm{em}} \mathcal{C}^2 S_\pi}{432\pi^2 F_\pi^2 \left(M_\pi^2-\delta^2\right)^2}+\frac{\alpha_{\mathrm{em}} \mathcal{C}^2 S_K}{864\pi^2 F_K^2 \left(M_K^2-\delta^2\right)^2},\\
	\alpha_E^{(c^\prime-g^\prime)}\left(\Sigma^{*0}\right)=&\frac{\alpha_{\mathrm{em}} \mathcal{C}^2 S_\pi}{864\pi^2 F_\pi^2 \left(M_\pi^2-\delta^2\right)^2}+\frac{\alpha_{\mathrm{em}} \mathcal{C}^2 S_K}{864\pi^2 F_K^2 \left(M_K^2-\delta^2\right)^2},\\
    	\alpha_E^{(c^\prime-g^\prime)}\left(\Sigma^{*-}\right)=&\frac{\alpha_{\mathrm{em}} \mathcal{C}^2 S_\pi}{432\pi^2 F_\pi^2 \left(M_\pi^2-\delta^2\right)^2}+\frac{\alpha_{\mathrm{em}} \mathcal{C}^2 S_K}{864\pi^2 F_K^2 \left(M_K^2-\delta^2\right)^2},\\
        	\alpha_E^{(c^\prime-g^\prime)}\left(\Xi^{*0}\right)=&\frac{\alpha_{\mathrm{em}} \mathcal{C}^2 S_\pi}{864\pi^2 F_\pi^2 \left(M_\pi^2-\delta^2\right)^2}+\frac{\alpha_{\mathrm{em}} \mathcal{C}^2 S_K}{864\pi^2 F_K^2 \left(M_K^2-\delta^2\right)^2},\\
            	\alpha_E^{(c^\prime-g^\prime)}\left(\Xi^{*-}\right)=&\frac{\alpha_{\mathrm{em}} \mathcal{C}^2 S_\pi}{864\pi^2 F_\pi^2 \left(M_\pi^2-\delta^2\right)}+\frac{\alpha_{\mathrm{em}} \mathcal{C}^2 S_K}{432\pi^2 F_K^2 \left(M_K^2-\delta^2\right)^2},\\
                	\alpha_E^{(c^\prime-g^\prime)}\left(\Omega^{-}\right)=&\frac{\alpha_{\mathrm{em}} \mathcal{C}^2 S_K}{288\pi^2 F_K^2 \left(M_K^2-\delta^2\right)^2}.\\
    \beta_M^{(c^\prime-g^\prime)}\left(\Delta^{++}\right)=&\frac{\alpha_{\mathrm{em}} \mathcal{C}^2 R_\pi}{288\pi^2 F_\pi^2 \left(M_\pi^2-\delta^2\right)}+\frac{\alpha_{\mathrm{em}} \mathcal{C}^2 R_K}{288\pi^2 F_K^2 \left(M_K^2-\delta^2\right)},\\
	\beta_M^{(c^\prime-g^\prime)}\left(\Delta^{+}\right)=&\frac{\alpha_{\mathrm{em}} \mathcal{C}^2 R_\pi}{864\pi^2 F_\pi^2 \left(M_\pi^2-\delta^2\right)}+\frac{\alpha_{\mathrm{em}} \mathcal{C}^2 R_K}{432\pi^2 F_K^2 \left(M_K^2-\delta^2\right)},\\
	\beta_M^{(c^\prime-g^\prime)}\left(\Delta^{0}\right)=&\frac{\alpha_{\mathrm{em}} \mathcal{C}^2 R_\pi}{864\pi^2 F_\pi^2 \left(M_\pi^2-\delta^2\right)}+\frac{\alpha_{\mathrm{em}} \mathcal{C}^2 R_K}{864\pi^2 F_K^2 \left(M_K^2-\delta^2\right)},\\
	\beta_M^{(c^\prime-g^\prime)}\left(\Delta^-\right)=&\frac{\alpha_{\mathrm{em}} \mathcal{C}^2 R_\pi}{288\pi^2 F_\pi^2 \left(M_\pi^2-\delta^2\right)},\\
	\beta_M^{(c^\prime-g^\prime)}\left(\Sigma^{*+}\right)=&\frac{\alpha_{\mathrm{em}} \mathcal{C}^2 R_\pi}{432\pi^2 F_\pi^2 \left(M_\pi^2-\delta^2\right)}+\frac{\alpha_{\mathrm{em}} \mathcal{C}^2 R_K}{864\pi^2 F_K^2 \left(M_K^2-\delta^2\right)},\\
	\beta_M^{(c^\prime-g^\prime)}\left(\Sigma^{*0}\right)=&\frac{\alpha_{\mathrm{em}} \mathcal{C}^2 R_\pi}{864\pi^2 F_\pi^2 \left(M_\pi^2-\delta^2\right)}+\frac{\alpha_{\mathrm{em}} \mathcal{C}^2 R_K}{864\pi^2 F_K^2 \left(M_K^2-\delta^2\right)},\\
    	\beta_M^{(c^\prime-g^\prime)}\left(\Sigma^{*-}\right)=&\frac{\alpha_{\mathrm{em}} \mathcal{C}^2 R_\pi}{432\pi^2 F_\pi^2 \left(M_\pi^2-\delta^2\right)}+\frac{\alpha_{\mathrm{em}} \mathcal{C}^2 R_K}{864\pi^2 F_K^2 \left(M_K^2-\delta^2\right)},\\
        	\beta_M^{(c^\prime-g^\prime)}\left(\Xi^{*0}\right)=&\frac{\alpha_{\mathrm{em}} \mathcal{C}^2 R_\pi}{864\pi^2 F_\pi^2 \left(M_\pi^2-\delta^2\right)}+\frac{\alpha_{\mathrm{em}} \mathcal{C}^2 R_K}{864\pi^2 F_K^2 \left(M_K^2-\delta^2\right)},\\
            	\beta_M^{(c^\prime-g^\prime)}\left(\Xi^{*-}\right)=&\frac{\alpha_{\mathrm{em}} \mathcal{C}^2 R_\pi}{864\pi^2 F_\pi^2 \left(M_\pi^2-\delta^2\right)}+\frac{\alpha_{\mathrm{em}} \mathcal{C}^2 R_K}{432\pi^2 F_K^2 \left(M_K^2-\delta^2\right)},\\
                	\beta_M^{(c^\prime-g^\prime)}\left(\Omega^{-}\right)=&\frac{\alpha_{\mathrm{em}} \mathcal{C}^2 R_K}{288\pi^2 F_K^2 \left(M_K^2-\delta^2\right)}
\end{align}
where we have defined
\begin{equation}
\begin{aligned}
	R_\chi&=\sqrt{M_\chi^2-\delta^2}\arccos\left[\frac{-\delta}{M_\chi}\right],\\
	S_\chi&=M_\pi^2\left(10R_\pi+9\delta\right)+\delta^2\left(-9\delta-R_\pi\right).
	\end{aligned}
\end{equation}

By summing all the contributions calculated above, we can obtain the total electromagnetic polarizabilities:
\begin{equation}
	\begin{aligned}
	\alpha_E^{\mathrm{Tot.}}(\xi) &=\alpha_E^{(c-g)}(\xi)+\alpha_E^{(c^\prime-g^\prime)}(\xi),\\
	\beta_M^{\mathrm{Tot.}}(\xi) &=\beta_M^{(b^\prime)}(\xi)+\beta_M^{(c-g)}(\xi)+\beta_M^{(c^\prime-g^\prime)}(\xi).
    \end{aligned}
\end{equation}
It is worth noting that the mass splitting $\delta = 294~\mathrm{MeV}$ is large enough to allow the decay $\mathcal{T} \to \mathcal{N} \pi$. Consequently, the electromagnetic polarizabilities arising from the $\mathcal{N}\phi$ loop diagrams are expected to have an imaginary component.

\subsection{Numerical Results}\label{sec:NUMERICAL_RESULTS_T}  

In the analytical expressions for the polarizabilities, there are low-energy constants (LECs) that need to be determined: the axial coupling constants $\mathcal{C},\,\mathcal{H}$ from $\mathcal{L}^{(1)}_{\mathcal{T/N}\phi}$, and the magnetic dipole transition parameter $b_2$ from $\mathcal{L}^{(2)}_{\mathcal{T}\phi}$. The values of $\mathcal{C}$ and $\mathcal{H}$ are extracted by fitting the strong decay widths of $\Delta$, $\Sigma^*$, and $\Xi^*$~\cite{Butler:1992pn}:  
\begin{equation}  
     |\mathcal{C}|=1.2\pm 0.1,\quad \mathcal{H}=-2.2 \pm 0.6.  
\end{equation}
Notably, identifying $\mathcal{H} = g_1$ as in the SU(2) case of Ref.~\cite{Fettes:2000bb}, the central values of $\mathcal{H}$ exhibit good agreement with the large-$N_c$ prediction, $\mathcal{H} = \frac{9}{5} g_A$. In contrast, the central values of $\mathcal{C}$ lie below the large-$N_c$ expectation of $\mathcal{C} = \frac{6}{5} g_A$.

The magnetic dipole transition parameter $b_2$ can be estimated by utilizing the electromagnetic decay widths of decuplet baryons. Assuming that the leading-order magnetic dipole transition dominates the electromagnetic decay, the decay width is given by~\cite{Wang:2018cre}
\begin{equation}
\Gamma(\mathcal{T} \rightarrow \mathcal{N} \gamma)  =\frac{\alpha}{2} \frac{\omega^3}{ M_\mathcal{N}M_\mathcal{T}} \left|\frac{\mu(\mathcal{T}\rightarrow \mathcal{N}\gamma)}{\mu_N}\right|^2, \qquad \omega=\frac{M_\mathcal{T}^2-M_\mathcal{N}^2}{2M_\mathcal{T}}.
\end{equation}
The transition magnetic moments of decuplet baryons, calculated within the leading-order HB$\chi$PT and extracted from experimental data, are listed in the Table~\ref{tab:magnetic_moments_TN}. To determine an optimal value of $b_2$, we perform a weighted least squares (WLS) fit to the experimental data:
\begin{equation}
	\chi^2=\frac{1}{\mathrm{d.o.f.}}\sum_{i=1}^\mathrm{d.o.f.}\left(\frac{\mu_i^{\mathrm{HB\chi PT}}-\mu_i^{\mathrm{exp}}}{\sigma_i^{\mathrm{exp}}}\right)^2.
\end{equation}
The resulting value is
\begin{equation}
    b_2=8.28 \pm 0.47.
\end{equation}
Using the LECs estimated above, the numerical results for the electromagnetic polarizabilities are calculated and presented in Table~\ref{tab:polarizabilities_numerical_results_T}.

\begin{table}[htbp]
    \centering
    \caption{Transition magnetic moments $\mu_{\mathcal{T}\rightarrow \mathcal{N}\gamma}$ obtained from the leading-order HB$\chi$PT calculations, and experimental results~\cite{ParticleDataGroup:2024cfk}. Magnetic moments are expressed in units of the nuclear magneton $\mu_N$.}
    \label{tab:magnetic_moments_TN}
    \setlength{\tabcolsep}{3mm}
    \begin{tabular}{c|c|c|c}
    \hline\hline
  & $\Delta\rightarrow N\gamma$ & $\Sigma^{*+}\rightarrow \Sigma^+\gamma$ & $\Sigma^{*0}\rightarrow\Lambda\gamma$  \\
\hline
HB$\chi$PT~  & -$\frac{\sqrt{2}}{3} b_2$ & $\frac{\sqrt{2}}{3} b_2$ & $\frac{1}{\sqrt{6}} b_2$\\
$|\mu_{exp}|$ & $3.57\pm0.30$ & $4.43\pm 0.54$ & $3.69\pm 0.37$ \\
\hline\hline
    \end{tabular}
\end{table}

\begin{table}[htbp]	\caption{The numerical results of spin-$\frac{3}{2}$ decuplet baryon electromagnetic polarizabilities (in unit of $10^{-4}~\mathrm{fm}^3$). The values in parentheses represent the uncertainties of the results.}
	\label{tab:polarizabilities_numerical_results_T}
    \centering
    \setlength{\tabcolsep}{1.5mm}
    \begin{tabular}{c|ccc|cccc}
    \hline\hline
         & $\alpha_E^{(c-g)}$ & $\alpha_E^{(c^\prime-g^\prime)}$ & $\alpha_E^{\text{Tot.}}$ & $\beta_M^{(c-g)}$ & $\beta_M^{(c^\prime-g^\prime)}$ & $\beta_M^{(b^\prime)}$ & $\beta_M^{\mathrm{Tot.}}$   \\ \hline
        $\Delta^{++}$ & $8.4(46)$ & $1.2(2)+0.61(10)i$ & $9.6(46)+0.61(10)i$ & $0.84(46)$ & $-0.06(1)-0.40(7)i$ & 0 & $0.79(45)-0.40(7)i$ \\ 
        $\Delta^{+}$ & $17.5(95)$ & $1.1(2)+0.20(3)i$ & $18.6(95)+0.20(3)i$ & $1.75(95)$ & $0.02(1)-0.13(2)i$ & $-8.3(9)$ & $-6.5(13)-0.13(2)i$ \\ 
        $\Delta^{0}$ & $17.0(93)$ & $0.39(6)+0.20(3)i$ & $17.4(93)+0.20(3)i$ & $1.70(93)$ & $-0.02(1)-0.13(2)i$ & $-8.3(9)$ & $-6.6(13)-0.13(2)i$ \\ 
        $\Delta^{-}$ & $7.1(39)$ & $-1.0(2)+0.61(10)i$ & $6.1(39)+0.61(10)i$ & $0.71(39)$ & $-0.18(3)-0.40(7)i$ & $0$ & $0.53(38)-0.40(7)i$\\ 
        $\Sigma^{*+}$ & $7.8(43)$ & $0.04(1)+0.40(7)i$ & $7.9(42)+0.40(7)i$ & $0.78(43)$ & $-0.08(1)-0.27(4)i$ & $-8.3(9)$ & $-7.6(10)-0.27(4)i$ \\ 
        $\Sigma^{*0}$ & $11.2(61)$ & $0.39(6)+0.20(3)i$ & $11.6(61)+0.20(3)i$ & $1.1(6)$ & $-0.02(1)-0.13(2)i$ & $-8.3(9)$ & $-7.2(11)-0.13(2)i$ \\ 
        $\Sigma^{*-}$ & $5.2(28)$ & $0.04(1)+0.40(7)i$ & $5.2(28)+0.40(7)i$ & $0.52(28)$ & $-0.08(1)-0.27(4)i$ & $0$ & $0.44(28)-0.27(4)i$ \\ 
        $\Xi^{*0}$ & $5.4(30)$ & $0.39(6)+0.20(3)i$ & $5.8(30)+0.20(3)i$ & $0.54(30)$ & $-0.02(1)-0.13(2)i$ & $-8.3(9)$ & $-7.7(9)-0.13(2)i$ \\ 
        $\Xi^{*-}$ & $3.2(17)$ & $1.1(2)+0.20(3)i$ & $4.4(18)+0.20(3)i$ & $0.32(18)$ & $0.02(1)-0.13(2)i$ & $0$ & $0.35(18)-0.13(2)i$ \\ 
        $\Omega^{-}$ & $1.3(7)$ & $2.2(4)$ & $3.5(8)$ & $0.13(7)$ & $0.12(2)$ & $0$ & $0.25(7)$ \\ 
        \hline\hline
    \end{tabular}
\end{table}

For the electric polarizabilities, the dominant contributions arise from the $\mathcal{T}\phi$-loop diagrams $\alpha_E^{(c-g)}$ rather than from the $\mathcal{N}\phi$-loop diagrams $\alpha_E^{\left(c^{\prime}-g^{\prime}\right)}$. This can be attributed to two main reasons. First, the $\mathcal{TT}\phi$ vertex involves a strong coupling, which is enhanced by flavor symmetry, as indicated in Table~\ref{tab:loop_cofficients_TT}. Second, according to quantum mechanical perturbation theory, both $\mathcal{T}\phi$-loops and $\mathcal{N}K$-loops correspond to higher energy intermediate states and thus yield positive, real contributions to the polarizabilities. In contrast, the $\mathcal{N}\pi$-loops contribute complex values with negative real parts. The $\mathcal{N}\pi$-loop and $\mathcal{N}K$-loop contributions partially cancel each other out.

Among the decuplet baryons, the $\Delta^{+}$ and $\Delta^0$ exhibit the largest electric polarizabilities, with values around $\alpha_E(\Delta^{0,+}) \simeq 17 \times 10^{-4} \, \mathrm{fm}^{3}$, significantly larger than typical nucleon values, $\alpha_E(n,p) \simeq 10 \times 10^{-4} \, \mathrm{fm}^{3}$~\cite{ParticleDataGroup:2024cfk,Bernard:1993bg}. This is primarily due to their large $\mathcal{T}\pi$-loop coefficients. Although U-spin symmetry in electromagnetic interactions leads to identical loop coefficients for $\Sigma^{*+}$ and $\Xi^{*0}$ in $\mathcal{T}K$-loops as for $\Delta^+$ and $\Delta^0$ in $\mathcal{T}\pi$-loops, the much larger kaon mass ($M_K \approx 4 M_\pi$) suppresses the $K$-loop contributions. Consequently, $\Sigma^{*+}$ and $\Xi^{*0}$ have comparatively smaller electromagnetic polarizabilities.

For magnetic polarizabilities, the small mass difference between decuplet baryons and octet baryons causes the magnetic dipole transition $\mathcal{T} \rightarrow \mathcal{N}\gamma$ to be dominant when present. This yields the negative contributions $\beta_M^{(b')}$ to the magnetic polarizabilities. Analogous to the electric polarizabilities, contributions from  $\mathcal{T}\phi$-loops $\beta_M^{(c-g)}$ are substantially larger than those from octet baryons $\beta_M^{(c'-g')}$. The total magnetic polarizabilities $\beta_M^{\text{Tot.}}$ display both positive and negative real parts.

\section{Electromagnetic polarizabilities of $\mathcal{B}_{6}^*$}\label{sec:B}
In Refs.~\cite{Meng:2018gan,Wang:2018gpl,Wang:2018cre}, the authors compute the magnetic moments of singly heavy baryons using heavy baryon chiral perturbation theory (HB$\chi$PT). These calculations show that decoupling $\mathcal{B}_{\bar{3}}$ from $\mathcal{B}_{6}^{(*)}$ improves chiral convergence due to the large mass splitting $\delta_{2,3}$. Conversely, in the heavy quark limit, $\mathcal{B}_{6}$ and $\mathcal{B}_6^*$ are nearly degenerate, and the smaller mass splitting $\delta_1$ has a negligible impact on chiral convergence. Moreover, when incorporating chiral fluctuations, it is essential to account for the coupling between $\mathcal{B}_{6}$ and $\mathcal{B}_6^*$, as dictated by heavy quark spin symmetry. In the subsequent discussion, we adopt this approach.  

\subsection{ANALYTICAL EXPRESSIONS}
\subsubsection{Tree diagrams}
The tree diagrams in Figs.~\ref{fig:fmdiagrams_full}($a$) and ($b$) yield the Thomson amplitude
\begin{equation}\label{eq:UV_B6_ab}
	U_{\xi}^{(a+b)}(\omega)=\frac{Q_\xi^2 e^2}{M_{\mathcal{B}_6^*}}, \quad V_{\xi}^{(a+b)}(\omega)=0,
\end{equation}
which does not contribute to the electromagnetic polarizabilities:
\begin{equation}
	\alpha_E^{(a+b)}\left(\xi\right)=\beta_M^{(a+b)}\left(\xi\right)=0.
\end{equation}

The tree diagram with $\mathcal{B}_6$ as the intermediate state, shown in Fig.~\ref{fig:fmdiagrams_full}($b^\prime$), yields:
\begin{equation}\label{eq:UV_fig_b_2}
	U_{\xi}^{(b^\prime)}(\omega)=\frac{e^2 C_\xi^2\omega^2}{24 M_N^2}\frac{\delta_1}{\delta_1^2-\omega^2},\quad
	V_{\xi}^{(b^\prime)}(\omega)=\frac{e^2 C_\xi^2}{24 M_N^2}\frac{\delta_1}{\delta_1^2-\omega^2},
\end{equation}
where $C_\xi$ represents the coefficients for different baryons, as listed in Table~\ref{tab:C_xi}. The coefficient $C_\xi$ is proportional to the transition magnetic moment obtained at leading order in HB$\chi$PT, following the same manner as in Eq.~\eqref{44},~\eqref{45} and \eqref{eq:transition_LO}:  
\begin{equation}\label{eq:transition_LO_B}  
	C_\xi=-\sqrt{6}~\frac{\mu^{\mathrm{LO}}_{\xi^*\to \xi +\gamma}}{\mu_N}.  
\end{equation}

The electromagnetic polarizabilities from Eq.~\eqref{eq:UV_fig_b_2} are:
\begin{equation}\label{eq:ab_b6_b2}
	\alpha^{(b^\prime)}_{E}(\xi)=0,\quad \beta^{(b^\prime)}_{M}(\xi)=-\frac{\alpha_{\mathrm{em}} C_\xi^2}{24 M_N^2} \frac{1}{\delta_1}.
\end{equation}
The mass splitting between $\mathcal{B}_6$ and $\mathcal{B}_6^*$ is significantly smaller than $M_{\mathcal{T}} - M_{\mathcal{N}}$, approaching zero in the heavy quark limit. As a result, Eq.~\eqref{eq:ab_b6_b2} provides a more substantial contribution than in the case of decuplet baryons.

\begin{table}[htbp]
    \centering
    \caption{The coefficients $C_\xi$ of the diagram in Fig.~\ref{fig:fmdiagrams_full}($b^\prime$)}
    \label{tab:C_xi}
    \setlength{\tabcolsep}{2.5mm}
    \begin{tabular}{c|cccccc}
    \hline\hline
         & $\Sigma_c^{*++}$ & $\Sigma_c^{*+}$ & $\Sigma_c^{*0}$ & $\Xi_{c}^{*+}$ & $\Xi_{c}^{*0}$ & $\Omega_c^{*0}$  \\ \hline
        $C_\xi$ & $f_7+\frac{2}{3}f_6$ & $f_7+\frac{1}{6}f_6$ & $f_7-\frac{1}{3}f_6$ & $f_7+\frac{1}{6}f_6$ & $f_7-\frac{1}{3}f_6$ & $f_7-\frac{1}{3}f_6$  \\
        \hline\hline
    \end{tabular}
\end{table}

\subsubsection{Loop diagrams}

Using the $\mathcal{J}$-function defined in Appendix~\ref{appendix:loop_integrals}, we can obtain the form factors of the $\mathcal{B}_6^*\phi$-loop diagrams in Figs.~\ref{fig:fmdiagrams_full}($c$)--($g$):
\begin{align}
U^{(c)}_{\xi}(\omega)= & A\sum_{\chi}\frac{D^{(c)}_{\xi,\chi}}{F_\chi^2}\mathcal{J}_0(\omega,0,M_{\chi}^2),\label{eq:U_6_c}\\
	U^{(d+e)}_{\xi}(\omega)= & A\left[\sum_{\chi} \frac{D_{\xi,\chi}^{(d+e)}}{F_\chi^2}\int_0^1 dx~\mathcal{J}^{\prime}_2(\omega x,0,M_{\chi}^2)\right],\\
	U^{(f)}_{\xi}(\omega)= & A\left[\sum_{\chi} \frac{D_{\xi,\chi}^{(f)}}{F_\chi^2}\int_0^1dx~ (1-x)(d+1)\mathcal{J}_6^{\prime\prime}(\omega x,0,M_{\chi}^2)\right.\notag \\
	&\quad ~ -\left. \sum_{\chi} \frac{D_{\xi,\chi}^{(f)}}{F_\chi^2} \int_0^1dx~\omega^2(1-x)x^2\mathcal{J}_2^{\prime\prime}(\omega x,0,M_{\chi}^2)\right],\\
	U^{(g)}_{\xi}(\omega)= &A\sum_{\chi}\frac{D^{(g)}_{\xi,\chi}}{F_\chi^2}(d-1)\mathcal{J}_2^{\prime}(0,0,M_{\chi}^2),\\
	V^{(c)}_{\xi}(\omega)= & 0,\\
	V^{(d+e)}_{\xi}(\omega)= & A\left[\sum_{\chi} -\frac{1}{2 F_\chi^2}D_{\xi,\chi}^{(d+e)}\int_0^1 dx~x(1-2x)\mathcal{J}^{\prime}_0(\omega x,0,M_{\chi}^2)\right],\\
	V^{(f)}_{\xi}(\omega)= & A\left[\sum_{\chi} \frac{1}{4 F_\chi^2}D^{(f)}_{\xi,\chi}\int_0^1dx~(1-x)\left[8x(2x-1)+(2x-1)^2(d-1)\right]\mathcal{J}_2^{\prime\prime}(\omega x,0,M_{\chi}^2)\right.\notag \\
	 &\quad ~ -\left. \sum_{\chi} \frac{1}{4 F_\chi^2}D^{(f)}_{\xi,\chi} \int_0^1 dx~ \omega^2(1-x)x^2(2x-1)^2 \mathcal{J}_0^{\prime\prime}(\omega x, 0,M_{\chi}^2)\right],\\
	 V^{(g)}_{\xi}(\omega)= & 0.\label{eq:V_6_g}
\end{align}
In Eqs.~\eqref{eq:U_6_c}--\eqref{eq:V_6_g}, we have considered the contributions from the crossed diagrams. $\chi$ denotes a specific meson in Eq.~\eqref{eq:octet_meson}, and $M_\chi$ represents its mass. $D_{\xi,\chi}$ are the coefficients of the loops, which are given in Table~\ref{tab:loop_cofficients}. $A$ is a common factor
\begin{equation}
	A=i\frac{(d^3-4d^2+d+6)e^2g_5^2}{4(d-1)^2},
\end{equation}
Expanding Eqs.~\eqref{eq:U_6_c}--\eqref{eq:V_6_g} into a power series in $\omega$ and combining with Eq.~\eqref{eq:def_alpha_beta}, we can obtain the electromagnetic polarizabilities from the $\mathcal{B}_6^*\phi$-loop diagrams in Figs.~\ref{fig:fmdiagrams_full}($c$)--($g$):
\begin{align}
	\alpha_E^{(c-g)}(\Sigma_c^{*++})=& \frac{25 \alpha_{\mathrm{em}}g_5^2}{1728 \pi M_\pi F_\pi^2}+ \frac{25 \alpha_{\mathrm{em}}g_5^2}{1728 \pi M_K F_K^2},\\
	\alpha_E^{(c-g)}(\Sigma_c^{*+})=&\frac{25 \alpha_{\mathrm{em}}g_5^2}{864 \pi M_\pi F_\pi^2}+ \frac{25 \alpha_{\mathrm{em}}g_5^2}{3456 \pi M_K F_K^2},\\
	\alpha_E^{(c-g)}(\Sigma_c^{*0})=& \frac{25 \alpha_{\mathrm{em}}g_5^2}{1728 \pi M_\pi F_\pi^2},\\
	\alpha_E^{(c-g)}(\Xi_c^{* +})=& \frac{25 \alpha_{\mathrm{em}}g_5^2}{3456 \pi M_\pi F_\pi^2}+ \frac{25 \alpha_{\mathrm{em}}g_5^2}{864 \pi M_K F_K^2},\\
	\alpha_E^{(c-g)}(\Xi_c^{* 0})=& \frac{25 \alpha_{\mathrm{em}}g_5^2}{3456 \pi M_\pi F_\pi^2}+ \frac{25 \alpha_{\mathrm{em}}g_5^2}{3456 \pi M_K F_K^2},\\
	\alpha_E^{(c-g)}(\Omega_c^{*0})=& \frac{25 \alpha_{\mathrm{em}}g_5^2}{1728 \pi M_K F_K^2},\\
	\beta_M^{(c-g)}(\xi) =& \frac{1}{10} \alpha_E^{(c-g)}(\xi).
\end{align}
The results diverge as $1/m_\chi$ in the chiral limit, similar to the behavior of decuplet baryons, which arise from long-range meson contributions.

\begin{table}[htbp]
    \centering
    \caption{The coefficients of the loop diagrams in Fig.~\ref{fig:fmdiagrams_full}.}
    \label{tab:loop_cofficients}
    \setlength{\tabcolsep}{3mm}
    \begin{tabular}{c|cccccc}
    \hline\hline
         & $\Sigma_c^{*++}$ & $\Sigma_c^{*+}$ & $\Sigma_c^{*0}$ & $\Xi_{c}^{*+}$ & $\Xi_{c}^{*0}$ & $\Omega_c^{*0}$  \\ \hline
        $D_{\pi}^{(c)}$ & $-\frac{1}{2}$ & $-1$ & $-\frac{1}{2}$ & $-\frac{1}{4}$ & $-\frac{1}{4}$ & $0$  \\ 
        $D_{K}^{(c)}$ & $-\frac{1}{2}$ & $-\frac{1}{4}$ & $0$ & $-1$ & $-\frac{1}{4}$ & $-\frac{1}{2}$  \\
        $D_{\pi}^{(d+e)}$ & $2$ & $4$ & $2$ & $1$ & $1$ & $0$  \\ 
        $D_{K}^{(d+e)}$ & $2$ & $1$ & $0$ & $4$ & $1$ & $2$  \\ 
        $D_{\pi}^{(f)}$ & $-2$ & $-4$ & $-2$ & $-1$ & $-1$ & $0$  \\ 
        $D_{K}^{(f)}$ & $-2$ & $-1$ & $0$ & $-4$ & $-1$ & $-2$  \\ 
        $D_{\pi}^{(g)}$ & $\frac{1}{2}$ & $1$ & $\frac{1}{2}$ & $\frac{1}{4}$ & $\frac{1}{4}$ & $0$  \\ 
        $D_{K}^{(g)}$ & $\frac{1}{2}$ & $\frac{1}{4}$ & $0$ & $1$ & $\frac{1}{4}$ & $\frac{1}{2}$ \\ 
        \hline\hline
    \end{tabular}
\end{table}

Similarly, the form factors of the $\mathcal{B}_6\phi$-loop diagrams in Figs.~\ref{fig:fmdiagrams_full}($c^\prime$)--($g^\prime$) are:
\begin{align}
	U^{(c^\prime)}_{\xi}(\omega)= & B\sum_{\chi}\frac{D^{(c)}_{\xi,\chi}}{F_\chi^2}\mathcal{J}_0(\omega,\delta_1,M_{\chi}^2)\label{eq:U_6s_c},\\
	U^{(d^\prime+e^\prime)}_{\xi}(\omega)= & B\left[\sum_{\chi} \frac{D_{\xi,\chi}^{(d+e)}}{F_\chi^2}\int_0^1 dx~\mathcal{J}^{\prime}_2(\omega x,\delta_1,M_{\chi}^2)\right],\\
	U^{(f^\prime)}_{\xi}(\omega)= & B\left[\sum_{\chi} \frac{D_{\xi,\chi}^{(f)}}{F_\chi^2}\int_0^1dx~ (1-x)(d+1)\mathcal{J}_6^{\prime\prime}(\omega x,\delta_1,M_{\chi}^2)\right.\notag \\
	&\quad ~ -\left. \sum_{\chi} \frac{D_{\xi,\chi}^{(f)}}{F_\chi^2} \int_0^1dx~\omega^2(1-x)x^2\mathcal{J}_2^{\prime\prime}(\omega x,\delta_1,M_{\chi}^2)\right],\\
	U^{(g^\prime)}_{\xi}(\omega)= &B\sum_{\chi}\frac{D^{(g)}_{\xi,\chi}}{F_\chi^2}(d-1)\mathcal{J}_2^{\prime}(0,\delta_1,M_{\chi}^2),\\
	V^{(c^\prime)}_{\xi}(\omega)= & 0,\\
	V^{(d^\prime+e^\prime)}_{\xi}(\omega)= & B\left[\sum_{\chi} -\frac{1}{2 F_\chi^2}D_{\xi,\chi}^{(d+e)}\int_0^1 dx~x(1-2x)\mathcal{J}^{\prime}_0(\omega x,\delta_1,M_{\chi}^2)\right],\\
	V^{(f^\prime)}_{\xi}(\omega)= & B\left[\sum_{\chi} \frac{1}{4 F_\chi^2}D^{(f)}_{\xi,\chi}\int_0^1dx~(1-x)\left[8x(2x-1)+(2x-1)^2(d-1)\right]\mathcal{J}_2^{\prime\prime}(\omega x,\delta_1,M_{\chi}^2)\right.\notag \\
	 &\quad ~ -\left. \sum_{\chi} \frac{1}{4 F_\chi^2}D^{(f)}_{\xi,\chi} \int_0^1 dx~ \omega^2(1-x)x^2(2x-1)^2 \mathcal{J}_0^{\prime\prime}(\omega x,\delta_1,M_{\chi}^2)\right],\\
	 V^{(g^\prime)}_{\xi}(\omega)= & 0,\label{eq:V_6s_g}
\end{align}
where $B$ is a common factor
\begin{equation}
	B=i\frac{e^2g_3^2}{4}\frac{d-2}{d-1}.
\end{equation}
Expanding Eq.~\eqref{eq:U_6s_c}--\eqref{eq:V_6s_g} into a power series in $\omega$ and combining with Eq.~\eqref{eq:def_alpha_beta}, we can obtain the electromagnetic polarizabilities from the $\mathcal{B}_6\phi$-loop diagrams in Figs.~\ref{fig:fmdiagrams_full}($c^\prime$)--($g^\prime$):

\begin{align}
	\alpha_E^{(c^\prime-g^\prime)}\left(\Sigma_c^{*++}\right)=&\frac{\alpha_{\mathrm{em}} g_3^2 S_\pi}{576\pi^2 F_\pi^2 \left(M_\pi^2-\delta_1^2\right)^2}+\frac{\alpha_{\mathrm{em}} g_3^2 S_K}{576\pi^2 F_K^2 \left(M_K^2-\delta_1^2\right)^2},\\
	\alpha_E^{(c^\prime-g^\prime)}\left(\Sigma_c^{*+}\right)=&\frac{\alpha_{\mathrm{em}} g_3^2 S_\pi}{288\pi^2 F_\pi^2 \left(M_\pi^2-\delta_1^2\right)^2}+\frac{\alpha_{\mathrm{em}} g_3^2 S_K}{1152\pi^2 F_K^2 \left(M_K^2-\delta_1^2\right)^2},\\
	\alpha_E^{(c^\prime-g^\prime)}\left(\Sigma_c^{*0}\right)=&\frac{\alpha_{\mathrm{em}} g_3^2 S_\pi}{576\pi^2 F_\pi^2 \left(M_\pi^2-\delta_1^2\right)^2},\\
	\alpha_E^{(c^\prime-g^\prime)}\left(\Xi_c^{* +}\right)=&\frac{\alpha_{\mathrm{em}} g_3^2 S_\pi}{1152\pi^2 F_\pi^2 \left(M_\pi^2-\delta_1^2\right)^2}+\frac{\alpha_{\mathrm{em}} g_3^2 S_K}{288\pi^2 F_K^2 \left(M_K^2-\delta_1^2\right)^2},\\
	\alpha_E^{(c^\prime-g^\prime)}\left(\Xi_c^{* 0}\right)=&\frac{\alpha_{\mathrm{em}} g_3^2 S_\pi}{1152\pi^2 F_\pi^2 \left(M_\pi^2-\delta_1^2\right)^2}+\frac{\alpha_{\mathrm{em}} g_3^2 S_K}{1152\pi^2 F_K^2 \left(M_K^2-\delta_1^2\right)^2},\\
	\alpha_E^{(c^\prime-g^\prime)}\left(\Omega_c^{*0}\right)=&\frac{\alpha_{\mathrm{em}} g_3^2 S_K}{576\pi^2 F_K^2 \left(M_K^2-\delta_1^2\right)^2}.\\
	\beta_M^{(c^\prime-g^\prime)}\left(\Sigma_c^{*++}\right)=&\frac{\alpha_{\mathrm{em}} g_3^2 R_\pi}{576 \pi^2 F_\pi^2 (M_\pi^2-\delta_1^2)}+\frac{\alpha_{\mathrm{em}} g_3^2 R_K}{576 \pi^2 F_K^2 (M_K^2-\delta_1^2)},\\
	\beta_M^{(c^\prime-g^\prime)}\left(\Sigma_c^{*+}\right)=&\frac{\alpha_{\mathrm{em}} g_3^2 R_\pi}{288 \pi^2 F_\pi^2 (M_\pi^2-\delta_1^2)}+\frac{\alpha_{\mathrm{em}} g_3^2 R_K}{1152 \pi^2 F_K^2 (M_K^2-\delta_1^2)},\\
	\beta_M^{(c^\prime-g^\prime)}\left(\Sigma_c^{*0}\right)=&\frac{\alpha_{\mathrm{em}} g_3^2 R_\pi}{576 \pi^2 F_\pi^2 (M_\pi^2-\delta_1^2)},\\
	\beta_M^{(c^\prime-g^\prime)}\left(\Xi_c^{*+}\right)=&\frac{\alpha_{\mathrm{em}} g_3^2 R_\pi}{1152 \pi^2 F_\pi^2 (M_\pi^2-\delta_1^2)}+\frac{\alpha_{\mathrm{em}} g_3^2 R_K}{288 \pi^2 F_K^2 (M_K^2-\delta_1^2)},\\
	\beta_M^{(c^\prime-g^\prime)}\left(\Xi_c^{* 0}\right)=&\frac{\alpha_{\mathrm{em}} g_3^2 R_\pi}{1152 \pi^2 F_\pi^2 (M_\pi^2-\delta_1^2)}+\frac{\alpha_{\mathrm{em}} g_3^2 R_K}{1152 \pi^2 F_K^2 (M_K^2-\delta_1^2)},\\
	\beta_M^{(c^\prime-g^\prime)}\left(\Omega_c^{*0}\right)=&\frac{\alpha_{\mathrm{em}} g_3^2 R_K}{576 \pi^2 F_K^2 (M_K^2-\delta_1^2)},
\end{align}
where we have defined
\begin{equation}
\begin{aligned}
	R_\chi&=\sqrt{M_\chi^2-\delta_1^2}\arccos\left[\frac{-\delta_1}{M_\chi}\right],\\
	S_\chi&=M_\pi^2\left(10R_\pi+9\delta_1\right)+\delta_1^2\left(-9\delta_1-R_\pi\right).
	\end{aligned}
\end{equation}

In the heavy quark limit with $g_5^2 = 3 g_3^2$~\cite{Cheng:1993kp,Cho:1992nt,Jiang:2015xqa} and  $\delta_1 = 0$ , we have
\begin{equation}
	R_\chi = \frac{\pi M_\chi}{2}, \quad S_\chi = 5 \pi M_\chi^3.
\end{equation}
It is easy to verify that in the heavy quark limit:
\begin{equation}\label{eq:relation_hqs}
	\alpha_E^{(c-g)}(\xi) =5 \alpha_E^{(c'-g')}(\xi),\quad \beta_M^{(c-g)}(\xi) = 5 \beta_M^{(c'-g')}(\xi).
\end{equation}
In realistic physical scenarios, $\delta_1$ represents a non-zero mass splitting that reduces the energy difference between $\mathcal{B}_6\phi$ and $\mathcal{B}_6^*$. Therefore, we expect the electromagnetic polarizabilities arising from the $\mathcal{B}_6\phi$-loop diagrams to be enhanced compared to those in the heavy quark limit.

Finally, by summing all the contributions calculated above, we can obtain the total electromagnetic polarizabilities:
\begin{equation}
	\begin{aligned}
	\alpha_E^{\mathrm{Tot.}}(\xi) &=\alpha_E^{(c-g)}(\xi)+\alpha_E^{(c^\prime-g^\prime)}(\xi),\\
	\beta_M^{\mathrm{Tot.}}(\xi) &=\beta_M^{(b^\prime)}(\xi)+\beta_M^{(c-g)}(\xi)+\beta_M^{(c^\prime-g^\prime)}(\xi).
    \end{aligned}
\end{equation}

\subsection{NUMERICAL RESULTS}\label{sec:NUMERICAL_RESULTS}

The low-energy constant $g_2$ can be determined by decay width of $\Sigma_c$,
\begin{equation}
\Gamma\left(\Sigma_c \rightarrow \Lambda_c \pi\right)=\frac{g_2^2}{4 \pi f_\pi^2} \frac{M_{\Lambda_c}}{M_{\Sigma_c}}\left|\vec{p}_\pi\right|^3,
\end{equation}
\begin{equation}
    g_2=-0.568 \pm 0.023.
\end{equation}
The other low energy constants $g_3,g_5$ can be extracted with the help of quark model~\cite{Yan:1992gz}. 
\begin{equation}
    g_3=-\sqrt{2} g_2=0.803\pm 0.080\pm0.032 , \quad g_5=\sqrt{6}g_2=-1.39 \pm0.14\pm 0.056,
\end{equation}
The first uncertainty arises from the assumed 10\% error introduced by using the quark model, while the second is due to the experimental uncertainty in the $\Sigma_c^{(*)}$ widths. 

In tree diagram Fig.~\ref{fig:fmdiagrams_full}($b'$), the the coefficients $C_\xi$ contain two remaining LECs ($f_6$ and $f_7$), as shown in Table~\ref{tab:C_xi}. For the lack of experimental data of electromagnetic decay of singly heavy barons, we fit the $C_\xi$'s from the quark model and Lattice QCD simulations, following the same approach in our previous work~\cite{Chen:2024xks}. This fit yields the following results (in units of $\mu_N$):
\begin{equation}
	\mu_u=-2\mu_d=1.078(88),\quad \mu_s=-0.456(23),\quad \mu_c=0.205(15),
\end{equation}
The values in parentheses represent the uncertainties originating from the lattice QCD results. It can be seen that the magnetic moment of the charm quark is suppressed by $1/m_c$. Then we calculate the transition magnetic moments for singly charmed baryons in quark model~\cite{Wang:2018gpl}, which are (in units of $\mu_N$):
\begin{align}
	&\mu_{\Sigma_c^{*++} \rightarrow \Sigma_c^{++} \gamma}  =\frac{2 \sqrt{2}}{3}\left(\mu_u-\mu_c\right)=0.82(10),\\
	&\mu_{\Sigma_c^{*+} \rightarrow \Sigma_c^{+} \gamma}  =\frac{ \sqrt{2}}{3}\left(\mu_u+\mu_d-2\mu_c\right)=0.06(3),\\
	&\mu_{\Sigma_c^{*0} \rightarrow \Sigma_c^{0} \gamma}  =\frac{ 2 \sqrt{2}}{3}\left(\mu_d-\mu_c\right)=-0.70(5),\\
	&\mu_{\Xi_c^{*+} \rightarrow \Xi_c^{\prime +} \gamma}  =\frac{ \sqrt{2}}{3}\left(\mu_u+\mu_s-2\mu_c\right)=0.10(6),\\
	&\mu_{\Xi_c^{*0} \rightarrow \Xi_c^{\prime 0} \gamma}  =\frac{ \sqrt{2}}{3}\left(\mu_d+\mu_s-2\mu_c\right)=-0.66(3),\\
	&\mu_{\Omega_c^{*0} \rightarrow \Omega_c^{0} \gamma}  =\frac{2 \sqrt{2}}{3}\left(\mu_s-\mu_c\right)=-0.62(3).
\end{align}

The numerical results for the electromagnetic polarizabilities of singly charmed baryons are listed in Table.~\ref{tab:polarizabilities_numerical_results}. We observe that
\begin{equation}
	\alpha_E^{(c-g)}(\xi)\approx 2\alpha_E^{(c'-g')}(\xi),\quad \beta_M^{(c-g)}(\xi)\approx 3\beta_M^{(c'-g')}(\xi).
\end{equation}
The deviation from Eq.~\eqref{eq:relation_hqs} aligns with expectations. The non-zero mass splitting decreases the energy difference between $\mathcal{B}_6^*$ and $\mathcal{B}_6\phi$, thereby enhancing the $\mathcal{B}_6\phi$-loop contributions. The results indicate that long-range chiral corrections make substantial contributions to the electric polarizabilities. Meanwhile, magnetic dipole transitions $\mathcal{B}_6^* \to \mathcal{B}_6 + \gamma$ for the other sextet baryons significantly contribute to the magnetic polarizabilities, except for $\Sigma_c^{*+}$ and $\Xi_c^{* +}$, whose transition magnetic moments $\mu_{\xi^*\to \xi+\gamma} \approx 0$. 

\begin{table}[htbp]	\caption{Numerical results for the electromagnetic polarizabilities of spin-$\frac{3}{2}$ singly charmed baryons (in units of $10^{-4}\mathrm{fm}^3$). The first two sets of columns show the electric and magnetic polarizabilities of the spin-$\frac{3}{2}$ states, respectively. For comparison, the last two columns list the total electric and magnetic polarizabilities of spin-$\frac{1}{2}$ singly charmed baryons, taken from Ref.\cite{Chen:2024xks}. Uncertainties are given in parentheses.}
	\label{tab:polarizabilities_numerical_results}
    \centering
    \setlength{\tabcolsep}{2.2mm}
    \begin{tabular}{c|ccc|cccc||c|c|c|}
    \hline\hline
         & $\alpha_E^{(c-g)}$ & $\alpha_E^{(c^\prime-g^\prime)}$ & $\alpha_E^{\text{Tot.}}$ & $\beta_M^{(c-g)}$ & $\beta_M^{(c^\prime-g^\prime)}$ & $\beta_M^{(b^\prime)}$ & $\beta_M^{\mathrm{Tot.}}$ &  & $\alpha_E^{\mathrm{Tot.}}$ &  $\beta_M^{\mathrm{Tot.}}$  \\ \hline
        $\Sigma_c^{*++}$ & $5.05(110)$ & $2.16(47)$ & $7.20(156)$ & $0.51(11)$ & $0.15(3)$ & $-1.61(38)$ & $-0.95(40)$ &$\Sigma_c^{++}$ &  $5.39(151)$ & $3.78(77)$ \\ 
        $\Sigma_c^{*+}$ & $8.91(193)$ & $4.04(87)$ & $12.94(280)$ & $0.89(19)$ & $0.27(6)$ & $-0.01(1)$ & $1.15(25)$ &  $\Sigma_c^{+}$ &  $9.42(264)$  & $1.01(28)$ \\ 
        $\Sigma_c^{*0}$ & $4.26(92)$ & $1.97(42)$ & $6.23(135)$ & $0.43(9)$ & $0.13(3)$ & $-1.17(16)$ & $-0.61(20)$ & $\Sigma_c^{0}$ & $4.48(126)$  & $2.81(36)$\\ 
        $\Xi_{c}^{*+}$ & $3.71(80)$ & $1.36(29)$ & $5.07(110)$ & $0.37(8)$ & $0.10(2)$ & $-0.03(3)$ & $0.45(11)$ &$\Xi_{c}^{\prime+}$ &  $4.05(114)$ & $0.47(13)$\\ 
        $\Xi_{c}^{*0}$ & $2.52(55)$ & $1.08(23)$ & $3.60(78)$ & $0.25(5)$ & $0.07(2)$ & $-1.04(10)$ & $-0.71(12)$ &$\Xi_{c}^{\prime0}$  & $2.69(76)$ & $2.36(20)$ \\ 
        $\Omega_c^{*0}$ & $0.79(17)$ & $0.19(4)$ & $0.98(21)$ & $0.08(2)$ & $0.02(1)$ & $-0.92(8)$ & $-0.82(8)$ & $\Omega_c^{0}$ & $0.90(26)$ & $1.94(17)$\\ \hline\hline
    \end{tabular}
\end{table}

For comparison, we present the electromagnetic polarizabilities of spin-$\frac{1}{2}$ sextet singly charmed baryons in the last two columns of  Table~\ref{tab:polarizabilities_numerical_results}. We observe that the electric polarizabilities of most spin-$\frac{3}{2}$ singly charmed baryons are comparable to those of spin-$\frac{1}{2}$ singly charmed baryons. For magnetic polarizabilities, the situation differs significantly. 
The magnetic polarizabilities $\beta_M^{(b')}$ arising from intermediate spin-$\frac{1}{2}$ singly heavy baryons become negative. As a result, the contributions to the magnetic polarizabilities from tree diagrams and loop diagrams tend to cancel each other out. Consequently, the magnetic polarizabilities of spin-$\frac{3}{2}$ singly charmed baryons exhibit both positive and negative values. The electromagnetic polarizabilities of bottom baryons are similar to those of charm baryons and are detailed in Appendix~\ref{appendix:b_baryon_polarizabilities}.


\section{Summary}\label{sec:DISCUSSION_AND_CONCULSION}

We calculate the electromagnetic polarizabilities of the spin-$\frac{3}{2}$ baryons. The analytical expressions are derived up to $\mathcal{O}(p^3)$. For the decuplet baryons, we utilize the strong and electromagnetic decay widths to determine several low-energy constants. Our results demonstrate that long-range chiral corrections make substantial contributions to the polarizabilities, with $\mathcal{T}\phi$-loop contributions overwhelming those from $\mathcal{N}\phi$-loops. Additionally, the non-zero M1 transitions dominate magnetic polarizabilities when present.

For the spin-$\frac{3}{2}$ singly heavy baryons, we employ the heavy quark symmetry and the quark model to estimate the low-energy constants. Our results indicate that both the $\mathcal{B}^*_6 \phi$-loop and the $\mathcal{B}_6 \phi$-loop play a crucial role in the polarizabilities. The electric polarizabilities of spin-$\frac{3}{2}$ singly heavy baryons are similar to those of spin-$\frac{1}{2}$ singly heavy baryons. In contrast, magnetic polarizabilities of spin-$\frac{3}{2}$ singly heavy baryons show both positive and negative values, differing significantly from spin-$\frac{1}{2}$ singly heavy baryons.

Our numerical results can be further refined with new experimental data and lattice QCD simulation results in the future. Meanwhile, our analytical expressions can facilitate chiral extrapolation in lattice QCD simulations.


\section*{ACKNOWLEDGMENTS}
We are grateful to Zi-Yang Lin for the helpful discussions. This project was supported by the National
Natural Science Foundation of China (12475137) and by ERC NuclearTheory (grant No. 885150). The computational resources were supported by the high-performance computing platform of Peking University.

\begin{appendix}
\section{General Spin-averaged Compton Tensor}\label{appendix:General_Spin-averaged_Compton_Tensor}
The general spin-averaged Compton tensor $\Theta_{\mu \nu}$ with gauge invariance and crossing symmetry can be expressed in terms of four independent basis tensors $\mathcal{L}_i^{\mu \nu}$ that are free from singularities~\cite{Bernabeu:1976jq},
\begin{equation}
\Theta^{\mu \nu}=\sum_{i=1}^4 B_i\left(q^2, Q \cdot K\right) \mathcal{L}_i^{\mu \nu},
\label{eq:general compton}
\end{equation}

\begin{equation}
\begin{aligned}
& L_1^{\mu \nu}=k \cdot k^{\prime} g^{\mu \nu}-k^\mu k^{\prime \nu}, \\
& L_2^{\mu \nu}=k \cdot k^{\prime} Q^\mu Q^\nu-Q \cdot K\left(Q^\mu k^{\prime \nu}+k^\mu Q^\nu\right)+(Q \cdot K)^2 g^{\mu \nu}, \\
&L_3^{\mu \nu}  =k \cdot k^{\prime}\left(Q^\mu k^\nu+k^{\prime \mu} Q^\nu\right)-Q \cdot K\left(k^\mu k^\nu+k^{\prime \mu} k^{\prime \nu}\right), \\
&L_4^{\mu \nu}  =k^{\prime \mu} k^\nu,
\end{aligned}
\end{equation}
where we denote the four-momenta of the initial (final) target particle and photon by p and k (p' and k'),  respectively, and define $Q=\frac{p+p^{\prime}}{2}$,$K=\frac{k+k^{\prime}}{2}$ and $q=p'-p$.

The pole-free part of the electromagnetic polarizabilities arises at the photon momentum order of $O(p^2)$. Consequently, the term $\mathcal{L}_3^{\mu \nu}$, beging of order $O(p^3)$ does not contribute to the polarizabilities. In contrast, the $\mathcal{L}_4^{\mu \nu}$ term does not contribute due to the kinematical zero of the associated factor  $B_4(0,0)$~\cite{Llanta:1979kj}. 

\section{Loop Integrals}\label{appendix:loop_integrals}
To combine propagator denominators, we introduce integrals over Feynman parameters:
\begin{equation}
	\frac{1}{A_1 A_2 \cdots A_n}=\int_0^1 d x_1 \cdots d x_n \delta\left(\sum x_i-1\right) \frac{(n-1)!}{\left[x_1 A_1+x_2 A_2+\cdots x_n A_n\right]^n}.
\end{equation}
To regularize divergent loop integrals, we use the dimensional regularization scheme and expand them around 4-dimensional spacetime. In this way, one can define the loop functions that frequently occur in calculations~\cite{Bernard:1995dp,Scherer:2002tk,Hemmert:1996rw}. Here, we list only those that we need:
\begin{equation}
\begin{aligned}
\frac{1}{i} &\int \frac{d^d \ell}{(2 \pi)^d} \frac{\left\{1, \ell_\mu \ell_\nu, \ell_\mu \ell_\nu \ell_\alpha \ell_\beta\right\}}{(v \cdot \ell-\omega-i \epsilon)\left(M_{\chi}^2-\ell^2-i \epsilon\right)}  \\
&=\left\{J_0\left(\omega, M_{\chi}^2\right)\right.,\quad g_{\mu \nu} J_2\left(\omega, M_{\chi}^2\right)+v_\mu v_\nu J_3\left(\omega, M_{\chi}^2\right),\quad \left.\left(g_{\mu \nu} g_{\alpha \beta}+\text { perm. }\right) J_6\left(\omega, M_{\chi}^2\right)+\ldots\right\}.
\end{aligned}
\end{equation}
All loop-integrals can be expressed via the basis-function $J_0$:
\begin{equation}\label{eq:J0_J2_J6}
\begin{aligned}
	J_0\left(\omega, M_{\chi}^2\right) & =-4 L \omega+\frac{\omega}{8 \pi^2}\left(1-2 \ln \frac{M_\chi}{\mu}\right)-\frac{1}{4 \pi^2} \sqrt{M_{\chi}^2-\omega^2} \arccos \frac{-\omega}{M_\chi}+\mathcal{O}(d-4),\\
J_2\left(\omega, M_{\chi}^2\right) & =\frac{1}{d-1}\left[\left(M_{\chi}^2-\omega^2\right) J_0\left(\omega, M_{\chi}^2\right)-\omega \Delta_{\chi}\right], \\
J_6\left(\omega, M_{\chi}^2\right) & =\frac{1}{d+1}\left[\left(M_{\chi}^2-\omega^2\right) J_2\left(\omega, M_{\chi}^2\right)-\frac{M_{\chi}^2 \omega}{d} \Delta_{\chi}\right].
\end{aligned}
\end{equation}
In Eq.~\eqref{eq:J0_J2_J6} we have used
\begin{equation}
\begin{aligned}
\Delta_{\chi} & =2 M_{\chi}^2\left(L+\frac{1}{16 \pi^2} \ln \frac{M_\chi}{\mu}\right) +\mathcal{O}(d-4),\\
L & =\frac{\mu^{d-4}}{16 \pi^2}\left[\frac{1}{d-4}+\frac{1}{2}\left(\gamma_E-1-\ln 4 \pi\right)\right],
\end{aligned}
\end{equation}
The $\gamma_E=0.557215$ is Euler constant. The scale $\mu$ is introduced in dimensional regularization.

For the spin-averaged forward Compton scattering amplitude, $J_i(-\omega+\delta)$ and $J_i(\omega+\delta)$ always appear symmetrically. Therefore, for simplicity, we define a new $\mathcal{J}$-function:
\begin{equation}
	\mathcal{J}_i(\omega,\delta,M_{\chi}^2)=J_i(\omega+\delta,M_{\chi}^2)+J_i(-\omega+\delta,M_{\chi}^2)
\end{equation}
With $\mathcal{J}_i^{\prime}$ and $\mathcal{J}_i^{\prime \prime}$ we define the first and second partial derivative with respect to $M_{\chi}^2$,

\begin{equation}
\begin{aligned}
\mathcal{J}_i^{\prime}\left(\omega,\delta ,M_{\chi}^2\right) & =\frac{\partial}{\partial\left(M_{\chi}^2\right)} \mathcal{J}_i\left(\omega, \delta,M_{\chi}^2\right) \\
\mathcal{J}_i^{\prime \prime}\left(\omega, \delta,M_{\chi}^2\right) & =\frac{\partial^2}{\partial\left(M_{\chi}^2\right)^2} \mathcal{J}_i\left(\omega,\delta, M_{\chi}^2\right)
\end{aligned}
\end{equation}

\section{The electromagnetic polarizabilities of singly bottom baryons}~\label{appendix:b_baryon_polarizabilities}

The analytical expression for the electromagnetic polarizabilities of singly bottom baryons is basically the same as that for singly charmed baryons. The only difference is that the parameters should be replaced as follows:
\begin{equation}
	g_1\to g_{1,b},\quad g_{3}\to g_{3,b},\quad \delta_1\to \delta_{1,b},\quad C_{\xi}\to C_{\xi,b}.
\end{equation}
Using the same approach as in Sec.~\ref{sec:NUMERICAL_RESULTS}, we determine the parameters as:
\begin{equation}
	g_{3,b}=0.707 \pm 0.071 \pm 0.025,\quad 
    g_{5,b}=1.225 \pm 0.123 \pm 0.043,\quad \delta_{1,b}=20~\mathrm{MeV}.
\end{equation}
The magnetic moment of the $b$ quark cannot be precisely determined due to the lack of experimental or lattice QCD data. However, since the bottom quark is extremely heavy, its magnetic moment should be very small and will not significantly affect the final results. We estimate it as:
\begin{equation}
	\mu_b=(-0.05\pm 0.05)\mu_N.
\end{equation}
The numerical results for the electromagnetic polarizabilities of singly bottom baryons are listed in Table ~\ref{tab:polarizabilities_numerical_results_b}.

\begin{table}[htbp]
	\caption{The numerical results of spin-$\frac{3}{2}$ singly bottom baryon electromagnetic polarizabilities (in unit of $10^{-4}~\mathrm{fm}^3$). The values in parentheses represent the uncertainties of the results.}
	\label{tab:polarizabilities_numerical_results_b}
    \centering
    \setlength{\tabcolsep}{3mm}
    \begin{tabular}{c|ccc|cccc}
    \hline\hline
         & $\alpha_E^{(c-g)}$ & $\alpha_E^{(c^\prime-g^\prime)}$ & $\alpha_E^{\text{Tot.}}$ & $\beta_M^{(c-g)}$ & $\beta_M^{(c^\prime-g^\prime)}$ & $\beta_M^{(b^\prime)}$ & $\beta_M^{\mathrm{Tot.}}$   \\ \hline
        $\Sigma_b^{*+}$ & $3.9(8)$ & $0.92(20)$ & $4.8(10)$ & $0.39(8)$ & $0.085(18)$ & $-9.1(18)$ & $-8.6(18)$ \\ 
        $\Sigma_b^{*0}$ & $6.9(15)$ & $1.7(4)$ & $8.5(18)$ & $0.69(15)$ & $0.15(3)$ & $-0.73(27)$ & $0.11(32)$ \\ 
        $\Sigma_b^{*-}$ & $3.3(7)$ & $0.8(2)$ & $4.1(9)$ & $0.33(7)$ & $0.073(15)$ & $-1.7(5)$ & $-1.3(5)$ \\ 
        $\Xi_{b}^{* 0}$ & $2.9(6)$ & $0.66(14)$ & $3.5(8)$ & $0.29(6)$ & $0.061(13)$ & $-0.93(39)$ & $-0.58(39)$\\ 
        $\Xi_{b}^{* -}$ & $2.0(4)$ & $0.46(10)$ & $2.4(5)$ & $0.20(4)$ & $0.043(9)$ & $-1.4(4)$ & $-1.2(4)$ \\ 
        $\Omega_b^{*-}$ & $0.61(13)$ & $0.13(3)$ & $0.74(16)$ & $0.061(13)$ & $0.013(3)$ & $-1.2(4)$ & $-1.1(4)$ \\ \hline\hline
    \end{tabular}
\end{table}

\end{appendix}
\bibliography{references}

\begin{thebibliography}{84}%
\makeatletter
\providecommand \@ifxundefined [1]{%
 \@ifx{#1\undefined}
}%
\providecommand \@ifnum [1]{%
 \ifnum #1\expandafter \@firstoftwo
 \else \expandafter \@secondoftwo
 \fi
}%
\providecommand \@ifx [1]{%
 \ifx #1\expandafter \@firstoftwo
 \else \expandafter \@secondoftwo
 \fi
}%
\providecommand \natexlab [1]{#1}%
\providecommand \enquote  [1]{``#1''}%
\providecommand \bibnamefont  [1]{#1}%
\providecommand \bibfnamefont [1]{#1}%
\providecommand \citenamefont [1]{#1}%
\providecommand \href@noop [0]{\@secondoftwo}%
\providecommand \href [0]{\begingroup \@sanitize@url \@href}%
\providecommand \@href[1]{\@@startlink{#1}\@@href}%
\providecommand \@@href[1]{\endgroup#1\@@endlink}%
\providecommand \@sanitize@url [0]{\catcode `\\12\catcode `\$12\catcode `\&12\catcode `\#12\catcode `\^12\catcode `\_12\catcode `\%12\relax}%
\providecommand \@@startlink[1]{}%
\providecommand \@@endlink[0]{}%
\providecommand \url  [0]{\begingroup\@sanitize@url \@url }%
\providecommand \@url [1]{\endgroup\@href {#1}{\urlprefix }}%
\providecommand \urlprefix  [0]{URL }%
\providecommand \Eprint [0]{\href }%
\providecommand \doibase [0]{https://doi.org/}%
\providecommand \selectlanguage [0]{\@gobble}%
\providecommand \bibinfo  [0]{\@secondoftwo}%
\providecommand \bibfield  [0]{\@secondoftwo}%
\providecommand \translation [1]{[#1]}%
\providecommand \BibitemOpen [0]{}%
\providecommand \bibitemStop [0]{}%
\providecommand \bibitemNoStop [0]{.\EOS\space}%
\providecommand \EOS [0]{\spacefactor3000\relax}%
\providecommand \BibitemShut  [1]{\csname bibitem#1\endcsname}%
\let\auto@bib@innerbib\@empty
\bibitem [{\citenamefont {Rosenbluth}(1950)}]{Rosenbluth:1950yq}%
  \BibitemOpen
  \bibfield  {author} {\bibinfo {author} {\bibfnamefont {M.~N.}\ \bibnamefont {Rosenbluth}},\ }\bibfield  {title} {\bibinfo {title} {{High Energy Elastic Scattering of Electrons on Protons}},\ }\href {https://doi.org/10.1103/PhysRev.79.615} {\bibfield  {journal} {\bibinfo  {journal} {Phys. Rev.}\ }\textbf {\bibinfo {volume} {79}},\ \bibinfo {pages} {615} (\bibinfo {year} {1950})}\BibitemShut {NoStop}%
\bibitem [{\citenamefont {Bezginov}\ \emph {et~al.}(2019)\citenamefont {Bezginov}, \citenamefont {Valdez}, \citenamefont {Horbatsch}, \citenamefont {Marsman}, \citenamefont {Vutha},\ and\ \citenamefont {Hessels}}]{Bezginov:2019mdi}%
  \BibitemOpen
  \bibfield  {author} {\bibinfo {author} {\bibfnamefont {N.}~\bibnamefont {Bezginov}}, \bibinfo {author} {\bibfnamefont {T.}~\bibnamefont {Valdez}}, \bibinfo {author} {\bibfnamefont {M.}~\bibnamefont {Horbatsch}}, \bibinfo {author} {\bibfnamefont {A.}~\bibnamefont {Marsman}}, \bibinfo {author} {\bibfnamefont {A.~C.}\ \bibnamefont {Vutha}},\ and\ \bibinfo {author} {\bibfnamefont {E.~A.}\ \bibnamefont {Hessels}},\ }\bibfield  {title} {\bibinfo {title} {{A measurement of the atomic hydrogen Lamb shift and the proton charge radius}},\ }\href {https://doi.org/10.1126/science.aau7807} {\bibfield  {journal} {\bibinfo  {journal} {Science}\ }\textbf {\bibinfo {volume} {365}},\ \bibinfo {pages} {1007} (\bibinfo {year} {2019})}\BibitemShut {NoStop}%
\bibitem [{\citenamefont {Xiong}\ \emph {et~al.}(2019)\citenamefont {Xiong} \emph {et~al.}}]{Xiong:2019umf}%
  \BibitemOpen
  \bibfield  {author} {\bibinfo {author} {\bibfnamefont {W.}~\bibnamefont {Xiong}} \emph {et~al.},\ }\bibfield  {title} {\bibinfo {title} {{A small proton charge radius from an electron\textendash{}proton scattering experiment}},\ }\href {https://doi.org/10.1038/s41586-019-1721-2} {\bibfield  {journal} {\bibinfo  {journal} {Nature}\ }\textbf {\bibinfo {volume} {575}},\ \bibinfo {pages} {147} (\bibinfo {year} {2019})}\BibitemShut {NoStop}%
\bibitem [{\citenamefont {Antognini}\ \emph {et~al.}(2013)\citenamefont {Antognini} \emph {et~al.}}]{Antognini:2013txn}%
  \BibitemOpen
  \bibfield  {author} {\bibinfo {author} {\bibfnamefont {A.}~\bibnamefont {Antognini}} \emph {et~al.},\ }\bibfield  {title} {\bibinfo {title} {{Proton Structure from the Measurement of $2S-2P$ Transition Frequencies of Muonic Hydrogen}},\ }\href {https://doi.org/10.1126/science.1230016} {\bibfield  {journal} {\bibinfo  {journal} {Science}\ }\textbf {\bibinfo {volume} {339}},\ \bibinfo {pages} {417} (\bibinfo {year} {2013})}\BibitemShut {NoStop}%
\bibitem [{\citenamefont {Schneider}\ \emph {et~al.}(2017)\citenamefont {Schneider} \emph {et~al.}}]{Schneider:2017lff}%
  \BibitemOpen
  \bibfield  {author} {\bibinfo {author} {\bibfnamefont {G.}~\bibnamefont {Schneider}} \emph {et~al.},\ }\bibfield  {title} {\bibinfo {title} {{Double-trap measurement of the proton magnetic moment at 0.3 parts per billion precision}},\ }\href {https://doi.org/10.1126/science.aan0207} {\bibfield  {journal} {\bibinfo  {journal} {Science}\ }\textbf {\bibinfo {volume} {358}},\ \bibinfo {pages} {1081} (\bibinfo {year} {2017})}\BibitemShut {NoStop}%
\bibitem [{\citenamefont {Lin}\ \emph {et~al.}(2021)\citenamefont {Lin}, \citenamefont {Hammer},\ and\ \citenamefont {Mei\ss{}ner}}]{Lin:2021umk}%
  \BibitemOpen
  \bibfield  {author} {\bibinfo {author} {\bibfnamefont {Y.-H.}\ \bibnamefont {Lin}}, \bibinfo {author} {\bibfnamefont {H.-W.}\ \bibnamefont {Hammer}},\ and\ \bibinfo {author} {\bibfnamefont {U.-G.}\ \bibnamefont {Mei\ss{}ner}},\ }\bibfield  {title} {\bibinfo {title} {{High-precision determination of the electric and magnetic radius of the proton}},\ }\href {https://doi.org/10.1016/j.physletb.2021.136254} {\bibfield  {journal} {\bibinfo  {journal} {Phys. Lett. B}\ }\textbf {\bibinfo {volume} {816}},\ \bibinfo {pages} {136254} (\bibinfo {year} {2021})},\ \Eprint {https://arxiv.org/abs/2102.11642} {arXiv:2102.11642 [hep-ph]} \BibitemShut {NoStop}%
\bibitem [{\citenamefont {Pascalutsa}\ \emph {et~al.}(2007)\citenamefont {Pascalutsa}, \citenamefont {Vanderhaeghen},\ and\ \citenamefont {Yang}}]{Pascalutsa:2006up}%
  \BibitemOpen
  \bibfield  {author} {\bibinfo {author} {\bibfnamefont {V.}~\bibnamefont {Pascalutsa}}, \bibinfo {author} {\bibfnamefont {M.}~\bibnamefont {Vanderhaeghen}},\ and\ \bibinfo {author} {\bibfnamefont {S.~N.}\ \bibnamefont {Yang}},\ }\bibfield  {title} {\bibinfo {title} {{Electromagnetic excitation of the Delta(1232)-resonance}},\ }\href {https://doi.org/10.1016/j.physrep.2006.09.006} {\bibfield  {journal} {\bibinfo  {journal} {Phys. Rept.}\ }\textbf {\bibinfo {volume} {437}},\ \bibinfo {pages} {125} (\bibinfo {year} {2007})},\ \Eprint {https://arxiv.org/abs/hep-ph/0609004} {arXiv:hep-ph/0609004} \BibitemShut {NoStop}%
\bibitem [{\citenamefont {Krivoruchenko}\ and\ \citenamefont {Giannini}(1991)}]{Krivoruchenko:1991pm}%
  \BibitemOpen
  \bibfield  {author} {\bibinfo {author} {\bibfnamefont {M.~I.}\ \bibnamefont {Krivoruchenko}}\ and\ \bibinfo {author} {\bibfnamefont {M.~M.}\ \bibnamefont {Giannini}},\ }\bibfield  {title} {\bibinfo {title} {{Quadrupole moments of the decuplet baryons}},\ }\href {https://doi.org/10.1103/PhysRevD.43.3763} {\bibfield  {journal} {\bibinfo  {journal} {Phys. Rev. D}\ }\textbf {\bibinfo {volume} {43}},\ \bibinfo {pages} {3763} (\bibinfo {year} {1991})}\BibitemShut {NoStop}%
\bibitem [{\citenamefont {Schlumpf}(1993)}]{Schlumpf:1993rm}%
  \BibitemOpen
  \bibfield  {author} {\bibinfo {author} {\bibfnamefont {F.}~\bibnamefont {Schlumpf}},\ }\bibfield  {title} {\bibinfo {title} {{Magnetic moments of the baryon decuplet in a relativistic quark model}},\ }\href {https://doi.org/10.1103/PhysRevD.48.4478} {\bibfield  {journal} {\bibinfo  {journal} {Phys. Rev. D}\ }\textbf {\bibinfo {volume} {48}},\ \bibinfo {pages} {4478} (\bibinfo {year} {1993})},\ \Eprint {https://arxiv.org/abs/hep-ph/9305293} {arXiv:hep-ph/9305293} \BibitemShut {NoStop}%
\bibitem [{\citenamefont {Buchmann}\ \emph {et~al.}(1997)\citenamefont {Buchmann}, \citenamefont {Hernandez},\ and\ \citenamefont {Faessler}}]{Buchmann:1996bd}%
  \BibitemOpen
  \bibfield  {author} {\bibinfo {author} {\bibfnamefont {A.~J.}\ \bibnamefont {Buchmann}}, \bibinfo {author} {\bibfnamefont {E.}~\bibnamefont {Hernandez}},\ and\ \bibinfo {author} {\bibfnamefont {A.}~\bibnamefont {Faessler}},\ }\bibfield  {title} {\bibinfo {title} {{Electromagnetic properties of the Delta (1232)}},\ }\href {https://doi.org/10.1103/PhysRevC.55.448} {\bibfield  {journal} {\bibinfo  {journal} {Phys. Rev. C}\ }\textbf {\bibinfo {volume} {55}},\ \bibinfo {pages} {448} (\bibinfo {year} {1997})},\ \Eprint {https://arxiv.org/abs/nucl-th/9610040} {arXiv:nucl-th/9610040} \BibitemShut {NoStop}%
\bibitem [{\citenamefont {Berger}\ \emph {et~al.}(2004)\citenamefont {Berger}, \citenamefont {Wagenbrunn},\ and\ \citenamefont {Plessas}}]{Berger:2004yi}%
  \BibitemOpen
  \bibfield  {author} {\bibinfo {author} {\bibfnamefont {K.}~\bibnamefont {Berger}}, \bibinfo {author} {\bibfnamefont {R.~F.}\ \bibnamefont {Wagenbrunn}},\ and\ \bibinfo {author} {\bibfnamefont {W.}~\bibnamefont {Plessas}},\ }\bibfield  {title} {\bibinfo {title} {{Covariant baryon charge radii and magnetic moments in a chiral constituent quark model}},\ }\href {https://doi.org/10.1103/PhysRevD.70.094027} {\bibfield  {journal} {\bibinfo  {journal} {Phys. Rev. D}\ }\textbf {\bibinfo {volume} {70}},\ \bibinfo {pages} {094027} (\bibinfo {year} {2004})},\ \Eprint {https://arxiv.org/abs/nucl-th/0407009} {arXiv:nucl-th/0407009} \BibitemShut {NoStop}%
\bibitem [{\citenamefont {Ramalho}\ and\ \citenamefont {Pena}(2009)}]{Ramalho:2008dc}%
  \BibitemOpen
  \bibfield  {author} {\bibinfo {author} {\bibfnamefont {G.}~\bibnamefont {Ramalho}}\ and\ \bibinfo {author} {\bibfnamefont {M.~T.}\ \bibnamefont {Pena}},\ }\bibfield  {title} {\bibinfo {title} {{Electromagnetic form factors of the Delta in a S-wave approach}},\ }\href {https://doi.org/10.1088/0954-3899/36/8/085004} {\bibfield  {journal} {\bibinfo  {journal} {J. Phys. G}\ }\textbf {\bibinfo {volume} {36}},\ \bibinfo {pages} {085004} (\bibinfo {year} {2009})},\ \Eprint {https://arxiv.org/abs/0807.2922} {arXiv:0807.2922 [hep-ph]} \BibitemShut {NoStop}%
\bibitem [{\citenamefont {Kim}\ \emph {et~al.}(1998)\citenamefont {Kim}, \citenamefont {Praszalowicz},\ and\ \citenamefont {Goeke}}]{Kim:1997ip}%
  \BibitemOpen
  \bibfield  {author} {\bibinfo {author} {\bibfnamefont {H.-C.}\ \bibnamefont {Kim}}, \bibinfo {author} {\bibfnamefont {M.}~\bibnamefont {Praszalowicz}},\ and\ \bibinfo {author} {\bibfnamefont {K.}~\bibnamefont {Goeke}},\ }\bibfield  {title} {\bibinfo {title} {{Magnetic moments of the SU(3) decuplet baryons in the chiral quark - soliton model}},\ }\href {https://doi.org/10.1103/PhysRevD.57.2859} {\bibfield  {journal} {\bibinfo  {journal} {Phys. Rev. D}\ }\textbf {\bibinfo {volume} {57}},\ \bibinfo {pages} {2859} (\bibinfo {year} {1998})},\ \Eprint {https://arxiv.org/abs/hep-ph/9706531} {arXiv:hep-ph/9706531} \BibitemShut {NoStop}%
\bibitem [{\citenamefont {Ledwig}\ \emph {et~al.}(2009)\citenamefont {Ledwig}, \citenamefont {Silva},\ and\ \citenamefont {Vanderhaeghen}}]{Ledwig:2008es}%
  \BibitemOpen
  \bibfield  {author} {\bibinfo {author} {\bibfnamefont {T.}~\bibnamefont {Ledwig}}, \bibinfo {author} {\bibfnamefont {A.}~\bibnamefont {Silva}},\ and\ \bibinfo {author} {\bibfnamefont {M.}~\bibnamefont {Vanderhaeghen}},\ }\bibfield  {title} {\bibinfo {title} {{Electromagnetic properties of the Delta(1232) and decuplet baryons in the self-consistent SU(3) chiral quark-soliton model}},\ }\href {https://doi.org/10.1103/PhysRevD.79.094025} {\bibfield  {journal} {\bibinfo  {journal} {Phys. Rev. D}\ }\textbf {\bibinfo {volume} {79}},\ \bibinfo {pages} {094025} (\bibinfo {year} {2009})},\ \Eprint {https://arxiv.org/abs/0811.3086} {arXiv:0811.3086 [hep-ph]} \BibitemShut {NoStop}%
\bibitem [{\citenamefont {Fu}\ \emph {et~al.}(2022)\citenamefont {Fu}, \citenamefont {Sun},\ and\ \citenamefont {Dong}}]{Fu:2022rkn}%
  \BibitemOpen
  \bibfield  {author} {\bibinfo {author} {\bibfnamefont {D.}~\bibnamefont {Fu}}, \bibinfo {author} {\bibfnamefont {B.-D.}\ \bibnamefont {Sun}},\ and\ \bibinfo {author} {\bibfnamefont {Y.}~\bibnamefont {Dong}},\ }\bibfield  {title} {\bibinfo {title} {{Electromagnetic and gravitational form factors of \ensuremath{\Delta} resonance in a covariant quark-diquark approach}},\ }\href {https://doi.org/10.1103/PhysRevD.105.096002} {\bibfield  {journal} {\bibinfo  {journal} {Phys. Rev. D}\ }\textbf {\bibinfo {volume} {105}},\ \bibinfo {pages} {096002} (\bibinfo {year} {2022})},\ \Eprint {https://arxiv.org/abs/2201.08059} {arXiv:2201.08059 [hep-ph]} \BibitemShut {NoStop}%
\bibitem [{\citenamefont {Wang}\ \emph {et~al.}(2024{\natexlab{a}})\citenamefont {Wang}, \citenamefont {Fu},\ and\ \citenamefont {Dong}}]{Wang:2023bjp}%
  \BibitemOpen
  \bibfield  {author} {\bibinfo {author} {\bibfnamefont {J.}~\bibnamefont {Wang}}, \bibinfo {author} {\bibfnamefont {D.}~\bibnamefont {Fu}},\ and\ \bibinfo {author} {\bibfnamefont {Y.}~\bibnamefont {Dong}},\ }\bibfield  {title} {\bibinfo {title} {{Form factors of decuplet baryons in a covariant quark\textendash{}diquark approach}},\ }\href {https://doi.org/10.1140/epjc/s10052-024-12406-4} {\bibfield  {journal} {\bibinfo  {journal} {Eur. Phys. J. C}\ }\textbf {\bibinfo {volume} {84}},\ \bibinfo {pages} {79} (\bibinfo {year} {2024}{\natexlab{a}})},\ \Eprint {https://arxiv.org/abs/2311.07149} {arXiv:2311.07149 [hep-ph]} \BibitemShut {NoStop}%
\bibitem [{\citenamefont {Kaelbermann}\ and\ \citenamefont {Eisenberg}(1983)}]{Kaelbermann:1983zb}%
  \BibitemOpen
  \bibfield  {author} {\bibinfo {author} {\bibfnamefont {G.}~\bibnamefont {Kaelbermann}}\ and\ \bibinfo {author} {\bibfnamefont {J.~M.}\ \bibnamefont {Eisenberg}},\ }\bibfield  {title} {\bibinfo {title} {{PION PHOTOPRODUCTION IN THE DELTA (1232) REGION AND CHIRAL BAG MODELS}},\ }\href {https://doi.org/10.1103/PhysRevD.28.71} {\bibfield  {journal} {\bibinfo  {journal} {Phys. Rev. D}\ }\textbf {\bibinfo {volume} {28}},\ \bibinfo {pages} {71} (\bibinfo {year} {1983})}\BibitemShut {NoStop}%
\bibitem [{\citenamefont {Lu}\ \emph {et~al.}(1997)\citenamefont {Lu}, \citenamefont {Thomas},\ and\ \citenamefont {Williams}}]{Lu:1996rj}%
  \BibitemOpen
  \bibfield  {author} {\bibinfo {author} {\bibfnamefont {D.-H.}\ \bibnamefont {Lu}}, \bibinfo {author} {\bibfnamefont {A.~W.}\ \bibnamefont {Thomas}},\ and\ \bibinfo {author} {\bibfnamefont {A.~G.}\ \bibnamefont {Williams}},\ }\bibfield  {title} {\bibinfo {title} {{A Chiral bag model approach to delta electroproduction}},\ }\href {https://doi.org/10.1103/PhysRevC.55.3108} {\bibfield  {journal} {\bibinfo  {journal} {Phys. Rev. C}\ }\textbf {\bibinfo {volume} {55}},\ \bibinfo {pages} {3108} (\bibinfo {year} {1997})},\ \Eprint {https://arxiv.org/abs/nucl-th/9612017} {arXiv:nucl-th/9612017} \BibitemShut {NoStop}%
\bibitem [{\citenamefont {Luty}\ \emph {et~al.}(1995)\citenamefont {Luty}, \citenamefont {March-Russell},\ and\ \citenamefont {White}}]{Luty:1994ub}%
  \BibitemOpen
  \bibfield  {author} {\bibinfo {author} {\bibfnamefont {M.~A.}\ \bibnamefont {Luty}}, \bibinfo {author} {\bibfnamefont {J.}~\bibnamefont {March-Russell}},\ and\ \bibinfo {author} {\bibfnamefont {M.~J.}\ \bibnamefont {White}},\ }\bibfield  {title} {\bibinfo {title} {{Baryon magnetic moments in a simultaneous expansion in 1/N and m(s)}},\ }\href {https://doi.org/10.1103/PhysRevD.51.2332} {\bibfield  {journal} {\bibinfo  {journal} {Phys. Rev. D}\ }\textbf {\bibinfo {volume} {51}},\ \bibinfo {pages} {2332} (\bibinfo {year} {1995})},\ \Eprint {https://arxiv.org/abs/hep-ph/9405272} {arXiv:hep-ph/9405272} \BibitemShut {NoStop}%
\bibitem [{\citenamefont {Jenkins}\ and\ \citenamefont {Manohar}(1994)}]{Jenkins:1994md}%
  \BibitemOpen
  \bibfield  {author} {\bibinfo {author} {\bibfnamefont {E.~E.}\ \bibnamefont {Jenkins}}\ and\ \bibinfo {author} {\bibfnamefont {A.~V.}\ \bibnamefont {Manohar}},\ }\bibfield  {title} {\bibinfo {title} {{Baryon magnetic moments in the 1 / N(c) expansion}},\ }\href {https://doi.org/10.1016/0370-2693(94)90377-8} {\bibfield  {journal} {\bibinfo  {journal} {Phys. Lett. B}\ }\textbf {\bibinfo {volume} {335}},\ \bibinfo {pages} {452} (\bibinfo {year} {1994})},\ \Eprint {https://arxiv.org/abs/hep-ph/9405431} {arXiv:hep-ph/9405431} \BibitemShut {NoStop}%
\bibitem [{\citenamefont {Oh}(1995)}]{Oh:1995hn}%
  \BibitemOpen
  \bibfield  {author} {\bibinfo {author} {\bibfnamefont {Y.-s.}\ \bibnamefont {Oh}},\ }\bibfield  {title} {\bibinfo {title} {{Electric quadrupole moments of the decuplet baryons in the Skyrme model}},\ }\href {https://doi.org/10.1142/S0217732395001137} {\bibfield  {journal} {\bibinfo  {journal} {Mod. Phys. Lett. A}\ }\textbf {\bibinfo {volume} {10}},\ \bibinfo {pages} {1027} (\bibinfo {year} {1995})},\ \Eprint {https://arxiv.org/abs/hep-ph/9506308} {arXiv:hep-ph/9506308} \BibitemShut {NoStop}%
\bibitem [{\citenamefont {Lee}(1998)}]{Lee:1997jk}%
  \BibitemOpen
  \bibfield  {author} {\bibinfo {author} {\bibfnamefont {F.~X.}\ \bibnamefont {Lee}},\ }\bibfield  {title} {\bibinfo {title} {{Determination of decuplet baryon magnetic moments from QCD sum rules}},\ }\href {https://doi.org/10.1103/PhysRevD.57.1801} {\bibfield  {journal} {\bibinfo  {journal} {Phys. Rev. D}\ }\textbf {\bibinfo {volume} {57}},\ \bibinfo {pages} {1801} (\bibinfo {year} {1998})},\ \Eprint {https://arxiv.org/abs/hep-ph/9708323} {arXiv:hep-ph/9708323} \BibitemShut {NoStop}%
\bibitem [{\citenamefont {Aliev}\ \emph {et~al.}(2000)\citenamefont {Aliev}, \citenamefont {Ozpineci},\ and\ \citenamefont {Savci}}]{Aliev:2000rc}%
  \BibitemOpen
  \bibfield  {author} {\bibinfo {author} {\bibfnamefont {T.~M.}\ \bibnamefont {Aliev}}, \bibinfo {author} {\bibfnamefont {A.}~\bibnamefont {Ozpineci}},\ and\ \bibinfo {author} {\bibfnamefont {M.}~\bibnamefont {Savci}},\ }\bibfield  {title} {\bibinfo {title} {{Magnetic moments of Delta baryons in light cone QCD sum rules}},\ }\href {https://doi.org/10.1016/S0375-9474(00)00329-8} {\bibfield  {journal} {\bibinfo  {journal} {Nucl. Phys. A}\ }\textbf {\bibinfo {volume} {678}},\ \bibinfo {pages} {443} (\bibinfo {year} {2000})},\ \Eprint {https://arxiv.org/abs/hep-ph/0002228} {arXiv:hep-ph/0002228} \BibitemShut {NoStop}%
\bibitem [{\citenamefont {Aliev}\ \emph {et~al.}(2009)\citenamefont {Aliev}, \citenamefont {Azizi},\ and\ \citenamefont {Savci}}]{Aliev:2009pd}%
  \BibitemOpen
  \bibfield  {author} {\bibinfo {author} {\bibfnamefont {T.~M.}\ \bibnamefont {Aliev}}, \bibinfo {author} {\bibfnamefont {K.}~\bibnamefont {Azizi}},\ and\ \bibinfo {author} {\bibfnamefont {M.}~\bibnamefont {Savci}},\ }\bibfield  {title} {\bibinfo {title} {{Electric Quadrupole and Magnetic Octupole Moments of the Light Decuplet Baryons Within Light Cone QCD Sum Rules}},\ }\href {https://doi.org/10.1016/j.physletb.2009.10.026} {\bibfield  {journal} {\bibinfo  {journal} {Phys. Lett. B}\ }\textbf {\bibinfo {volume} {681}},\ \bibinfo {pages} {240} (\bibinfo {year} {2009})},\ \Eprint {https://arxiv.org/abs/0904.2485} {arXiv:0904.2485 [hep-ph]} \BibitemShut {NoStop}%
\bibitem [{\citenamefont {Buchmann}\ and\ \citenamefont {Henley}(2002)}]{Buchmann:2002xq}%
  \BibitemOpen
  \bibfield  {author} {\bibinfo {author} {\bibfnamefont {A.~J.}\ \bibnamefont {Buchmann}}\ and\ \bibinfo {author} {\bibfnamefont {E.~M.}\ \bibnamefont {Henley}},\ }\bibfield  {title} {\bibinfo {title} {{Quadrupole moments of baryons}},\ }\href {https://doi.org/10.1103/PhysRevD.65.073017} {\bibfield  {journal} {\bibinfo  {journal} {Phys. Rev. D}\ }\textbf {\bibinfo {volume} {65}},\ \bibinfo {pages} {073017} (\bibinfo {year} {2002})},\ \Eprint {https://arxiv.org/abs/1908.09910} {arXiv:1908.09910 [hep-ph]} \BibitemShut {NoStop}%
\bibitem [{\citenamefont {Butler}\ \emph {et~al.}(1994)\citenamefont {Butler}, \citenamefont {Savage},\ and\ \citenamefont {Springer}}]{Butler:1993ej}%
  \BibitemOpen
  \bibfield  {author} {\bibinfo {author} {\bibfnamefont {M.~N.}\ \bibnamefont {Butler}}, \bibinfo {author} {\bibfnamefont {M.~J.}\ \bibnamefont {Savage}},\ and\ \bibinfo {author} {\bibfnamefont {R.~P.}\ \bibnamefont {Springer}},\ }\bibfield  {title} {\bibinfo {title} {{Electromagnetic moments of the baryon decuplet}},\ }\href {https://doi.org/10.1103/PhysRevD.49.3459} {\bibfield  {journal} {\bibinfo  {journal} {Phys. Rev. D}\ }\textbf {\bibinfo {volume} {49}},\ \bibinfo {pages} {3459} (\bibinfo {year} {1994})},\ \Eprint {https://arxiv.org/abs/hep-ph/9308317} {arXiv:hep-ph/9308317} \BibitemShut {NoStop}%
\bibitem [{\citenamefont {Banerjee}\ and\ \citenamefont {Milana}(1996)}]{Banerjee:1995wz}%
  \BibitemOpen
  \bibfield  {author} {\bibinfo {author} {\bibfnamefont {M.~K.}\ \bibnamefont {Banerjee}}\ and\ \bibinfo {author} {\bibfnamefont {J.}~\bibnamefont {Milana}},\ }\bibfield  {title} {\bibinfo {title} {{The Decuplet revisited in chi(PT)}},\ }\href {https://doi.org/10.1103/PhysRevD.54.5804} {\bibfield  {journal} {\bibinfo  {journal} {Phys. Rev. D}\ }\textbf {\bibinfo {volume} {54}},\ \bibinfo {pages} {5804} (\bibinfo {year} {1996})},\ \Eprint {https://arxiv.org/abs/hep-ph/9508340} {arXiv:hep-ph/9508340} \BibitemShut {NoStop}%
\bibitem [{\citenamefont {Arndt}\ and\ \citenamefont {Tiburzi}(2003)}]{Arndt:2003we}%
  \BibitemOpen
  \bibfield  {author} {\bibinfo {author} {\bibfnamefont {D.}~\bibnamefont {Arndt}}\ and\ \bibinfo {author} {\bibfnamefont {B.~C.}\ \bibnamefont {Tiburzi}},\ }\bibfield  {title} {\bibinfo {title} {{Electromagnetic properties of the baryon decuplet in quenched and partially quenched chiral perturbation theory}},\ }\href {https://doi.org/10.1103/PhysRevD.69.059904} {\bibfield  {journal} {\bibinfo  {journal} {Phys. Rev. D}\ }\textbf {\bibinfo {volume} {68}},\ \bibinfo {pages} {114503} (\bibinfo {year} {2003})},\ \bibinfo {note} {[Erratum: Phys.Rev.D 69, 059904 (2004)]},\ \Eprint {https://arxiv.org/abs/hep-lat/0308001} {arXiv:hep-lat/0308001} \BibitemShut {NoStop}%
\bibitem [{\citenamefont {Pascalutsa}\ and\ \citenamefont {Vanderhaeghen}(2005)}]{Pascalutsa:2004je}%
  \BibitemOpen
  \bibfield  {author} {\bibinfo {author} {\bibfnamefont {V.}~\bibnamefont {Pascalutsa}}\ and\ \bibinfo {author} {\bibfnamefont {M.}~\bibnamefont {Vanderhaeghen}},\ }\bibfield  {title} {\bibinfo {title} {{Magnetic moment of the Delta(1232)-resonance in chiral effective field theory}},\ }\href {https://doi.org/10.1103/PhysRevLett.94.102003} {\bibfield  {journal} {\bibinfo  {journal} {Phys. Rev. Lett.}\ }\textbf {\bibinfo {volume} {94}},\ \bibinfo {pages} {102003} (\bibinfo {year} {2005})},\ \Eprint {https://arxiv.org/abs/nucl-th/0412113} {arXiv:nucl-th/0412113} \BibitemShut {NoStop}%
\bibitem [{\citenamefont {Hacker}\ \emph {et~al.}(2006)\citenamefont {Hacker}, \citenamefont {Wies}, \citenamefont {Gegelia},\ and\ \citenamefont {Scherer}}]{Hacker:2006gu}%
  \BibitemOpen
  \bibfield  {author} {\bibinfo {author} {\bibfnamefont {C.}~\bibnamefont {Hacker}}, \bibinfo {author} {\bibfnamefont {N.}~\bibnamefont {Wies}}, \bibinfo {author} {\bibfnamefont {J.}~\bibnamefont {Gegelia}},\ and\ \bibinfo {author} {\bibfnamefont {S.}~\bibnamefont {Scherer}},\ }\bibfield  {title} {\bibinfo {title} {{Magnetic dipole moment of the Delta(1232) in chiral perturbation theory}},\ }\href {https://doi.org/10.1140/epja/i2006-10043-7} {\bibfield  {journal} {\bibinfo  {journal} {Eur. Phys. J. A}\ }\textbf {\bibinfo {volume} {28}},\ \bibinfo {pages} {5} (\bibinfo {year} {2006})},\ \Eprint {https://arxiv.org/abs/hep-ph/0603267} {arXiv:hep-ph/0603267} \BibitemShut {NoStop}%
\bibitem [{\citenamefont {Tiburzi}(2009)}]{Tiburzi:2009yd}%
  \BibitemOpen
  \bibfield  {author} {\bibinfo {author} {\bibfnamefont {B.~C.}\ \bibnamefont {Tiburzi}},\ }\bibfield  {title} {\bibinfo {title} {{Connected Parts of Decuplet Electromagnetic Properties}},\ }\href {https://doi.org/10.1103/PhysRevD.79.077501} {\bibfield  {journal} {\bibinfo  {journal} {Phys. Rev. D}\ }\textbf {\bibinfo {volume} {79}},\ \bibinfo {pages} {077501} (\bibinfo {year} {2009})},\ \Eprint {https://arxiv.org/abs/0903.0359} {arXiv:0903.0359 [hep-lat]} \BibitemShut {NoStop}%
\bibitem [{\citenamefont {Geng}\ \emph {et~al.}(2009)\citenamefont {Geng}, \citenamefont {Martin~Camalich},\ and\ \citenamefont {Vicente~Vacas}}]{Geng:2009ys}%
  \BibitemOpen
  \bibfield  {author} {\bibinfo {author} {\bibfnamefont {L.~S.}\ \bibnamefont {Geng}}, \bibinfo {author} {\bibfnamefont {J.}~\bibnamefont {Martin~Camalich}},\ and\ \bibinfo {author} {\bibfnamefont {M.~J.}\ \bibnamefont {Vicente~Vacas}},\ }\bibfield  {title} {\bibinfo {title} {{Electromagnetic structure of the lowest-lying decuplet resonances in covariant chiral perturbation theory}},\ }\href {https://doi.org/10.1103/PhysRevD.80.034027} {\bibfield  {journal} {\bibinfo  {journal} {Phys. Rev. D}\ }\textbf {\bibinfo {volume} {80}},\ \bibinfo {pages} {034027} (\bibinfo {year} {2009})},\ \Eprint {https://arxiv.org/abs/0907.0631} {arXiv:0907.0631 [hep-ph]} \BibitemShut {NoStop}%
\bibitem [{\citenamefont {Li}\ \emph {et~al.}(2017)\citenamefont {Li}, \citenamefont {Liu}, \citenamefont {Chen}, \citenamefont {Deng},\ and\ \citenamefont {Zhu}}]{Li:2016ezv}%
  \BibitemOpen
  \bibfield  {author} {\bibinfo {author} {\bibfnamefont {H.-S.}\ \bibnamefont {Li}}, \bibinfo {author} {\bibfnamefont {Z.-W.}\ \bibnamefont {Liu}}, \bibinfo {author} {\bibfnamefont {X.-L.}\ \bibnamefont {Chen}}, \bibinfo {author} {\bibfnamefont {W.-Z.}\ \bibnamefont {Deng}},\ and\ \bibinfo {author} {\bibfnamefont {S.-L.}\ \bibnamefont {Zhu}},\ }\bibfield  {title} {\bibinfo {title} {{Magnetic moments and electromagnetic form factors of the decuplet baryons in chiral perturbation theory}},\ }\href {https://doi.org/10.1103/PhysRevD.95.076001} {\bibfield  {journal} {\bibinfo  {journal} {Phys. Rev. D}\ }\textbf {\bibinfo {volume} {95}},\ \bibinfo {pages} {076001} (\bibinfo {year} {2017})},\ \Eprint {https://arxiv.org/abs/1608.04617} {arXiv:1608.04617 [hep-ph]} \BibitemShut {NoStop}%
\bibitem [{\citenamefont {Nozawa}\ and\ \citenamefont {Leinweber}(1990)}]{Nozawa:1990gt}%
  \BibitemOpen
  \bibfield  {author} {\bibinfo {author} {\bibfnamefont {S.}~\bibnamefont {Nozawa}}\ and\ \bibinfo {author} {\bibfnamefont {D.~B.}\ \bibnamefont {Leinweber}},\ }\bibfield  {title} {\bibinfo {title} {{Electromagnetic form-factors of spin 3/2 baryons}},\ }\href {https://doi.org/10.1103/PhysRevD.42.3567} {\bibfield  {journal} {\bibinfo  {journal} {Phys. Rev. D}\ }\textbf {\bibinfo {volume} {42}},\ \bibinfo {pages} {3567} (\bibinfo {year} {1990})}\BibitemShut {NoStop}%
\bibitem [{\citenamefont {Leinweber}\ \emph {et~al.}(1992)\citenamefont {Leinweber}, \citenamefont {Draper},\ and\ \citenamefont {Woloshyn}}]{Leinweber:1992hy}%
  \BibitemOpen
  \bibfield  {author} {\bibinfo {author} {\bibfnamefont {D.~B.}\ \bibnamefont {Leinweber}}, \bibinfo {author} {\bibfnamefont {T.}~\bibnamefont {Draper}},\ and\ \bibinfo {author} {\bibfnamefont {R.~M.}\ \bibnamefont {Woloshyn}},\ }\bibfield  {title} {\bibinfo {title} {{Decuplet baryon structure from lattice QCD}},\ }\href {https://doi.org/10.1103/PhysRevD.46.3067} {\bibfield  {journal} {\bibinfo  {journal} {Phys. Rev. D}\ }\textbf {\bibinfo {volume} {46}},\ \bibinfo {pages} {3067} (\bibinfo {year} {1992})},\ \Eprint {https://arxiv.org/abs/hep-lat/9208025} {arXiv:hep-lat/9208025} \BibitemShut {NoStop}%
\bibitem [{\citenamefont {Cloet}\ \emph {et~al.}(2003)\citenamefont {Cloet}, \citenamefont {Leinweber},\ and\ \citenamefont {Thomas}}]{Cloet:2003jm}%
  \BibitemOpen
  \bibfield  {author} {\bibinfo {author} {\bibfnamefont {I.~C.}\ \bibnamefont {Cloet}}, \bibinfo {author} {\bibfnamefont {D.~B.}\ \bibnamefont {Leinweber}},\ and\ \bibinfo {author} {\bibfnamefont {A.~W.}\ \bibnamefont {Thomas}},\ }\bibfield  {title} {\bibinfo {title} {{Delta baryon magnetic moments from lattice QCD}},\ }\href {https://doi.org/10.1016/S0370-2693(03)00418-0} {\bibfield  {journal} {\bibinfo  {journal} {Phys. Lett. B}\ }\textbf {\bibinfo {volume} {563}},\ \bibinfo {pages} {157} (\bibinfo {year} {2003})},\ \Eprint {https://arxiv.org/abs/hep-lat/0302008} {arXiv:hep-lat/0302008} \BibitemShut {NoStop}%
\bibitem [{\citenamefont {Lee}\ \emph {et~al.}(2005)\citenamefont {Lee}, \citenamefont {Kelly}, \citenamefont {Zhou},\ and\ \citenamefont {Wilcox}}]{Lee:2005ds}%
  \BibitemOpen
  \bibfield  {author} {\bibinfo {author} {\bibfnamefont {F.~X.}\ \bibnamefont {Lee}}, \bibinfo {author} {\bibfnamefont {R.}~\bibnamefont {Kelly}}, \bibinfo {author} {\bibfnamefont {L.}~\bibnamefont {Zhou}},\ and\ \bibinfo {author} {\bibfnamefont {W.}~\bibnamefont {Wilcox}},\ }\bibfield  {title} {\bibinfo {title} {{Baryon magnetic moments in the background field method}},\ }\href {https://doi.org/10.1016/j.physletb.2005.08.106} {\bibfield  {journal} {\bibinfo  {journal} {Phys. Lett. B}\ }\textbf {\bibinfo {volume} {627}},\ \bibinfo {pages} {71} (\bibinfo {year} {2005})},\ \Eprint {https://arxiv.org/abs/hep-lat/0509067} {arXiv:hep-lat/0509067} \BibitemShut {NoStop}%
\bibitem [{\citenamefont {Alexandrou}\ \emph {et~al.}(2010)\citenamefont {Alexandrou}, \citenamefont {Korzec}, \citenamefont {Koutsou}, \citenamefont {Negele},\ and\ \citenamefont {Proestos}}]{Alexandrou:2010jv}%
  \BibitemOpen
  \bibfield  {author} {\bibinfo {author} {\bibfnamefont {C.}~\bibnamefont {Alexandrou}}, \bibinfo {author} {\bibfnamefont {T.}~\bibnamefont {Korzec}}, \bibinfo {author} {\bibfnamefont {G.}~\bibnamefont {Koutsou}}, \bibinfo {author} {\bibfnamefont {J.~W.}\ \bibnamefont {Negele}},\ and\ \bibinfo {author} {\bibfnamefont {Y.}~\bibnamefont {Proestos}},\ }\bibfield  {title} {\bibinfo {title} {{The Electromagnetic form factors of the $\Omega^-$ in lattice QCD}},\ }\href {https://doi.org/10.1103/PhysRevD.82.034504} {\bibfield  {journal} {\bibinfo  {journal} {Phys. Rev. D}\ }\textbf {\bibinfo {volume} {82}},\ \bibinfo {pages} {034504} (\bibinfo {year} {2010})},\ \Eprint {https://arxiv.org/abs/1006.0558} {arXiv:1006.0558 [hep-lat]} \BibitemShut {NoStop}%
\bibitem [{\citenamefont {Boinepalli}\ \emph {et~al.}(2009)\citenamefont {Boinepalli}, \citenamefont {Leinweber}, \citenamefont {Moran}, \citenamefont {Williams}, \citenamefont {Zanotti},\ and\ \citenamefont {Zhang}}]{Boinepalli:2009sq}%
  \BibitemOpen
  \bibfield  {author} {\bibinfo {author} {\bibfnamefont {S.}~\bibnamefont {Boinepalli}}, \bibinfo {author} {\bibfnamefont {D.~B.}\ \bibnamefont {Leinweber}}, \bibinfo {author} {\bibfnamefont {P.~J.}\ \bibnamefont {Moran}}, \bibinfo {author} {\bibfnamefont {A.~G.}\ \bibnamefont {Williams}}, \bibinfo {author} {\bibfnamefont {J.~M.}\ \bibnamefont {Zanotti}},\ and\ \bibinfo {author} {\bibfnamefont {J.~B.}\ \bibnamefont {Zhang}},\ }\bibfield  {title} {\bibinfo {title} {{Precision electromagnetic structure of decuplet baryons in the chiral regime}},\ }\href {https://doi.org/10.1103/PhysRevD.80.054505} {\bibfield  {journal} {\bibinfo  {journal} {Phys. Rev. D}\ }\textbf {\bibinfo {volume} {80}},\ \bibinfo {pages} {054505} (\bibinfo {year} {2009})},\ \Eprint {https://arxiv.org/abs/0902.4046} {arXiv:0902.4046 [hep-lat]} \BibitemShut {NoStop}%
\bibitem [{\citenamefont {Aubin}\ \emph {et~al.}(2009)\citenamefont {Aubin}, \citenamefont {Orginos}, \citenamefont {Pascalutsa},\ and\ \citenamefont {Vanderhaeghen}}]{Aubin:2008qp}%
  \BibitemOpen
  \bibfield  {author} {\bibinfo {author} {\bibfnamefont {C.}~\bibnamefont {Aubin}}, \bibinfo {author} {\bibfnamefont {K.}~\bibnamefont {Orginos}}, \bibinfo {author} {\bibfnamefont {V.}~\bibnamefont {Pascalutsa}},\ and\ \bibinfo {author} {\bibfnamefont {M.}~\bibnamefont {Vanderhaeghen}},\ }\bibfield  {title} {\bibinfo {title} {{Magnetic Moments of Delta and Omega- Baryons with Dynamical Clover Fermions}},\ }\href {https://doi.org/10.1103/PhysRevD.79.051502} {\bibfield  {journal} {\bibinfo  {journal} {Phys. Rev. D}\ }\textbf {\bibinfo {volume} {79}},\ \bibinfo {pages} {051502} (\bibinfo {year} {2009})},\ \Eprint {https://arxiv.org/abs/0811.2440} {arXiv:0811.2440 [hep-lat]} \BibitemShut {NoStop}%
\bibitem [{\citenamefont {Kotulla}\ \emph {et~al.}(2002)\citenamefont {Kotulla} \emph {et~al.}}]{Kotulla:2002cg}%
  \BibitemOpen
  \bibfield  {author} {\bibinfo {author} {\bibfnamefont {M.}~\bibnamefont {Kotulla}} \emph {et~al.},\ }\bibfield  {title} {\bibinfo {title} {{The Reaction gamma p ---\ensuremath{>} pi zero gamma-prime p and the magnetic dipole moment of the delta+(1232) resonance}},\ }\href {https://doi.org/10.1103/PhysRevLett.89.272001} {\bibfield  {journal} {\bibinfo  {journal} {Phys. Rev. Lett.}\ }\textbf {\bibinfo {volume} {89}},\ \bibinfo {pages} {272001} (\bibinfo {year} {2002})},\ \Eprint {https://arxiv.org/abs/nucl-ex/0210040} {arXiv:nucl-ex/0210040} \BibitemShut {NoStop}%
\bibitem [{\citenamefont {Bosshard}\ \emph {et~al.}(1991)\citenamefont {Bosshard} \emph {et~al.}}]{Bosshard:1991zp}%
  \BibitemOpen
  \bibfield  {author} {\bibinfo {author} {\bibfnamefont {A.}~\bibnamefont {Bosshard}} \emph {et~al.},\ }\bibfield  {title} {\bibinfo {title} {{Analyzing power in pion proton bremsstrahlung, and the Delta++ (1232) magnetic moment}},\ }\href {https://doi.org/10.1103/PhysRevD.44.1962} {\bibfield  {journal} {\bibinfo  {journal} {Phys. Rev. D}\ }\textbf {\bibinfo {volume} {44}},\ \bibinfo {pages} {1962} (\bibinfo {year} {1991})}\BibitemShut {NoStop}%
\bibitem [{\citenamefont {Endrodi}(2025)}]{Endrodi:2024cqn}%
  \BibitemOpen
  \bibfield  {author} {\bibinfo {author} {\bibfnamefont {G.}~\bibnamefont {Endrodi}},\ }\bibfield  {title} {\bibinfo {title} {{QCD with background electromagnetic fields on the lattice: A review}},\ }\href {https://doi.org/10.1016/j.ppnp.2024.104153} {\bibfield  {journal} {\bibinfo  {journal} {Prog. Part. Nucl. Phys.}\ }\textbf {\bibinfo {volume} {141}},\ \bibinfo {pages} {104153} (\bibinfo {year} {2025})},\ \Eprint {https://arxiv.org/abs/2406.19780} {arXiv:2406.19780 [hep-lat]} \BibitemShut {NoStop}%
\bibitem [{\citenamefont {Can}(2021)}]{Can:2021ehb}%
  \BibitemOpen
  \bibfield  {author} {\bibinfo {author} {\bibfnamefont {K.~U.}\ \bibnamefont {Can}},\ }\bibfield  {title} {\bibinfo {title} {{Lattice QCD study of the elastic and transition form factors of charmed baryons}},\ }\href {https://doi.org/10.1142/S0217751X21300131} {\bibfield  {journal} {\bibinfo  {journal} {Int. J. Mod. Phys. A}\ }\textbf {\bibinfo {volume} {36}},\ \bibinfo {pages} {2130013} (\bibinfo {year} {2021})},\ \Eprint {https://arxiv.org/abs/2107.13159} {arXiv:2107.13159 [hep-lat]} \BibitemShut {NoStop}%
\bibitem [{\citenamefont {Meng}\ \emph {et~al.}(2023)\citenamefont {Meng}, \citenamefont {Wang}, \citenamefont {Wang},\ and\ \citenamefont {Zhu}}]{Meng:2022ozq}%
  \BibitemOpen
  \bibfield  {author} {\bibinfo {author} {\bibfnamefont {L.}~\bibnamefont {Meng}}, \bibinfo {author} {\bibfnamefont {B.}~\bibnamefont {Wang}}, \bibinfo {author} {\bibfnamefont {G.-J.}\ \bibnamefont {Wang}},\ and\ \bibinfo {author} {\bibfnamefont {S.-L.}\ \bibnamefont {Zhu}},\ }\bibfield  {title} {\bibinfo {title} {{Chiral perturbation theory for heavy hadrons and chiral effective field theory for heavy hadronic molecules}},\ }\href {https://doi.org/10.1016/j.physrep.2023.04.003} {\bibfield  {journal} {\bibinfo  {journal} {Phys. Rept.}\ }\textbf {\bibinfo {volume} {1019}},\ \bibinfo {pages} {1} (\bibinfo {year} {2023})},\ \Eprint {https://arxiv.org/abs/2204.08716} {arXiv:2204.08716 [hep-ph]} \BibitemShut {NoStop}%
\bibitem [{\citenamefont {Jenkins}\ and\ \citenamefont {Manohar}(1991)}]{Jenkins:1990jv}%
  \BibitemOpen
  \bibfield  {author} {\bibinfo {author} {\bibfnamefont {E.~E.}\ \bibnamefont {Jenkins}}\ and\ \bibinfo {author} {\bibfnamefont {A.~V.}\ \bibnamefont {Manohar}},\ }\bibfield  {title} {\bibinfo {title} {{Baryon chiral perturbation theory using a heavy fermion Lagrangian}},\ }\href {https://doi.org/10.1016/0370-2693(91)90266-S} {\bibfield  {journal} {\bibinfo  {journal} {Phys. Lett. B}\ }\textbf {\bibinfo {volume} {255}},\ \bibinfo {pages} {558} (\bibinfo {year} {1991})}\BibitemShut {NoStop}%
\bibitem [{\citenamefont {Hemmert}\ \emph {et~al.}(1998)\citenamefont {Hemmert}, \citenamefont {Holstein},\ and\ \citenamefont {Kambor}}]{Hemmert:1997ye}%
  \BibitemOpen
  \bibfield  {author} {\bibinfo {author} {\bibfnamefont {T.~R.}\ \bibnamefont {Hemmert}}, \bibinfo {author} {\bibfnamefont {B.~R.}\ \bibnamefont {Holstein}},\ and\ \bibinfo {author} {\bibfnamefont {J.}~\bibnamefont {Kambor}},\ }\bibfield  {title} {\bibinfo {title} {{Chiral Lagrangians and delta(1232) interactions: Formalism}},\ }\href {https://doi.org/10.1088/0954-3899/24/10/003} {\bibfield  {journal} {\bibinfo  {journal} {J. Phys. G}\ }\textbf {\bibinfo {volume} {24}},\ \bibinfo {pages} {1831} (\bibinfo {year} {1998})},\ \Eprint {https://arxiv.org/abs/hep-ph/9712496} {arXiv:hep-ph/9712496} \BibitemShut {NoStop}%
\bibitem [{\citenamefont {Becher}\ and\ \citenamefont {Leutwyler}(1999)}]{Becher:1999he}%
  \BibitemOpen
  \bibfield  {author} {\bibinfo {author} {\bibfnamefont {T.}~\bibnamefont {Becher}}\ and\ \bibinfo {author} {\bibfnamefont {H.}~\bibnamefont {Leutwyler}},\ }\bibfield  {title} {\bibinfo {title} {{Baryon chiral perturbation theory in manifestly Lorentz invariant form}},\ }\href {https://doi.org/10.1007/PL00021673} {\bibfield  {journal} {\bibinfo  {journal} {Eur. Phys. J.}\ }\textbf {\bibinfo {volume} {9}},\ \bibinfo {pages} {643} (\bibinfo {year} {1999})},\ \Eprint {https://arxiv.org/abs/hep-ph/9901384} {arXiv:hep-ph/9901384} \BibitemShut {NoStop}%
\bibitem [{\citenamefont {Gegelia}\ and\ \citenamefont {Japaridze}(1999)}]{Gegelia:1999gf}%
  \BibitemOpen
  \bibfield  {author} {\bibinfo {author} {\bibfnamefont {J.}~\bibnamefont {Gegelia}}\ and\ \bibinfo {author} {\bibfnamefont {G.}~\bibnamefont {Japaridze}},\ }\bibfield  {title} {\bibinfo {title} {{Matching heavy particle approach to relativistic theory}},\ }\href {https://doi.org/10.1103/PhysRevD.60.114038} {\bibfield  {journal} {\bibinfo  {journal} {Phys. Rev. D}\ }\textbf {\bibinfo {volume} {60}},\ \bibinfo {pages} {114038} (\bibinfo {year} {1999})},\ \Eprint {https://arxiv.org/abs/hep-ph/9908377} {arXiv:hep-ph/9908377} \BibitemShut {NoStop}%
\bibitem [{\citenamefont {Fuchs}\ \emph {et~al.}(2003)\citenamefont {Fuchs}, \citenamefont {Gegelia}, \citenamefont {Japaridze},\ and\ \citenamefont {Scherer}}]{Fuchs:2003qc}%
  \BibitemOpen
  \bibfield  {author} {\bibinfo {author} {\bibfnamefont {T.}~\bibnamefont {Fuchs}}, \bibinfo {author} {\bibfnamefont {J.}~\bibnamefont {Gegelia}}, \bibinfo {author} {\bibfnamefont {G.}~\bibnamefont {Japaridze}},\ and\ \bibinfo {author} {\bibfnamefont {S.}~\bibnamefont {Scherer}},\ }\bibfield  {title} {\bibinfo {title} {{Renormalization of relativistic baryon chiral perturbation theory and power counting}},\ }\href {https://doi.org/10.1103/PhysRevD.68.056005} {\bibfield  {journal} {\bibinfo  {journal} {Phys. Rev. D}\ }\textbf {\bibinfo {volume} {68}},\ \bibinfo {pages} {056005} (\bibinfo {year} {2003})},\ \Eprint {https://arxiv.org/abs/hep-ph/0302117} {arXiv:hep-ph/0302117} \BibitemShut {NoStop}%
\bibitem [{\citenamefont {Bernard}\ \emph {et~al.}(1991)\citenamefont {Bernard}, \citenamefont {Kaiser},\ and\ \citenamefont {Meissner}}]{Bernard:1991rq}%
  \BibitemOpen
  \bibfield  {author} {\bibinfo {author} {\bibfnamefont {V.}~\bibnamefont {Bernard}}, \bibinfo {author} {\bibfnamefont {N.}~\bibnamefont {Kaiser}},\ and\ \bibinfo {author} {\bibfnamefont {U.~G.}\ \bibnamefont {Meissner}},\ }\bibfield  {title} {\bibinfo {title} {{Chiral expansion of the nucleon's electromagnetic polarizabilities}},\ }\href {https://doi.org/10.1103/PhysRevLett.67.1515} {\bibfield  {journal} {\bibinfo  {journal} {Phys. Rev. Lett.}\ }\textbf {\bibinfo {volume} {67}},\ \bibinfo {pages} {1515} (\bibinfo {year} {1991})}\BibitemShut {NoStop}%
\bibitem [{\citenamefont {Bernard}\ \emph {et~al.}(1992{\natexlab{a}})\citenamefont {Bernard}, \citenamefont {Kaiser},\ and\ \citenamefont {Meissner}}]{Bernard:1991ru}%
  \BibitemOpen
  \bibfield  {author} {\bibinfo {author} {\bibfnamefont {V.}~\bibnamefont {Bernard}}, \bibinfo {author} {\bibfnamefont {N.}~\bibnamefont {Kaiser}},\ and\ \bibinfo {author} {\bibfnamefont {U.~G.}\ \bibnamefont {Meissner}},\ }\bibfield  {title} {\bibinfo {title} {{Nucleons with chiral loops: Electromagnetic polarizabilities}},\ }\href {https://doi.org/10.1016/0550-3213(92)90436-F} {\bibfield  {journal} {\bibinfo  {journal} {Nucl. Phys. B}\ }\textbf {\bibinfo {volume} {373}},\ \bibinfo {pages} {346} (\bibinfo {year} {1992}{\natexlab{a}})}\BibitemShut {NoStop}%
\bibitem [{\citenamefont {Bernard}\ \emph {et~al.}(1993)\citenamefont {Bernard}, \citenamefont {Kaiser}, \citenamefont {Schmidt},\ and\ \citenamefont {Meissner}}]{Bernard:1993bg}%
  \BibitemOpen
  \bibfield  {author} {\bibinfo {author} {\bibfnamefont {V.}~\bibnamefont {Bernard}}, \bibinfo {author} {\bibfnamefont {N.}~\bibnamefont {Kaiser}}, \bibinfo {author} {\bibfnamefont {A.}~\bibnamefont {Schmidt}},\ and\ \bibinfo {author} {\bibfnamefont {U.~G.}\ \bibnamefont {Meissner}},\ }\bibfield  {title} {\bibinfo {title} {{Consistent calculation of the nucleon electromagnetic polarizabilities in chiral perturbation theory beyond next-to-leading order}},\ }\href {https://doi.org/10.1016/0370-2693(93)90813-W} {\bibfield  {journal} {\bibinfo  {journal} {Phys. Lett. B}\ }\textbf {\bibinfo {volume} {319}},\ \bibinfo {pages} {269} (\bibinfo {year} {1993})},\ \Eprint {https://arxiv.org/abs/hep-ph/9309211} {arXiv:hep-ph/9309211} \BibitemShut {NoStop}%
\bibitem [{\citenamefont {Bernard}\ \emph {et~al.}(1994)\citenamefont {Bernard}, \citenamefont {Kaiser}, \citenamefont {Meissner},\ and\ \citenamefont {Schmidt}}]{Bernard:1993ry}%
  \BibitemOpen
  \bibfield  {author} {\bibinfo {author} {\bibfnamefont {V.}~\bibnamefont {Bernard}}, \bibinfo {author} {\bibfnamefont {N.}~\bibnamefont {Kaiser}}, \bibinfo {author} {\bibfnamefont {U.~G.}\ \bibnamefont {Meissner}},\ and\ \bibinfo {author} {\bibfnamefont {A.}~\bibnamefont {Schmidt}},\ }\bibfield  {title} {\bibinfo {title} {{Aspects of nucleon Compton scattering}},\ }\href {https://doi.org/10.1007/BF01305891} {\bibfield  {journal} {\bibinfo  {journal} {Z. Phys. A}\ }\textbf {\bibinfo {volume} {348}},\ \bibinfo {pages} {317} (\bibinfo {year} {1994})},\ \Eprint {https://arxiv.org/abs/hep-ph/9311354} {arXiv:hep-ph/9311354} \BibitemShut {NoStop}%
\bibitem [{\citenamefont {Wang}\ \emph {et~al.}(2024{\natexlab{b}})\citenamefont {Wang}, \citenamefont {Zhang}, \citenamefont {Cao}, \citenamefont {Fan}, \citenamefont {Feng}, \citenamefont {Gao}, \citenamefont {Jin},\ and\ \citenamefont {Liu}}]{Wang:2023omf}%
  \BibitemOpen
  \bibfield  {author} {\bibinfo {author} {\bibfnamefont {X.-H.}\ \bibnamefont {Wang}}, \bibinfo {author} {\bibfnamefont {Z.-L.}\ \bibnamefont {Zhang}}, \bibinfo {author} {\bibfnamefont {X.-H.}\ \bibnamefont {Cao}}, \bibinfo {author} {\bibfnamefont {C.-L.}\ \bibnamefont {Fan}}, \bibinfo {author} {\bibfnamefont {X.}~\bibnamefont {Feng}}, \bibinfo {author} {\bibfnamefont {Y.-S.}\ \bibnamefont {Gao}}, \bibinfo {author} {\bibfnamefont {L.-C.}\ \bibnamefont {Jin}},\ and\ \bibinfo {author} {\bibfnamefont {C.}~\bibnamefont {Liu}},\ }\bibfield  {title} {\bibinfo {title} {{Nucleon Electric Polarizabilities and Nucleon-Pion Scattering at the Physical Pion Mass}},\ }\href {https://doi.org/10.1103/PhysRevLett.133.141901} {\bibfield  {journal} {\bibinfo  {journal} {Phys. Rev. Lett.}\ }\textbf {\bibinfo {volume} {133}},\ \bibinfo {pages} {141901} (\bibinfo {year} {2024}{\natexlab{b}})},\ \Eprint {https://arxiv.org/abs/2310.01168} {arXiv:2310.01168 [hep-lat]} \BibitemShut {NoStop}%
\bibitem [{\citenamefont {Butler}\ and\ \citenamefont {Savage}(1992)}]{Butler:1992ci}%
  \BibitemOpen
  \bibfield  {author} {\bibinfo {author} {\bibfnamefont {M.~N.}\ \bibnamefont {Butler}}\ and\ \bibinfo {author} {\bibfnamefont {M.~J.}\ \bibnamefont {Savage}},\ }\bibfield  {title} {\bibinfo {title} {{Electromagnetic polarizability of the nucleon in chiral perturbation theory}},\ }\href {https://doi.org/10.1016/0370-2693(92)91535-H} {\bibfield  {journal} {\bibinfo  {journal} {Phys. Lett. B}\ }\textbf {\bibinfo {volume} {294}},\ \bibinfo {pages} {369} (\bibinfo {year} {1992})},\ \Eprint {https://arxiv.org/abs/hep-ph/9209204} {arXiv:hep-ph/9209204} \BibitemShut {NoStop}%
\bibitem [{\citenamefont {Babusci}\ \emph {et~al.}(1997)\citenamefont {Babusci}, \citenamefont {Giordano},\ and\ \citenamefont {Matone}}]{Babusci:1996jr}%
  \BibitemOpen
  \bibfield  {author} {\bibinfo {author} {\bibfnamefont {D.}~\bibnamefont {Babusci}}, \bibinfo {author} {\bibfnamefont {G.}~\bibnamefont {Giordano}},\ and\ \bibinfo {author} {\bibfnamefont {G.}~\bibnamefont {Matone}},\ }\bibfield  {title} {\bibinfo {title} {{Chiral perturbation theory and nucleon polarizabilities}},\ }\href {https://doi.org/10.1103/PhysRevC.55.R1645} {\bibfield  {journal} {\bibinfo  {journal} {Phys. Rev. C}\ }\textbf {\bibinfo {volume} {55}},\ \bibinfo {pages} {R1645} (\bibinfo {year} {1997})}\BibitemShut {NoStop}%
\bibitem [{\citenamefont {Beane}\ \emph {et~al.}(2005)\citenamefont {Beane}, \citenamefont {Malheiro}, \citenamefont {McGovern}, \citenamefont {Phillips},\ and\ \citenamefont {van Kolck}}]{Beane:2004ra}%
  \BibitemOpen
  \bibfield  {author} {\bibinfo {author} {\bibfnamefont {S.~R.}\ \bibnamefont {Beane}}, \bibinfo {author} {\bibfnamefont {M.}~\bibnamefont {Malheiro}}, \bibinfo {author} {\bibfnamefont {J.~A.}\ \bibnamefont {McGovern}}, \bibinfo {author} {\bibfnamefont {D.~R.}\ \bibnamefont {Phillips}},\ and\ \bibinfo {author} {\bibfnamefont {U.}~\bibnamefont {van Kolck}},\ }\bibfield  {title} {\bibinfo {title} {{Compton scattering on the proton, neutron, and deuteron in chiral perturbation theory to O(Q**4)}},\ }\href {https://doi.org/10.1016/j.nuclphysa.2004.09.068} {\bibfield  {journal} {\bibinfo  {journal} {Nucl. Phys. A}\ }\textbf {\bibinfo {volume} {747}},\ \bibinfo {pages} {311} (\bibinfo {year} {2005})},\ \Eprint {https://arxiv.org/abs/nucl-th/0403088} {arXiv:nucl-th/0403088} \BibitemShut {NoStop}%
\bibitem [{\citenamefont {Choudhury}\ \emph {et~al.}(2007)\citenamefont {Choudhury}, \citenamefont {Nogga},\ and\ \citenamefont {Phillips}}]{Choudhury:2007qiz}%
  \BibitemOpen
  \bibfield  {author} {\bibinfo {author} {\bibfnamefont {D.}~\bibnamefont {Choudhury}}, \bibinfo {author} {\bibfnamefont {A.}~\bibnamefont {Nogga}},\ and\ \bibinfo {author} {\bibfnamefont {D.~R.}\ \bibnamefont {Phillips}},\ }\bibfield  {title} {\bibinfo {title} {{Investigating neutron polarizabilities through Compton scattering on $^{3}$He}},\ }\href {https://doi.org/10.1103/PhysRevLett.98.232303} {\bibfield  {journal} {\bibinfo  {journal} {Phys. Rev. Lett.}\ }\textbf {\bibinfo {volume} {98}},\ \bibinfo {pages} {232303} (\bibinfo {year} {2007})},\ \bibinfo {note} {[Erratum: Phys.Rev.Lett. 120, 249901 (2018)]},\ \Eprint {https://arxiv.org/abs/1804.01206} {arXiv:1804.01206 [nucl-th]} \BibitemShut {NoStop}%
\bibitem [{\citenamefont {Lensky}\ and\ \citenamefont {McGovern}(2014)}]{Lensky:2014efa}%
  \BibitemOpen
  \bibfield  {author} {\bibinfo {author} {\bibfnamefont {V.}~\bibnamefont {Lensky}}\ and\ \bibinfo {author} {\bibfnamefont {J.~A.}\ \bibnamefont {McGovern}},\ }\bibfield  {title} {\bibinfo {title} {{Proton polarizabilities from Compton data using covariant chiral effective field theory}},\ }\href {https://doi.org/10.1103/PhysRevC.89.032202} {\bibfield  {journal} {\bibinfo  {journal} {Phys. Rev. C}\ }\textbf {\bibinfo {volume} {89}},\ \bibinfo {pages} {032202} (\bibinfo {year} {2014})},\ \Eprint {https://arxiv.org/abs/1401.3320} {arXiv:1401.3320 [nucl-th]} \BibitemShut {NoStop}%
\bibitem [{\citenamefont {Lensky}\ \emph {et~al.}(2015)\citenamefont {Lensky}, \citenamefont {McGovern},\ and\ \citenamefont {Pascalutsa}}]{Lensky:2015awa}%
  \BibitemOpen
  \bibfield  {author} {\bibinfo {author} {\bibfnamefont {V.}~\bibnamefont {Lensky}}, \bibinfo {author} {\bibfnamefont {J.}~\bibnamefont {McGovern}},\ and\ \bibinfo {author} {\bibfnamefont {V.}~\bibnamefont {Pascalutsa}},\ }\bibfield  {title} {\bibinfo {title} {{Predictions of covariant chiral perturbation theory for nucleon polarisabilities and polarised Compton scattering}},\ }\href {https://doi.org/10.1140/epjc/s10052-015-3791-0} {\bibfield  {journal} {\bibinfo  {journal} {Eur. Phys. J. C}\ }\textbf {\bibinfo {volume} {75}},\ \bibinfo {pages} {604} (\bibinfo {year} {2015})},\ \Eprint {https://arxiv.org/abs/1510.02794} {arXiv:1510.02794 [hep-ph]} \BibitemShut {NoStop}%
\bibitem [{\citenamefont {Th\"urmann}\ \emph {et~al.}(2021)\citenamefont {Th\"urmann}, \citenamefont {Epelbaum}, \citenamefont {Gasparyan},\ and\ \citenamefont {Krebs}}]{Thurmann:2020mog}%
  \BibitemOpen
  \bibfield  {author} {\bibinfo {author} {\bibfnamefont {M.}~\bibnamefont {Th\"urmann}}, \bibinfo {author} {\bibfnamefont {E.}~\bibnamefont {Epelbaum}}, \bibinfo {author} {\bibfnamefont {A.~M.}\ \bibnamefont {Gasparyan}},\ and\ \bibinfo {author} {\bibfnamefont {H.}~\bibnamefont {Krebs}},\ }\bibfield  {title} {\bibinfo {title} {{Nucleon polarizabilities in covariant baryon chiral perturbation theory with explicit $\Delta$ degrees of freedom}},\ }\href {https://doi.org/10.1103/PhysRevC.103.035201} {\bibfield  {journal} {\bibinfo  {journal} {Phys. Rev. C}\ }\textbf {\bibinfo {volume} {103}},\ \bibinfo {pages} {035201} (\bibinfo {year} {2021})},\ \Eprint {https://arxiv.org/abs/2007.08438} {arXiv:2007.08438 [nucl-th]} \BibitemShut {NoStop}%
\bibitem [{\citenamefont {Chen}\ \emph {et~al.}(2025)\citenamefont {Chen}, \citenamefont {Wen}, \citenamefont {Meng},\ and\ \citenamefont {Zhu}}]{Chen:2024xks}%
  \BibitemOpen
  \bibfield  {author} {\bibinfo {author} {\bibfnamefont {Y.-K.}\ \bibnamefont {Chen}}, \bibinfo {author} {\bibfnamefont {L.-Z.}\ \bibnamefont {Wen}}, \bibinfo {author} {\bibfnamefont {L.}~\bibnamefont {Meng}},\ and\ \bibinfo {author} {\bibfnamefont {S.-L.}\ \bibnamefont {Zhu}},\ }\bibfield  {title} {\bibinfo {title} {{Electromagnetic polarizabilities of the spin-12 singly heavy baryons in heavy baryon chiral perturbation theory}},\ }\href {https://doi.org/10.1103/PhysRevD.111.054019} {\bibfield  {journal} {\bibinfo  {journal} {Phys. Rev. D}\ }\textbf {\bibinfo {volume} {111}},\ \bibinfo {pages} {054019} (\bibinfo {year} {2025})},\ \Eprint {https://arxiv.org/abs/2412.02297} {arXiv:2412.02297 [hep-ph]} \BibitemShut {NoStop}%
\bibitem [{\citenamefont {Rarita}\ and\ \citenamefont {Schwinger}(1941)}]{Rarita:1941mf}%
  \BibitemOpen
  \bibfield  {author} {\bibinfo {author} {\bibfnamefont {W.}~\bibnamefont {Rarita}}\ and\ \bibinfo {author} {\bibfnamefont {J.}~\bibnamefont {Schwinger}},\ }\bibfield  {title} {\bibinfo {title} {{On a theory of particles with half integral spin}},\ }\href {https://doi.org/10.1103/PhysRev.60.61} {\bibfield  {journal} {\bibinfo  {journal} {Phys. Rev.}\ }\textbf {\bibinfo {volume} {60}},\ \bibinfo {pages} {61} (\bibinfo {year} {1941})}\BibitemShut {NoStop}%
\bibitem [{\citenamefont {Hemmert}\ \emph {et~al.}(1997)\citenamefont {Hemmert}, \citenamefont {Holstein},\ and\ \citenamefont {Kambor}}]{Hemmert:1996rw}%
  \BibitemOpen
  \bibfield  {author} {\bibinfo {author} {\bibfnamefont {T.~R.}\ \bibnamefont {Hemmert}}, \bibinfo {author} {\bibfnamefont {B.~R.}\ \bibnamefont {Holstein}},\ and\ \bibinfo {author} {\bibfnamefont {J.}~\bibnamefont {Kambor}},\ }\bibfield  {title} {\bibinfo {title} {{Delta (1232) and the polarizabilities of the nucleon}},\ }\href {https://doi.org/10.1103/PhysRevD.55.5598} {\bibfield  {journal} {\bibinfo  {journal} {Phys. Rev. D}\ }\textbf {\bibinfo {volume} {55}},\ \bibinfo {pages} {5598} (\bibinfo {year} {1997})},\ \Eprint {https://arxiv.org/abs/hep-ph/9612374} {arXiv:hep-ph/9612374} \BibitemShut {NoStop}%
\bibitem [{\citenamefont {Bernabeu}\ and\ \citenamefont {Tarrach}(1976)}]{Bernabeu:1976jq}%
  \BibitemOpen
  \bibfield  {author} {\bibinfo {author} {\bibfnamefont {J.}~\bibnamefont {Bernabeu}}\ and\ \bibinfo {author} {\bibfnamefont {R.}~\bibnamefont {Tarrach}},\ }\bibfield  {title} {\bibinfo {title} {{Long Range Potentials and the Electromagnetic Polarizabilities}},\ }\href {https://doi.org/10.1016/0003-4916(76)90265-7} {\bibfield  {journal} {\bibinfo  {journal} {Annals Phys.}\ }\textbf {\bibinfo {volume} {102}},\ \bibinfo {pages} {323} (\bibinfo {year} {1976})}\BibitemShut {NoStop}%
\bibitem [{\citenamefont {Llanta}\ and\ \citenamefont {Tarrach}(1980)}]{Llanta:1979kj}%
  \BibitemOpen
  \bibfield  {author} {\bibinfo {author} {\bibfnamefont {E.}~\bibnamefont {Llanta}}\ and\ \bibinfo {author} {\bibfnamefont {R.}~\bibnamefont {Tarrach}},\ }\bibfield  {title} {\bibinfo {title} {{Pion Electromagnetic Polarizabilities and Quarks}},\ }\href {https://doi.org/10.1016/0370-2693(80)90678-4} {\bibfield  {journal} {\bibinfo  {journal} {Phys. Lett. B}\ }\textbf {\bibinfo {volume} {91}},\ \bibinfo {pages} {132} (\bibinfo {year} {1980})}\BibitemShut {NoStop}%
\bibitem [{\citenamefont {Bernard}\ \emph {et~al.}(1995)\citenamefont {Bernard}, \citenamefont {Kaiser},\ and\ \citenamefont {Meissner}}]{Bernard:1995dp}%
  \BibitemOpen
  \bibfield  {author} {\bibinfo {author} {\bibfnamefont {V.}~\bibnamefont {Bernard}}, \bibinfo {author} {\bibfnamefont {N.}~\bibnamefont {Kaiser}},\ and\ \bibinfo {author} {\bibfnamefont {U.-G.}\ \bibnamefont {Meissner}},\ }\bibfield  {title} {\bibinfo {title} {{Chiral dynamics in nucleons and nuclei}},\ }\href {https://doi.org/10.1142/S0218301395000092} {\bibfield  {journal} {\bibinfo  {journal} {Int. J. Mod. Phys. E}\ }\textbf {\bibinfo {volume} {4}},\ \bibinfo {pages} {193} (\bibinfo {year} {1995})},\ \Eprint {https://arxiv.org/abs/hep-ph/9501384} {arXiv:hep-ph/9501384} \BibitemShut {NoStop}%
\bibitem [{\citenamefont {Scherer}(2003)}]{Scherer:2002tk}%
  \BibitemOpen
  \bibfield  {author} {\bibinfo {author} {\bibfnamefont {S.}~\bibnamefont {Scherer}},\ }\bibfield  {title} {\bibinfo {title} {{Introduction to chiral perturbation theory}},\ }\href@noop {} {\bibfield  {journal} {\bibinfo  {journal} {Adv. Nucl. Phys.}\ }\textbf {\bibinfo {volume} {27}},\ \bibinfo {pages} {277} (\bibinfo {year} {2003})},\ \Eprint {https://arxiv.org/abs/hep-ph/0210398} {arXiv:hep-ph/0210398} \BibitemShut {NoStop}%
\bibitem [{\citenamefont {Yan}\ \emph {et~al.}(1992)\citenamefont {Yan}, \citenamefont {Cheng}, \citenamefont {Cheung}, \citenamefont {Lin}, \citenamefont {Lin},\ and\ \citenamefont {Yu}}]{Yan:1992gz}%
  \BibitemOpen
  \bibfield  {author} {\bibinfo {author} {\bibfnamefont {T.-M.}\ \bibnamefont {Yan}}, \bibinfo {author} {\bibfnamefont {H.-Y.}\ \bibnamefont {Cheng}}, \bibinfo {author} {\bibfnamefont {C.-Y.}\ \bibnamefont {Cheung}}, \bibinfo {author} {\bibfnamefont {G.-L.}\ \bibnamefont {Lin}}, \bibinfo {author} {\bibfnamefont {Y.~C.}\ \bibnamefont {Lin}},\ and\ \bibinfo {author} {\bibfnamefont {H.-L.}\ \bibnamefont {Yu}},\ }\bibfield  {title} {\bibinfo {title} {{Heavy quark symmetry and chiral dynamics}},\ }\href {https://doi.org/10.1103/PhysRevD.46.1148} {\bibfield  {journal} {\bibinfo  {journal} {Phys. Rev. D}\ }\textbf {\bibinfo {volume} {46}},\ \bibinfo {pages} {1148} (\bibinfo {year} {1992})},\ \bibinfo {note} {[Erratum: Phys.Rev.D 55, 5851 (1997)]}\BibitemShut {NoStop}%
\bibitem [{\citenamefont {Ou-Yang}\ and\ \citenamefont {Huang}(2024)}]{Ou-Yang:2024neq}%
  \BibitemOpen
  \bibfield  {author} {\bibinfo {author} {\bibfnamefont {J.}~\bibnamefont {Ou-Yang}}\ and\ \bibinfo {author} {\bibfnamefont {B.-L.}\ \bibnamefont {Huang}},\ }\bibfield  {title} {\bibinfo {title} {{Pion-nucleon scattering with decuplet contribution in heavy baryon SU(3) chiral perturbation theory}},\ }\href {https://doi.org/10.1103/PhysRevD.110.116008} {\bibfield  {journal} {\bibinfo  {journal} {Phys. Rev. D}\ }\textbf {\bibinfo {volume} {110}},\ \bibinfo {pages} {116008} (\bibinfo {year} {2024})},\ \Eprint {https://arxiv.org/abs/2409.09388} {arXiv:2409.09388 [nucl-th]} \BibitemShut {NoStop}%
\bibitem [{\citenamefont {Wang}\ \emph {et~al.}(2018)\citenamefont {Wang}, \citenamefont {Meng}, \citenamefont {Li}, \citenamefont {Liu},\ and\ \citenamefont {Zhu}}]{Wang:2018gpl}%
  \BibitemOpen
  \bibfield  {author} {\bibinfo {author} {\bibfnamefont {G.-J.}\ \bibnamefont {Wang}}, \bibinfo {author} {\bibfnamefont {L.}~\bibnamefont {Meng}}, \bibinfo {author} {\bibfnamefont {H.-S.}\ \bibnamefont {Li}}, \bibinfo {author} {\bibfnamefont {Z.-W.}\ \bibnamefont {Liu}},\ and\ \bibinfo {author} {\bibfnamefont {S.-L.}\ \bibnamefont {Zhu}},\ }\bibfield  {title} {\bibinfo {title} {{Magnetic moments of the spin-$\frac{1}{2}$ singly charmed baryons in chiral perturbation theory}},\ }\href {https://doi.org/10.1103/PhysRevD.98.054026} {\bibfield  {journal} {\bibinfo  {journal} {Phys. Rev. D}\ }\textbf {\bibinfo {volume} {98}},\ \bibinfo {pages} {054026} (\bibinfo {year} {2018})},\ \Eprint {https://arxiv.org/abs/1803.00229} {arXiv:1803.00229 [hep-ph]} \BibitemShut {NoStop}%
\bibitem [{\citenamefont {Bernard}\ \emph {et~al.}(1992{\natexlab{b}})\citenamefont {Bernard}, \citenamefont {Kaiser}, \citenamefont {Kambor},\ and\ \citenamefont {Meissner}}]{Bernard:1992qa}%
  \BibitemOpen
  \bibfield  {author} {\bibinfo {author} {\bibfnamefont {V.}~\bibnamefont {Bernard}}, \bibinfo {author} {\bibfnamefont {N.}~\bibnamefont {Kaiser}}, \bibinfo {author} {\bibfnamefont {J.}~\bibnamefont {Kambor}},\ and\ \bibinfo {author} {\bibfnamefont {U.~G.}\ \bibnamefont {Meissner}},\ }\bibfield  {title} {\bibinfo {title} {{Chiral structure of the nucleon}},\ }\href {https://doi.org/10.1016/0550-3213(92)90615-I} {\bibfield  {journal} {\bibinfo  {journal} {Nucl. Phys. B}\ }\textbf {\bibinfo {volume} {388}},\ \bibinfo {pages} {315} (\bibinfo {year} {1992}{\natexlab{b}})}\BibitemShut {NoStop}%
\bibitem [{\citenamefont {Schumacher}(2005)}]{Schumacher:2005an}%
  \BibitemOpen
  \bibfield  {author} {\bibinfo {author} {\bibfnamefont {M.}~\bibnamefont {Schumacher}},\ }\bibfield  {title} {\bibinfo {title} {{Polarizability of the nucleon and Compton scattering}},\ }\href {https://doi.org/10.1016/j.ppnp.2005.01.033} {\bibfield  {journal} {\bibinfo  {journal} {Prog. Part. Nucl. Phys.}\ }\textbf {\bibinfo {volume} {55}},\ \bibinfo {pages} {567} (\bibinfo {year} {2005})},\ \Eprint {https://arxiv.org/abs/hep-ph/0501167} {arXiv:hep-ph/0501167} \BibitemShut {NoStop}%
\bibitem [{\citenamefont {Holstein}\ and\ \citenamefont {Scherer}(2014)}]{Holstein:2013kia}%
  \BibitemOpen
  \bibfield  {author} {\bibinfo {author} {\bibfnamefont {B.~R.}\ \bibnamefont {Holstein}}\ and\ \bibinfo {author} {\bibfnamefont {S.}~\bibnamefont {Scherer}},\ }\bibfield  {title} {\bibinfo {title} {{Hadron Polarizabilities}},\ }\href {https://doi.org/10.1146/annurev-nucl-102313-025555} {\bibfield  {journal} {\bibinfo  {journal} {Ann. Rev. Nucl. Part. Sci.}\ }\textbf {\bibinfo {volume} {64}},\ \bibinfo {pages} {51} (\bibinfo {year} {2014})},\ \Eprint {https://arxiv.org/abs/1401.0140} {arXiv:1401.0140 [hep-ph]} \BibitemShut {NoStop}%
\bibitem [{\citenamefont {Pascalutsa}\ and\ \citenamefont {Phillips}(2003)}]{Pascalutsa:2002pi}%
  \BibitemOpen
  \bibfield  {author} {\bibinfo {author} {\bibfnamefont {V.}~\bibnamefont {Pascalutsa}}\ and\ \bibinfo {author} {\bibfnamefont {D.~R.}\ \bibnamefont {Phillips}},\ }\bibfield  {title} {\bibinfo {title} {{Effective theory of the delta(1232) in Compton scattering off the nucleon}},\ }\href {https://doi.org/10.1103/PhysRevC.67.055202} {\bibfield  {journal} {\bibinfo  {journal} {Phys. Rev. C}\ }\textbf {\bibinfo {volume} {67}},\ \bibinfo {pages} {055202} (\bibinfo {year} {2003})},\ \Eprint {https://arxiv.org/abs/nucl-th/0212024} {arXiv:nucl-th/0212024} \BibitemShut {NoStop}%
\bibitem [{\citenamefont {Butler}\ \emph {et~al.}(1993)\citenamefont {Butler}, \citenamefont {Savage},\ and\ \citenamefont {Springer}}]{Butler:1992pn}%
  \BibitemOpen
  \bibfield  {author} {\bibinfo {author} {\bibfnamefont {M.~N.}\ \bibnamefont {Butler}}, \bibinfo {author} {\bibfnamefont {M.~J.}\ \bibnamefont {Savage}},\ and\ \bibinfo {author} {\bibfnamefont {R.~P.}\ \bibnamefont {Springer}},\ }\bibfield  {title} {\bibinfo {title} {{Strong and electromagnetic decays of the baryon decuplet}},\ }\href {https://doi.org/10.1016/0550-3213(93)90617-X} {\bibfield  {journal} {\bibinfo  {journal} {Nucl. Phys. B}\ }\textbf {\bibinfo {volume} {399}},\ \bibinfo {pages} {69} (\bibinfo {year} {1993})},\ \Eprint {https://arxiv.org/abs/hep-ph/9211247} {arXiv:hep-ph/9211247} \BibitemShut {NoStop}%
\bibitem [{\citenamefont {Fettes}\ and\ \citenamefont {Meissner}(2001)}]{Fettes:2000bb}%
  \BibitemOpen
  \bibfield  {author} {\bibinfo {author} {\bibfnamefont {N.}~\bibnamefont {Fettes}}\ and\ \bibinfo {author} {\bibfnamefont {U.~G.}\ \bibnamefont {Meissner}},\ }\bibfield  {title} {\bibinfo {title} {{Pion - nucleon scattering in an effective chiral field theory with explicit spin 3/2 fields}},\ }\href {https://doi.org/10.1016/S0375-9474(00)00368-7} {\bibfield  {journal} {\bibinfo  {journal} {Nucl. Phys. A}\ }\textbf {\bibinfo {volume} {679}},\ \bibinfo {pages} {629} (\bibinfo {year} {2001})},\ \Eprint {https://arxiv.org/abs/hep-ph/0006299} {arXiv:hep-ph/0006299} \BibitemShut {NoStop}%
\bibitem [{\citenamefont {Wang}\ \emph {et~al.}(2019)\citenamefont {Wang}, \citenamefont {Meng},\ and\ \citenamefont {Zhu}}]{Wang:2018cre}%
  \BibitemOpen
  \bibfield  {author} {\bibinfo {author} {\bibfnamefont {G.-J.}\ \bibnamefont {Wang}}, \bibinfo {author} {\bibfnamefont {L.}~\bibnamefont {Meng}},\ and\ \bibinfo {author} {\bibfnamefont {S.-L.}\ \bibnamefont {Zhu}},\ }\bibfield  {title} {\bibinfo {title} {{Radiative decays of the singly heavy baryons in chiral perturbation theory}},\ }\href {https://doi.org/10.1103/PhysRevD.99.034021} {\bibfield  {journal} {\bibinfo  {journal} {Phys. Rev. D}\ }\textbf {\bibinfo {volume} {99}},\ \bibinfo {pages} {034021} (\bibinfo {year} {2019})},\ \Eprint {https://arxiv.org/abs/1811.06208} {arXiv:1811.06208 [hep-ph]} \BibitemShut {NoStop}%
\bibitem [{\citenamefont {Navas}\ \emph {et~al.}(2024)\citenamefont {Navas} \emph {et~al.}}]{ParticleDataGroup:2024cfk}%
  \BibitemOpen
  \bibfield  {author} {\bibinfo {author} {\bibfnamefont {S.}~\bibnamefont {Navas}} \emph {et~al.} (\bibinfo {collaboration} {Particle Data Group}),\ }\bibfield  {title} {\bibinfo {title} {{Review of particle physics}},\ }\href {https://doi.org/10.1103/PhysRevD.110.030001} {\bibfield  {journal} {\bibinfo  {journal} {Phys. Rev. D}\ }\textbf {\bibinfo {volume} {110}},\ \bibinfo {pages} {030001} (\bibinfo {year} {2024})}\BibitemShut {NoStop}%
\bibitem [{\citenamefont {Meng}\ \emph {et~al.}(2018)\citenamefont {Meng}, \citenamefont {Wang}, \citenamefont {Leng}, \citenamefont {Liu},\ and\ \citenamefont {Zhu}}]{Meng:2018gan}%
  \BibitemOpen
  \bibfield  {author} {\bibinfo {author} {\bibfnamefont {L.}~\bibnamefont {Meng}}, \bibinfo {author} {\bibfnamefont {G.-J.}\ \bibnamefont {Wang}}, \bibinfo {author} {\bibfnamefont {C.-Z.}\ \bibnamefont {Leng}}, \bibinfo {author} {\bibfnamefont {Z.-W.}\ \bibnamefont {Liu}},\ and\ \bibinfo {author} {\bibfnamefont {S.-L.}\ \bibnamefont {Zhu}},\ }\bibfield  {title} {\bibinfo {title} {{Magnetic moments of the spin-${3\over 2}$ singly heavy baryons}},\ }\href {https://doi.org/10.1103/PhysRevD.98.094013} {\bibfield  {journal} {\bibinfo  {journal} {Phys. Rev. D}\ }\textbf {\bibinfo {volume} {98}},\ \bibinfo {pages} {094013} (\bibinfo {year} {2018})},\ \Eprint {https://arxiv.org/abs/1805.09580} {arXiv:1805.09580 [hep-ph]} \BibitemShut {NoStop}%
\bibitem [{\citenamefont {Cheng}\ \emph {et~al.}(1994)\citenamefont {Cheng}, \citenamefont {Cheung}, \citenamefont {Lin}, \citenamefont {Lin}, \citenamefont {Yan},\ and\ \citenamefont {Yu}}]{Cheng:1993kp}%
  \BibitemOpen
  \bibfield  {author} {\bibinfo {author} {\bibfnamefont {H.-Y.}\ \bibnamefont {Cheng}}, \bibinfo {author} {\bibfnamefont {C.-Y.}\ \bibnamefont {Cheung}}, \bibinfo {author} {\bibfnamefont {G.-L.}\ \bibnamefont {Lin}}, \bibinfo {author} {\bibfnamefont {Y.~C.}\ \bibnamefont {Lin}}, \bibinfo {author} {\bibfnamefont {T.-M.}\ \bibnamefont {Yan}},\ and\ \bibinfo {author} {\bibfnamefont {H.-L.}\ \bibnamefont {Yu}},\ }\bibfield  {title} {\bibinfo {title} {{Corrections to chiral dynamics of heavy hadrons: SU(3) symmetry breaking}},\ }\href {https://doi.org/10.1103/PhysRevD.49.5857} {\bibfield  {journal} {\bibinfo  {journal} {Phys. Rev. D}\ }\textbf {\bibinfo {volume} {49}},\ \bibinfo {pages} {5857} (\bibinfo {year} {1994})},\ \bibinfo {note} {[Erratum: Phys.Rev.D 55, 5851--5852 (1997)]},\ \Eprint {https://arxiv.org/abs/hep-ph/9312304} {arXiv:hep-ph/9312304} \BibitemShut {NoStop}%
\bibitem [{\citenamefont {Cho}\ and\ \citenamefont {Georgi}(1992)}]{Cho:1992nt}%
  \BibitemOpen
  \bibfield  {author} {\bibinfo {author} {\bibfnamefont {P.~L.}\ \bibnamefont {Cho}}\ and\ \bibinfo {author} {\bibfnamefont {H.}~\bibnamefont {Georgi}},\ }\bibfield  {title} {\bibinfo {title} {{Electromagnetic interactions in heavy hadron chiral theory}},\ }\href {https://doi.org/10.1016/0370-2693(92)91340-F} {\bibfield  {journal} {\bibinfo  {journal} {Phys. Lett. B}\ }\textbf {\bibinfo {volume} {296}},\ \bibinfo {pages} {408} (\bibinfo {year} {1992})},\ \bibinfo {note} {[Erratum: Phys.Lett.B 300, 410 (1993)]},\ \Eprint {https://arxiv.org/abs/hep-ph/9209239} {arXiv:hep-ph/9209239} \BibitemShut {NoStop}%
\bibitem [{\citenamefont {Jiang}\ \emph {et~al.}(2015)\citenamefont {Jiang}, \citenamefont {Chen},\ and\ \citenamefont {Zhu}}]{Jiang:2015xqa}%
  \BibitemOpen
  \bibfield  {author} {\bibinfo {author} {\bibfnamefont {N.}~\bibnamefont {Jiang}}, \bibinfo {author} {\bibfnamefont {X.-L.}\ \bibnamefont {Chen}},\ and\ \bibinfo {author} {\bibfnamefont {S.-L.}\ \bibnamefont {Zhu}},\ }\bibfield  {title} {\bibinfo {title} {{Electromagnetic decays of the charmed and bottom baryons in chiral perturbation theory}},\ }\href {https://doi.org/10.1103/PhysRevD.92.054017} {\bibfield  {journal} {\bibinfo  {journal} {Phys. Rev. D}\ }\textbf {\bibinfo {volume} {92}},\ \bibinfo {pages} {054017} (\bibinfo {year} {2015})},\ \Eprint {https://arxiv.org/abs/1505.02999} {arXiv:1505.02999 [hep-ph]} \BibitemShut {NoStop}%
\end{thebibliography}%
\end{document}